\documentclass[useAMS,usenatbib,usegraphicx]{mn2e}
\usepackage{psfig}
\usepackage{times}
\usepackage{aas_macros}

\newcommand{\sersic}{S\'ersic}
\newcommand{\magarc}{{mag\ arcsec$^{-2}$}}

\title[STAGES:  the Space Telescope A901/2 Galaxy Evolution
  Survey]{STAGES:  the Space Telescope A901/2 Galaxy Evolution Survey}
\author[M. E. Gray et al.]{Meghan E. Gray$^1$\thanks{email:  Meghan.Gray@nottingham.ac.uk},
Christian Wolf$^2$,
Marco Barden$^3$,
Chien Y. Peng$^{4,5}$,
Boris H\"au\ss ler$^1$,\newauthor
Eric F. Bell$^6$,
Daniel H. McIntosh$^{7,8}$,
Yicheng Guo$^7$,
John A.R. Caldwell$^9$,
David Bacon$^{10}$,\newauthor 
Michael Balogh$^{11}$, 
Fabio D. Barazza$^{12}$, 
Asmus B\"ohm$^{13}$, 
Catherine Heymans$^{14,15}$,\newauthor 
Knud Jahnke$^{6}$,
Shardha Jogee$^{16}$,
Eelco van Kampen$^{3,17}$,
Kyle Lane$^1$,
Klaus Meisenheimer$^{6}$,\newauthor
Sebastian F. S\'anchez$^{18}$,
Andy Taylor$^{19}$,
Lutz Wisotzki$^{14}$,
Xianzhong Zheng$^{19}$,
David A. Green$^{20}$,\newauthor
R.J. Beswick$^{21}$,
D.J.Saikia$^{22}$,
Rachel Gilmour$^{23}$, 
Benjamin D. Johnson$^{24}$, \&
Casey Papovich$^{25}$\\
$^1$School of Physics and Astronomy, The University of Nottingham,
University Park, Nottingham NG7 2RD, UK.\\
$^2$Department of Astrophysics, Denys Wilkinson Building, University of
    Oxford, Keble Road, Oxford, OX1 3RH, UK.\\
$^3$Institute for Astro- and Particle Physics, University of
Innsbruck, Technikerstr. 25/8, A-6020 Innsbruck,
   Austria. \\
$^4$NRC Herzberg Institute of Astrophysics, 5071 West Saanich Road,
Victoria, V9E 2E7, Canada.\\
$^5$Space Telescope Science Institute, 3700 San Martin Drive, Baltimore, MD
    21218, USA.\\
$^6$Max-Planck-Institut f\"{u}r Astronomie, K\"{o}nigstuhl 17, D-69117,
Heidelberg, Germany.\\
$^7$Department of Astronomy, University of Massachusetts, 710 North
Pleasant Street, Amherst, MA 01003, USA.\\
$^8$Department of Physics, 5110 Rockhill Road, University of
Missouri-Kansas City, Kansas City, MO 64110, USA\\
$^9$University of Texas, McDonald Observatory, Fort Davis, TX 79734,
USA.\\
$^{10}$Institute of Cosmology and Gravitation, University of Portsmouth,
Hampshire Terrace, Portsmouth, PO1 2EG, UK. \\
$^{11}$Department of Physics and Astronomy, University Of Waterloo, Waterloo,
Ontario, N2L 3G1, Canada.\\
$^{12}$Laboratoire d'Astrophysique, \'Ecole Polytechnique F\'ed\'erale de
Lausanne (EPFL), Observatoire de Sauverny, CH-1290 Versoix, Switzerland\\
$^{13}$Astrophysikalisches Insitut Potsdam, An der Sternwarte 16, D-14482
Potsdam, Germany.\\
$^{14}$Department of Physics and Astronomy, University of British Columbia, 6224
Agricultural Road, Vancouver, V6T 1Z1, Canada.\\
$^{15}$The Scottish Universities Physics Alliance (SUPA), Institute for
Astronomy, University of Edinburgh, Blackford Hill, Edinburgh, EH9 3HJ,
UK.\\
$^{16}$Department of Astronomy, University of Texas at Austin, 1 University
    Station, C1400 Austin, TX 78712-0259, USA.\\
$^{17}$European Southern Observatory, Karl-Schwarzschild-Strasse 2,  D-85748 Garching bei Muenchen, Germany\\
$^{18}$Centro Hispano Aleman de Calar Alto, C/Jesus Durban Remon 2-2, E-04004
    Ameria, Spain.\\
$^{19}$Purple Mountain Observatory, National Astronomical Observatories,
Chinese Academy of Sciences, Nanjing 210008, PR China.\\
$^{20}$Cavendish Laboratory, 19 J.J. Thomson Avenue, Cambridge, CB3 0HE, UK\\
$^{21}$ Jodrell Bank Centre for Astrophysics, Department of Physics \& Astronomy, The University of
  Manchester, Oxford Road, Manchester, M13 9PL, UK\\
$^{22}$National Centre for Radio Astrophysics, TIFR, Pune University
  Campus, Post Bag 3, Pune 411 007, India\\
$^{23}$European Southern Observatory, Alonso de Cordova 3107,
  Vitacura, Casilla 19001, Santiago 19, Chile\\
$^{24}$Institute of Astronomy, Madingley Road, Cambridge CB3 0HA, UK\\
$^{25}$Department of Physics, Texas A\&M University, College Station,
  TX 77843 USA}
\begin{document}

\date{}

\pagerange{\pageref{firstpage}--\pageref{lastpage}} \pubyear{}

\maketitle

\label{firstpage}

\begin{abstract}
We present an overview of the Space Telescope A901/2 Galaxy Evolution
Survey (STAGES).  STAGES is a multiwavelength project designed to
probe physical drivers of galaxy evolution across a wide range of
environments and luminosity.  A complex multi-cluster system at
$z\sim0.165$ has been the subject of an 80-orbit F606W HST/ACS mosaic
covering the full $0.5\degr \times 0.5\degr$ ($\sim$5$\times$5
Mpc$^{2}$) span of the supercluster.  Extensive multiwavelength
observations with XMM-Newton, GALEX, Spitzer, 2dF, GMRT, and the
17-band COMBO-17 photometric redshift survey complement the HST
imaging.  Our survey goals include simultaneously linking galaxy
morphology with other observables such as age, star-formation rate,
nuclear activity, and stellar mass.  In addition, with the
multiwavelength dataset and new high resolution mass maps from
gravitational lensing, we are able to disentangle the large-scale
structure of the system.  By examining all aspects of environment we
will be able to evaluate the relative importance of the dark matter
halos, the local galaxy density, and the hot X-ray gas in driving
galaxy transformation.  This paper describes the HST imaging, data
reduction, and creation of a master catalogue. We perform \sersic\
fitting on the HST images and conduct associated simulations to
quantify completeness.  In addition, we present the COMBO-17
photometric redshift catalogue and estimates of stellar masses and
star-formation rates for this field.  We define galaxy and cluster
sample selection criteria which will be the basis for forthcoming
science analyses, and present a compilation of notable objects in the
field.  Finally, we describe the further multiwavelength observations
and announce public access to the data and catalogues.

\end{abstract}


\section{Survey motivation}

\subsection{A multiwavelength approach to galaxy evolution as a function of environment}

The precise role that environment plays in shaping galaxy evolution is
a hotly debated topic.  Trends to passive and/or more spheroidal
populations in dense environments are widely observed: galaxy
morphology \citep{dressler80,dressler97,goto03,treu03},
colour \citep{kodama01,blanton05,baldry06}, star-formation rate
\citep{gomez03,lewis02}, and stellar age and AGN fraction
\citep{kauffmann04} all correlate with measurements of the local
galaxy density.  Furthermore, these relations persist over a wide
range of redshift \citep{smith05, cooper07} and density
\citep{balogh04}.

Disentangling the relative importance of internal and external
physical mechanisms responsible for these relations is challenging.
It is natural to expect that high density environments will
preferentially host older stellar populations.  Hierarchical models of
galaxy formation \citep[e.g.][]{delucia06} suggest that galaxies in
the highest density peaks started forming stars and assembling mass
earlier: in essence they have a head-start.  Simultaneously, galaxies
forming in high-density environments will have more time to experience
the {\it external} influence of their local environment.  Those
processes will also act on infalling galaxies as they are continuously
accreted into larger haloes.  There are many plausible physical
mechanisms by which a galaxy could be transformed by its environment:
removal of the hot \citep{larson80} or cold \citep{gunn72} gas supply
through ram-pressure stripping; tidal effects leading to halo
truncation \citep{bekki99} or triggered star formation through gas
compression \citep{fujita98}; interactions between galaxies themselves
via low-speed major mergers \citep{barnes92} or frequent impulsive
encounters termed `harrassment' \citep{moore98}.

 Though some of the above mechanisms are largely cluster-specific
(e.g. ram-pressure stripping requires interaction with a hot
intracluster medium), it is also increasingly clear that low density
environments such as galaxy groups are important sites for galaxy
evolution \citep{balogh04,Zabludoff96}.  Additionally, luminosity (or
more directly, mass) is also critical in regulating how susceptible a
galaxy is to external influences.  For example, \citet{haines06} find
that in low density environments in the SDSS the fraction of passive
galaxies is a strong function of luminosity.  They find a complete
absence of passive dwarf galaxies in the lowest density regions (i.e.,
while luminous passive galaxies can occur in all environments,
low-luminosity passive galaxies can only occur in dense environments).

Understanding the full degree of transformation is further complicated
by the amount of dust-obscured star formation that may or may not be
present.  Many studies in the radio and MIR
\citep{millerowen03,coia05,gallazzi08} have shown that an optical
census of star formation can underestimate the true
rate. Cluster-cluster variations are strong, with induced star
formation linked to dynamically-disturbed large-scale structure
\citep{geach06}.  Nor are changes in morphology necessarily equivalent
to changes in star formation.  There is no guarantee that external
processes causing an increase or decrease in the star-formation rate
act on the same timescale, to the same degree, or in the same regime
as those responsible for structural changes.  A full census of
star-formation, AGN activity, and morphology therefore requires a
comprehensive view of galaxies, including multiwavelength coverage and
high resolution imaging. These are the aims of the STAGES project
described in this paper, targetting the Abell 901(a,b)/902 multiple
cluster system (hearafter A901/2) at $z\sim 0.165$.

In addition to the STAGES coverage of A901/2, there are several other
multiwavelength projects taking a similar approach to targeting
large-scale structures.  While we will argue below that STAGES
occupies a particular niche, the following is a (non-exhaustive) list
of surveys of large-scale structure including substantial HST imaging.
All are complementary to STAGES by way of the redshift range or
dynamical state probed.  The COSMOS survey has examined the evolution
of the morphology-density relation to $z=1.2$ \citep{capak07}, paying
particular attention to a large structure at $z=0.7$ \citep{guzzo07}.
Relevant to this work, in \citet{Smolcic07} they identify a complex of
small clusters at $z\sim 0.2$ via a wide-angle tail
radio galaxy.  At intermediate redshift, an extensive comparison
project has been undertaken targeting the two contrasting clusters
CL0024+17 and MS0451-03 at $z\sim0.5$ to compare the low- and
high-luminosity X-ray cluster environment \citep{moran07,geach06}.
Locally, the Coma cluster has also been extensively used as a
laboratory for galaxy evolution \citep{poggianti04,
carter01,carter08}.  There are many other examples of cluster-focused
environmental studies covering a range of redshifts, including the
large sample of EDisCS clusters at $z>0.5$
\citep{white05,poggianti06,desai07}; and the ACS GTO cluster program
of 7 clusters at $z\sim 1$
\citep{postman05,goto05,blakeslee06,homeier05}.

We summarize the motivation for our survey design as follows. In order
to successfully penetrate the environmental processes at work in
shaping galaxy evolution, several areas must be simultaneously
addressed: a wide range of environments; a wide range in galaxy
luminosity; and sensitivity to both obscured and unobscured star
formation, stellar masses, AGN, and detailed morphologies.
Furthermore, it is essential to use not just a single proxy for
`environment' but to understand directly the relative influences of
the local galaxy density, the hot ICM and the dark matter on galaxy
transformation.  A further advantage is given by examining systems
that are not simply massive clusters already in equilibrium.  By
including systems in the process of formation (when extensive mixing
has not yet erased the memory of early timescales), the various
environmental proxies listed above might still be disentangled.

Therefore, the goal of STAGES is to focus attention on a single
large-scale structure to understand the detailed aspects of galaxy
evolution as a function of environment.  While no single study will
provide a definitive answer to the question of environment and galaxy
evolution, we argue that STAGES occupies a unique vantage point in
this field, to be complemented by other studies locally and at higher
redshift.

\subsection{Galaxy evolution as a function of redshift:  STAGES and
GEMS}

In addition to science focused on the narrow redshift slice containing
the multiple cluster system, the multiwavelength data presented here
provide a valuable resource for those wishing to study the evolution
of the galaxy population since $z=1$.  With the advent of the HST and
multiwavelength data for this field, it is possible to quantify better
the sample variance and investigate rare subsamples using the
combination of the STAGES field together with the Galaxy Evolution and
Morphologies \citep[GEMS;][]{rix04} coverage of the Extended Chandra
Deep Field South (CDFS). In particular, the HST data were chosen to
have the same passband for both GEMS (F606W and 850LP) and STAGES
(F606W only, to allow study at optimum S/N of the cluster
subpopulation and to optimise the weak lensing analysis).  While the
choice of F606W means that the data probe above the 4000{\AA} break
for $z<0.5$ only, for a number of purposes the data can also be used
at higher redshift (although in those cases one needs to be
particularly cognizant of the effects of bandpass shifting and surface
brightness dimming; such effects can be understood and calibrated
using the GEMS 850LP and GOODS 850LP data).  Furthermore, the
24{\micron} observations (\S\ref{sec-mir}) are well-matched in depth
with the first Cycle GTO observations of the CDFS; analyses of the
CDFS and A901/2 fields have been presented by \citet{zheng07} and
\citet{Bell07}.  Several projects are already exploiting this
combined dataset (see \S\ref{sec-summary} for details), and with the
publicly-available data in the CDFS, these samples provide a valuable
starting point for many investigations of galaxy evolution.

\subsection{The Abell 901(a,b)/902 supercluster:  a laboratory for galaxy evolution}\label{sec-root2}

The A901/2 system is an exceptional testing ground with which to
address environmental influences on galaxy evolution.  Consisting of
three clusters and related groups at $z\sim0.165$, all within
$0.5\degr\times0.5\degr$, this region has been the target of extensive
ground- and space-based observations.  We have used the resulting
dataset to build up a comprehensive view of each of the main
components of the large-scale structure: the galaxies, the dark
matter, and the hot X-ray gas.  The moderate redshift is advantageous
as it enables us to study a large number of galaxies, yet the
structure is contained within a tractable field-of-view and probes a
volume with more gas and more star formation in general than in the
local universe.

The A901/2 region, centred at $(\alpha,\delta)_{\rm J2000}$ = $(9^{\rm
h}56^{\rm m}17\fs3$, $-10\degr01\arcmin11\arcsec)$, was originally one
of three fields targeted by the COMBO-17 survey \citep{wolf03}.  It
was specifically chosen as a known overdensity due to the multiple
Abell clusters present.  These included two clusters (A901a and A901b)
with X-ray luminosities sufficient to be included in the X-ray
Brightest Abell-type Cluster Survey \citep[XBACS;][]{ebeling96} of the
ROSAT All-Sky Survey, though pointed ROSAT HRI observations by
\cite{schindler00} subsequently revealed that the emission from A901a
suffers from confusion with several point sources in its vicinity.
The extended X-ray emission in the field is further resolved by our
deep XMM-Newton imaging (see \S\ref{sec-xray}).  Additional structures at
$z\sim 0.165$ in the field include A902 and a collection of galaxies
referred to as the Southwest Group (SWG).

The five broad- and 12 medium-band observations from COMBO-17 provide
high-quality photometric redshifts and spectral energy distributions
(SEDs).  Together with the high-quality imaging for ground-based
gravitational lensing, the A901/2 data have been used in a variety of
papers to date.  COMBO-17 derived results include 2D and 3D
reconstructions of the mass distribution \citep{gray02, taylor04}; the
star-formation--density relation \citep{gray04}; the discovery of a
substantial population of intermediate-age, dusty red cluster
galaxies \citep[][here-after WGM05]{wgm05}; and the morphology-density
\citep{lane07} and morphology-age-density \citep{wolf07} relations.

Further afield, the clusters are also known to be part of a larger
structure together with neighbouring clusters Abell 907 and Abell 868
(1.5 degrees and 2.6 degrees away, respectively). Nowak et al. (in
prep.) used a percolation (also called `friends-of-friends') algorithm
on the REFLEX cluster catalogue \citep{boehringer04} to produce a
catalogue of 79 X-ray superclusters.  Entry 33 is the
A868/A901a/A901b/A902/A907 supercluster, which also contains an
additional, but not very bright, non-Abell cluster.  Though not
observed as part of the STAGES study, these clusters are included in
the constrained N-body simulations used to understand the formation
history of the large-scale structure (\S\ref{sec-mocks}).

The plan of this paper is as follows: in \S\ref{sec-hst} we outline
the observations taken to construct the 80-tile mosaic with the
Advanced Camera for Surveys on HST.  We discuss data reduction, object
detection, and \sersic\ profile fitting.  In \S\ref{sec-c17} we
present the COMBO-17 catalogue for the A901/2 field and discuss how
the two catalogues are matched.  In \S\ref{sec-multi} we present a
summary of the further multiwavelength data for the field and derived
quantities such as stellar masses and star-formation rates.  We finish
with describing ongoing science goals, future prospects, and
instructions for public access to the data and catalogues described
within.  Appendix~\ref{app-notes} contains details on ten individual
objects of particular interest within the field.

Throughout this paper we adopt a concordance cosmology with
$\Omega_m=0.3,\Omega_\lambda=0.7$, and $H_0=70$ km s$^{-1}$
Mpc$^{-1}$.  In this cosmology, $1\arcsec = 2.83$ kpc at the redshift
of the supercluster ($z\sim0.165$), and the COMBO-17 field-of-view
covers $5.3\times5.1$ Mpc$^{2}$.  Magnitudes derived from the HST
imaging (\S\ref{sec-hst}) in the F606W ($V$-band) filter are on the AB
system,\footnote{For F606W, $m_{\rm AB}-m_{\rm Vega}=0.085$.} while
magnitudes from COMBO-17 (\S\ref{sec-c17}) in all filters are on the
Vega system.

\section{HST data}\label{sec-hst}

\subsection{Observations}

The primary goal of the STAGES HST imaging was to obtain morphologies
and structural parameters for all cluster galaxies down to $R=24$
($M_V\sim-16$ at $z\sim0.165$).  The full area of the COMBO-17
observations was targeted to sample a wide range of environments.
Secondary goals included obtaining accurate shape measurements of
faint background galaxies for the purposes of weak lensing, and
measuring morphologies and structural parameters for all remaining
foreground and background galaxies to $R=24$.  As discussed in
\S\ref{sec-root2}, the survey design and filter was chosen to match
that of the GEMS survey \citep{rix04} of the Chandra Deep Field South
(CDFS).  The CDFS is another field with both COMBO-17 and HST
coverage, but in contrast to the A901/2 field is known to contain
little significant large-scale structure.  It will therefore serve as
a matched control sample for comparing cluster and field environments
at similar epochs.

To this end we constructed an 80-tile mosaic with ACS in Cycle 13 to
cover an area of roughly 29.5$\arcmin\times$29.5$\arcmin$ in the F606W
filter, with a mean overlap of 100 pixels between tiles.  Scheduling
constraints forced the roll angle to be 125 degrees for the majority
of observations, and one gap in the northeast corner was imposed on
the otherwise contiguous region due to a bright ($V=9$) star.  A
4-point parallelogram-shaped dithering pattern was employed, with
shifts of 2.5 pixels in each direction.  An additional shift of 60.5
pixels in the y-direction was included between dithers two and three
in order to bridge the chip gap.

\begin{figure}
\centerline{\psfig{file=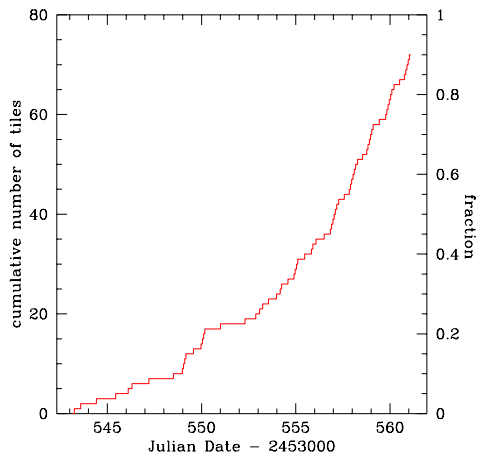,width=0.75\columnwidth}}
\caption{Cumulative plot of ACS data acquisition.  In order to
  minimize the effects of a time-vary PSF on weak lensing
  applications, 50\% of tiles were taken within 5 days and 90\% within 21
  days.  The remaining 7 tiles were observed 6 months
  later.}\label{fig-cumulative}
\end{figure}

Concerns about a time-varying PSF and possible effects on the weak
lensing measurements drove the requirement for the observations to be
taken in as short a time frame as possible.  In practice this was
largely successful, with $>50\%$ of tiles observed in a single
five-day period (Fig.~\ref{fig-cumulative}), and $>90\%$ within 21
days.  Six tiles (29, 75, 76, 77, 79, 80) were unobservable in that cycle
and were re-observed six months later, with a 180 degree rotation.
Furthermore, tile 46 was also re-observed at this orientation as the
original observation failed due to a lack of guide stars.  These seven
tiles were observed following the transition to  two-gyro mode  with no
adverse consequences in image quality. 

\begin{figure*}
\centerline{\psfig{file=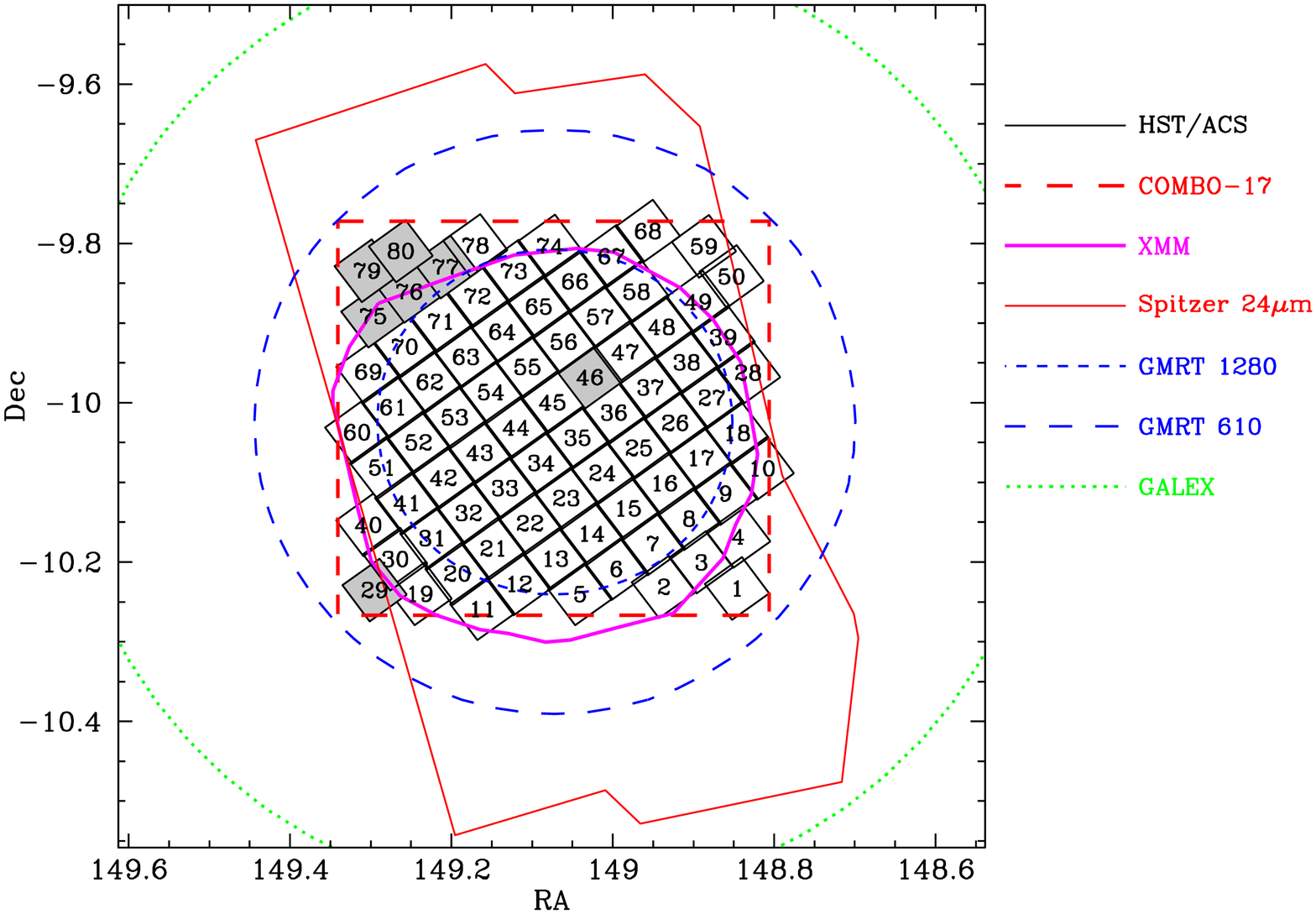,width=0.95\textwidth,angle=270}}
\caption{Layout of multiwavelength observations of the A901/2 field.
The numbered tiles represent the 80-orbit STAGES mosaic with HST/ACS,
which overlaps the 31.5$\times$30 arcmin COMBO-17 field-of-view
(long-dashed square). The seven shaded tiles were observed $\sim$6
months after the bulk of the observations and with a 180\degr
rotation.  The centres of A901a/A901b/A902/SWG are found in tiles
55/36/21/8 respectively.  Interior to the STAGES region are the
XMM-Newton coverage (heavy solid polygon) and the GMRT 1280 MHz
observations (short-dashed circle, indicating half-power beam
width). The STAGES area is also overlapped by the field-of-view of the
Spitzer 24$\micron$ imaging (solid polygon), the GMRT 610 MHz
observations (long-dashed circle), and the GALEX imaging (dotted
circle).}
\label{fig-field}
\end{figure*}

Details of the observations are listed in Table~\ref{tab-obs}.  A
schematic of the field showing the ACS tiles and the multiwavelength
observations is shown in Fig.~\ref{fig-field}.  Additionally, four
parallel observations with WFPC2 (F450W) and NICMOS3 (F110W and F160W)
were obtained simultaneously for each ACS pointing.  Due to the
separation of different instruments on the HST focal plane, most but
not all parallel images overlap with the ACS mosaic (52/10/18
WFPC images and 42/9/29 NICMOS3 images have full/partial/no overlap
with the ACS mosaic; most NICMOS3 images have partial overlap with a
WFPC2 image).  In this paper we restrict ourselves to a discussion of
the primary ACS data, analysis of the parallels will follow in a
future publication.

\begin{table*}
\begin{tabular}{rrrrrrrr}
\hline
\hline
\multicolumn{1}{c}{Tile} & 
\multicolumn{1}{c}{Date} &
\multicolumn{1}{c}{$\alpha$} & 
\multicolumn{1}{c}{$\delta$} &
\multicolumn{1}{c}{Exposure} & 
\multicolumn{1}{c}{N$_{\rm hot}$} &
\multicolumn{1}{c}{N$_{\rm cold}$} & 
\multicolumn{1}{c}{N$_{\rm good}$}\\

\multicolumn{1}{c}{} & 
\multicolumn{1}{c}{[dd/mm/yyyy]} &
\multicolumn{1}{c}{[J2000]} & 
\multicolumn{1}{c}{[J2000]} &
\multicolumn{1}{c}{[s]} & 
\multicolumn{1}{c}{} &
\multicolumn{1}{c}{} & 
\multicolumn{1}{c}{}\\
\hline
 1 & 09 07 2005 & 09:55:22.8 & -10:14:01 & 1960 & 851 & 173 & 796 \\  
 2 & 07 07 2005 & 09:55:44.5 & -10:13:54 & 1960 & 1082 & 209 & 982 \\  
 3 & 08 07 2005 & 09:55:33.4 & -10:12:03 & 1960 & 1157 & 233 & 1008 \\  
 4 & 07 07 2005 & 09:55:22.4 & -10:10:06 & 1960 & 1051 & 199 & 927 \\  
 5 & 04 07 2005 & 09:56:09.5 & -10:14:26 & 1950 & 1069 & 195 & 973 \\  
 6 & 03 07 2005 & 09:55:58.7 & -10:12:33 & 1950 & 1151 & 219 & 1027 \\  
 7 & 04 07 2005 & 09:55:47.9 & -10:10:39 & 1950 & 1038 & 237 & 905 \\  
 8 & 04 07 2005 & 09:55:37.0 & -10:08:45 & 1950 & 1095 & 262 & 938 \\  
 9 & 04 07 2005 & 09:55:26.2 & -10:06:52 & 1950 & 1020 & 188 & 876 \\  
10 & 05 07 2005 & 09:55:15.4 & -10:04:58 & 1950 & 1014 & 184 & 938 \\  
11 & 07 07 2005 & 09:56:38.6 & -10:15:34 & 1960 & 989 & 219 & 876 \\  
12 & 04 07 2005 & 09:56:27.8 & -10:13:40 & 1950 & 1020 & 226 & 885 \\  
13 & 28 06 2005 & 09:56:16.9 & -10:11:46 & 2120 & 1193 & 256 & 1037 \\  
14 & 28 06 2005 & 09:56:06.1 & -10:09:53 & 2120 & 1391 & 254 & 1111 \\  
15 & 28 06 2005 & 09:55:55.3 & -10:07:59 & 2120 & 1182 & 253 & 1052 \\  
16 & 29 06 2005 & 09:55:44.5 & -10:06:06 & 2120 & 1109 & 208 & 940 \\  
17 & 29 06 2005 & 09:55:33.7 & -10:04:12 & 1960 & 1116 & 250 & 888 \\  
18 & 04 07 2005 & 09:55:22.9 & -10:02:18 & 1950 & 995 & 178 & 868 \\  
19 & 09 07 2005 & 09:56:57.3 & -10:14:25 & 1960 & 963 & 180 & 786 \\  
20 & 07 07 2005 & 09:56:46.0 & -10:12:54 & 1960 & 979 & 222 & 829 \\  
21 & 30 06 2005 & 09:56:35.2 & -10:11:00 & 1960 & 1166 & 288 & 1005 \\  
22 & 28 06 2005 & 09:56:24.4 & -10:09:07 & 2120 & 1193 & 263 & 1012 \\  
23 & 25 06 2005 & 09:56:13.6 & -10:07:13 & 2120 & 1143 & 241 & 1000 \\  
24 & 25 06 2005 & 09:56:02.8 & -10:05:19 & 2120 & 1244 & 254 & 1128 \\  
25 & 22 06 2005 & 09:55:52.0 & -10:03:26 & 2120 & 1274 & 248 & 1051 \\  
26 & 29 06 2005 & 09:55:41.1 & -10:01:32 & 1960 & 1214 & 275 & 1063 \\  
27 & 05 07 2005 & 09:55:30.3 & -09:59:39 & 1950 & 1258 & 279 & 1068 \\  
28 & 08 07 2005 & 09:55:19.5 & -09:57:45 & 1960 & 1161 & 220 & 1052 \\  
29 & 04 01 2006 & 09:57:10.7 & -10:14:08 & 2120 & 1274 & 272 & 1123 \\  
30 & 09 07 2005 & 09:57:04.5 & -10:11:48 & 1960 & 943 & 209 & 781 \\  
31 & 08 07 2005 & 09:56:53.5 & -10:10:14 & 1960 & 900 & 200 & 713 \\  
32 & 03 07 2005 & 09:56:42.7 & -10:08:20 & 1950 & 1023 & 214 & 884 \\  
33 & 28 06 2005 & 09:56:31.9 & -10:06:27 & 2120 & 1150 & 223 & 955 \\  
34 & 22 06 2005 & 09:56:21.0 & -10:04:33 & 2120 & 1318 & 243 & 1111 \\  
35 & 22 06 2005 & 09:56:10.2 & -10:02:40 & 2120 & 1220 & 244 & 1028 \\  
36 & 24 06 2005 & 09:55:59.4 & -10:00:46 & 2120 & 1320 & 287 & 1101 \\  
37 & 29 06 2005 & 09:55:48.6 & -09:58:53 & 1960 & 1150 & 239 & 974 \\  
38 & 05 07 2005 & 09:55:37.8 & -09:56:59 & 1950 & 1123 & 205 & 951 \\  
39 & 08 07 2005 & 09:55:27.0 & -09:55:05 & 1960 & 1094 & 210 & 965 \\  
40 & 09 07 2005 & 09:57:12.5 & -10:09:14 & 1960 & 1062 & 198 & 916 \\  
41 & 07 07 2005 & 09:57:00.9 & -10:07:34 & 1960 & 962 & 176 & 828 \\  
42 & 03 07 2005 & 09:56:50.1 & -10:05:41 & 1950 & 1090 & 205 & 928 \\  
43 & 27 06 2005 & 09:56:39.3 & -10:03:47 & 2120 & 1198 & 202 & 1052 \\  
44 & 27 06 2005 & 09:56:28.5 & -10:01:54 & 2120 & 1266 & 230 & 1046 \\  
45 & 23 06 2005 & 09:56:17.7 & -10:00:00 & 2120 & 1280 & 285 & 1064 \\  
46 & 01 01 2006 & 09:56:05.4 & -09:57:47 & 2120 & 1438 & 355 & 1235 \\  
47 & 01 07 2005 & 09:55:56.0 & -09:56:13 & 1960 & 1198 & 273 & 972 \\  
48 & 06 07 2005 & 09:55:45.2 & -09:54:19 & 1950 & 989 & 176 & 852 \\  
49 & 06 07 2005 & 09:55:34.4 & -09:52:26 & 1960 & 1054 & 223 & 901 \\  
50 & 09 07 2005 & 09:55:24.4 & -09:50:31 & 1960 & 984 & 212 & 832 \\  
51 & 07 07 2005 & 09:57:08.4 & -10:04:55 & 1960 & 1050 & 189 & 923 \\  
52 & 03 07 2005 & 09:56:57.6 & -10:03:01 & 1960 & 1142 & 209 & 941 \\  
53 & 03 07 2005 & 09:56:46.8 & -10:01:07 & 1950 & 1135 & 211 & 920 \\  
54 & 02 07 2005 & 09:56:36.0 & -09:59:14 & 1950 & 1131 & 228 & 921 \\  
55 & 02 07 2005 & 09:56:25.1 & -09:57:20 & 1960 & 1205 & 311 & 974 \\  
56 & 02 07 2005 & 09:56:14.3 & -09:55:27 & 1960 & 1097 & 242 & 891 \\  
57 & 01 07 2005 & 09:56:03.5 & -09:53:33 & 1960 & 1090 & 210 & 911 \\  
58 & 06 07 2005 & 09:55:52.7 & -09:51:40 & 1950 & 1130 & 201 & 975 \\  
59 & 08 07 2005 & 09:55:32.7 & -09:48:15 & 1960 & 1075 & 204 & 900 \\  
60 & 07 07 2005 & 09:57:15.8 & -10:02:15 & 1950 & 1028 & 183 & 912 \\  
\hline
\end{tabular}
\caption{Details of STAGES HST/ACS observations. Only the second
  (successful) acquisition of tile 46 is listed.  `Hot',`cold', and
  `good' SExtractor configurations are described in
  \S\ref{sec-detect}.  Tiles 29, 46, 75, 76, 77, 79, and 80 are
  oriented at 180\degr\ with respect to the rest of the mosaic. 
  The exposure time varied according to the maximum window of
  visibility available in each orbit.}\label{tab-obs}
\end{table*}

\begin{table*}
\contcaption{}
\begin{tabular}{rrrrrrrr}
\hline
\hline
\multicolumn{1}{c}{Tile} & 
\multicolumn{1}{c}{Date} &
\multicolumn{1}{c}{$\alpha$} & 
\multicolumn{1}{c}{$\delta$} &
\multicolumn{1}{c}{Exposure} & 
\multicolumn{1}{c}{N$_{\rm hot}$} &
\multicolumn{1}{c}{N$_{\rm cold}$} & 
\multicolumn{1}{c}{N$_{\rm good}$}\\

\multicolumn{1}{c}{} & 
\multicolumn{1}{c}{[dd/mm/yyyy]} &
\multicolumn{1}{c}{[J2000]} & 
\multicolumn{1}{c}{[J2000]} &
\multicolumn{1}{c}{[s]} & 
\multicolumn{1}{c}{} &
\multicolumn{1}{c}{} & 
\multicolumn{1}{c}{}\\
\hline
61 & 07 07 2005 & 09:57:05.0 & -10:00:21 & 1950 & 971 & 183 & 826 \\  
62 & 07 07 2005 & 09:56:54.2 & -09:58:28 & 1950 & 1052 & 184 & 901 \\  
63 & 06 07 2005 & 09:56:43.4 & -09:56:34 & 1950 & 1141 & 217 & 930 \\  
64 & 06 07 2005 & 09:56:32.6 & -09:54:41 & 1950 & 1069 & 222 & 890 \\  
65 & 06 07 2005 & 09:56:21.8 & -09:52:47 & 1950 & 1071 & 227 & 908 \\  
66 & 06 07 2005 & 09:56:11.0 & -09:50:53 & 1950 & 1014 & 222 & 859 \\  
67 & 06 07 2005 & 09:56:00.1 & -09:48:60 & 1950 & 1046 & 226 & 922 \\  
68 & 08 07 2005 & 09:55:49.3 & -09:47:06 & 1960 & 967 & 179 & 851 \\  
69 & 10 07 2005 & 09:57:12.5 & -09:57:42 & 1960 & 876 & 145 & 784 \\  
70 & 09 07 2005 & 09:57:01.7 & -09:55:48 & 1960 & 934 & 183 & 798 \\  
71 & 09 07 2005 & 09:56:50.9 & -09:53:54 & 1960 & 1032 & 182 & 888 \\  
72 & 10 07 2005 & 09:56:40.0 & -09:52:01 & 1960 & 1118 & 212 & 950 \\  
73 & 09 07 2005 & 09:56:29.2 & -09:50:07 & 1960 & 910 & 168 & 773 \\  
74 & 08 07 2005 & 09:56:18.4 & -09:48:14 & 1960 & 907 & 192 & 822 \\  
75 & 04 01 2006 & 09:57:11.0 & -09:53:30 & 2120 & 1708 & 260 & 1140 \\  
76 & 05 01 2006 & 09:57:00.3 & -09:51:39 & 2120 & 1444 & 275 & 1134 \\  
77 & 05 01 2006 & 09:56:49.5 & -09:49:48 & 2120 & 1324 & 287 & 1094 \\  
78 & 05 07 2005 & 09:56:40.6 & -09:48:11 & 1960 & 1031 & 184 & 842 \\  
79 & 05 01 2006 & 09:57:12.9 & -09:50:05 & 2120 & 1357 & 302 & 1019 \\  
80 & 05 01 2006 & 09:57:02.8 & -09:48:36 & 2120 & 1255 & 246 & 973 \\  

\hline
\end{tabular}
\end{table*}

\subsection{ACS data reduction}

We retrieve the reduced STAGES images processed by the CALNICA
pipeline of STScI, which corrects for bias subtraction and
flat-fielding.  However, as the ACS camera is located 6 arcmin off the
centre of the HST optical axis, the images from the telescope have a
field-of-view with a parallelogram keystone distortion.  To produce a
final science image from the reduced pipeline data, we therefore also
have to remove the geometric distortion before combining the
individual dithered sub-exposures.  The removal of the image
distortion is now fairly routine through the use of the MULTIDRIZZLE
software \citep{koekemoer07}.  However, our particular science goals
motivated us to make several changes when optimizing the default
settings and combining the raw images.  These changes are discussed
below.

\subsubsection {Image Distortion Correction}

In STAGES, the science driver that demands the highest quality data
reduction in terms of producing the most consistent and stable PSF
from image to image, and across the field of view, is weak lensing
\citep{heymans08}.  With this goal in mind, we benefit from the
experience of \cite{rhodes07}, who conducted detailed studies of how
the pixel values are re-binned when the images are corrected for image
distortion.  Briefly speaking, to transform an image that is sampled
on a geometrically distorted grid onto one that is a uniform Cartesian
grid fundamentally involves rebinning, i.e. interpolating, the
original pixel values into the new grid.  Doing so is not a
straightforward process since the original ACS pixel scale samples the
telescope diffraction limit below Nyquist frequency, i.e. the
telescope PSF is undersampled.  When a PSF is undersampled, aliasing
of the pixel fluxes occurs, the result of which is that the recorded
structure of the PSF appears to change with position, depending on the
exact sub-pixel centroid of the PSF.  This variability effectively
produces a change in the ellipticity of the PSF as a function of
sub-pixel position, even if the PSF should be identical everywhere.
Because stellar PSFs are randomly centred about a pixel the
intrinsic ellipticity one then measures has a non-zero scatter.  So,
as weak lensing relies heavily on measuring the ellipticities of
galaxies, which are convolved by the PSF, the scatter in the PSF
ellipticity contributes significant noise to weak lensing
measurements.

An additional issue with non-Nyquist sampled images is that the process
of interpolating pixel values necessarily degrades the original image
resolution.  While the intrinsic resolution can in principle be
recovered by dithering the images while making observations, strictly
speaking this inversion is only possible when the image is on a
perfect Cartesian grid at the start, i.e.  with no image distortion.
Otherwise, there would be a residual ``beating frequency'' in the
sampling of the reconstituted image, such that some pixels would be
better sampled than others.  Because of this, recovering the intrinsic
resolution of the telescope when the field is distorted is not a well
posed problem, and cannot easily be solved by a small number of image
dithers.  Some resolution loss will necessarily occur in some parts of
the image.  This is especially true if the final images are combined
{\it after} having been geometrically corrected, as is currently the
process in MULTIDRIZZLE.  One last, unavoidable, side effect of
interpolating a non-Nyquist sampled image is that the pixel values
become necessarily correlated.  However, the degree of resolution loss
and noise correlation can be balanced by a suitable choice of
interpolation kernels: whereas square top-hat kernels effectively
amounts to linear interpolation and correlates only the immediate
neighbour pixels but cause high interpolation (pixellation) noise,
bell-shaped kernels (e.g.  Gaussian and Sinc) correlate more pixels
but better preserve the image resolution.

In light of these issues, it is clear that the goal of an optimal HST
data reduction should be a dataset where the PSF structure is stable
across the field of view and reproducible from image tile to tile.
The contribution to the PSF variation by the stochastic aliasing
of the PSF that necessarily occurs during `drizzling' can be reduced
by appropriate choices of drizzling kernel and output pixel scale.
\cite{rhodes07} characterize PSF stability in terms of the scatter in
the apparent ellipticity of the PSF in the ACS field of view.  After
experimenting, they determine that the optimal set of parameters in
MULTIDRIZZLE to use is a Gaussian drizzling kernel,
\texttt{pixfrac}=0.8, and an output pixel scale of $0\farcs03$.  We
thus follow their approach by adopting those parameters for our own
reduction, while keeping all the other default parameters unchanged.
However, they note, as we do, that a Gaussian kernel causes more
correlated pixels than tophat kernels.  Nonetheless because the choice
of interpolation kernel amounts effectively to a smoothing kernel,
correlated noise should in principle not have an impact on photometry
statistics since the flux is conserved.  Moreover, the same
interpolation (smoothing) kernel propagates into the PSF, thus the
choice of kernel should also not impact galaxy fitting analyses.

\subsubsection {Sky pedestal and further image flattening correction}

The images obtained from the HST archive have been bias subtracted and
flatfielded.  However, large-scale non-flatness on the order of 2-4\%
remains in the images, and there are slight but noticeable pedestal
offsets that remain between the four quadrants.  These large scale
patterns and pedestals are both stationary and consistent in images
that are observed closely in time.  And even though MULTIDRIZZLE tries
to equalize the pedestals before combining the final images, the
correction is not always perfect due to object contamination when
computing the sky pedestal.  These effects are small, and the sky
pedestal issue only affect large objects situated right on image
boundaries, so that the effects on the entire survey itself may only
be cosmetic.  Nevertheless, we try to correct for the effects by
producing a median image of data observed closely in time, after first
rejecting the brightest 30\% and faintest 20\% of the images (to
avoid over-subtraction).  Then, for each of the four CCD quadrants,
we fit a low order 2-D cubic-spline surface (IRAF/imsurfit)
individually to model the large scale non-uniformity in the median sky
image, and to remove noise.  The noiseless model of the sky is then
subtracted from all the data observed closely in time.  After
correction, the mean background in the four quadrants is essentially
equal, and the residual non-flatness is $\ll 1\%$.

\subsection{Object Detection}
\label{sec-detect}
Object detection and cataloguing were carried out automatically on the
STAGES F606W imaging data using the SExtractor V2.5.0 software
\citep{bertin96}. An optimized, dual (`cold' and `hot') configuration
was used, following the strategy developed for HST/ACS data of similar
depth for GEMS \citep{caldwell08}.  The main challenge to extracting
sources from the STAGES ACS data is the tradeoff between deblending
high-surface brightness cluster members that are close on the sky in
projection, and avoiding spurious splitting (`shredding') of highly
structured spiral galaxies into multiple sources. In addition, we
desire high detection completeness for faint, and often low-surface
brightness, background galaxies.  To optimize the detection
completeness and deblending reliability for counterparts to $R_{\rm
ap}\leq24$ mag galaxies\footnote{COMBO-17 redshifts are mostly useful
at $R_{\rm ap}\leq24$ for reasons discussed in detail in \S\ref{sec-c17},
and so we adopt this cut for our main science sample.} from the
COMBO-17 catalogue, we fine-tuned the combination of cold and hot
configuration parameters using three representative STAGES tiles (21,
39, and 55).  For STAGES, we converged on the parameters given in
Table \ref{sex_config}, which successfully detected 99.5\% (650/653)
of the $R_{\rm ap}\leq24$ mag COMBO-17 galaxies on these tiles, with
reliable deblending for 98.0\%.

SExtractor produces a list of source positions and basic photometric
parameters for each astrometrically/photometrically calibrated image,
and produces a segmentation map that parses the image into source and
background pixels, which is necessary for subsequent galaxy fitting
with GALFIT \citep{peng02} described in \S\ref{sec-fitting}. For both
configurations, a weight map ($\propto{\rm variance}^{-1}$) and a
three-pixel (FWHM) top-hat filtering kernel were used.  The former
suppresses spurious detections on low-weight pixels, and the latter
discriminates against noise peaks, which statistically have smaller
extent than real sources as convolved by the instrumental PSF.  Our
final catalogue contains 75\,805 {\it unique} F606W sources uniformly
and automatically identified from 17\,978 objects detected in the cold
run, and 89\,464 `good' sources found in the hot run (before
rejection of the unwanted hot detections that fell within the
isophotal area of any cold detection).  A total of 5\,921 objects
were manually removed from the catalogue after the detection
stage. These detections are mainly over-deblended galaxies or image
defects like cosmic rays. Another set of 658 detections were included
in fitting the sample galaxies to ensure the accurate fitting of real
objects, but excluded from the final catalogue. These were also mainly
cosmic ray hits or stellar diffraction spikes.  Although the main
analysis was performed on a tile-by-tile basis, rather than
mosaic-wise, the main catalogue only contains {\it unique} sources.
Objects detected on two tiles enter the catalogue only once. The most
interior-located was selected for entry into the catalogue.  The
breakdown of cold, hot, and good sources per ACS frame is given in
Table~\ref{tab-obs}.

\begin{figure}
\centerline{\psfig{file=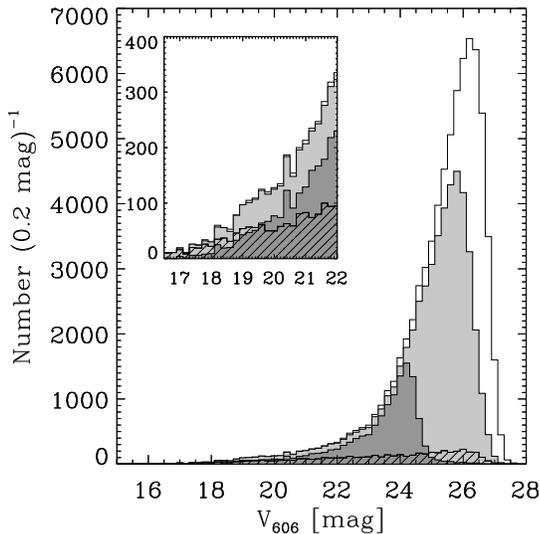,width=\columnwidth}}
\caption{ Source detections in the HST mosaic (overlap region with
STAGES and COMBO-17 coverage). The solid line represents all
SExtractor detected sources (74\,534 objects). The grey histograms
shows all objects with a corresponding match in the COMBO-17 catalogue
(light grey; 50\,701 sources) and extended sources with
$R_{\textrm{ap}}<24$ (dark grey; 12\,748 sources). In addition, the
hashed region indicates stars as defined by our star-galaxy separation
criterion (Equation~\ref{eqn-stargal}; 4\,969 stars in total). In the
inset we highlight the bright magnitude end where the total number of
stars dominates the source population.}\label{fig-hist}
\end{figure}

In Fig.~\ref{fig-hist} we show a histogram of various object samples
in the region of the HST-mosaic that overlaps with COMBO-17. The HST
data start to become incomplete at $V_{606}\sim 26$ (solid
line). Stars (hashed histogram) only make up a significant fraction of
all detections at the brightest magnitudes. A histogram of
counterparts from a cross-correlation with COMBO-17 is shown in light
grey. When the match is restricted to extended objects with
$R_{\rm ap}<24$ (ie. the primary 'galaxy' sample for which we have
reliable photometric redshifts), the HST sources largely have $V_{606}<24$.

\begin{figure}
\centerline{\psfig{file=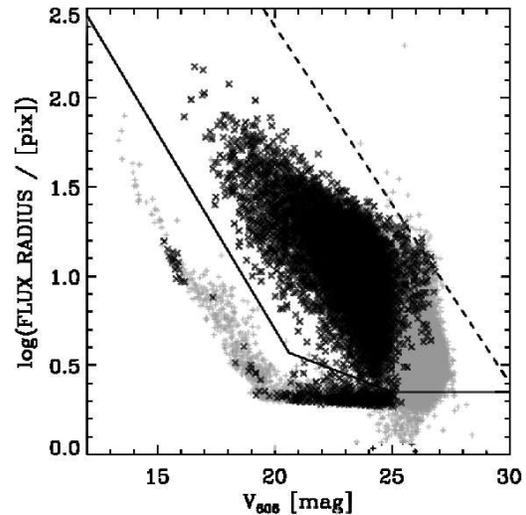,width=0.8\columnwidth}}
\caption{ Star-galaxy separation. We define a line in the
magnitude-size plane to separate stars and galaxies (solid line).
Objects above this line are extended galaxies; objects below are other
compact objects (including most AGN). Grey pluses indicate all
detections; black crosses only those with a COMBO-17 cross-match and
$R_{\textrm{ap}}<24$ and a redshift $z>0$. Note, a significant number
of mostly late-type stars are misidentified as galaxies by COMBO-17
photometry alone. The dashed line shows a line of constant surface
brightness, which is almost parallel to our selection line at the
bright end. 
}\label{fig-hstmorph}
\end{figure}

Star-galaxy separation is performed in the apparent magnitude -- 
size plane spanned by the SExtractor parameters MAG\_BEST ($V_{606}$) and
FLUX\_RADIUS ($r_f$). Objects with

\begin{equation}\label{eqn-stargal}
\log(r_f) < {\rm max}\left( 0.35; 1.60-0.05 V_{606}; 5.10-0.22 V_{606} \right)
\end{equation}
are classified as point sources; sources above that line are
identified as extended sources (galaxies). This plane is shown in
Fig.~\ref{fig-hstmorph}. The separation line clearly delineates
compact and extended sources, in particular when inspecting the
COMBO-17 sources only (crosses). Note that those AGN for which the
point source dominates are also found on the point-source locus and
therefore are removed from the galaxy sample by this selection.

\begin{figure}
\centerline{\psfig{file=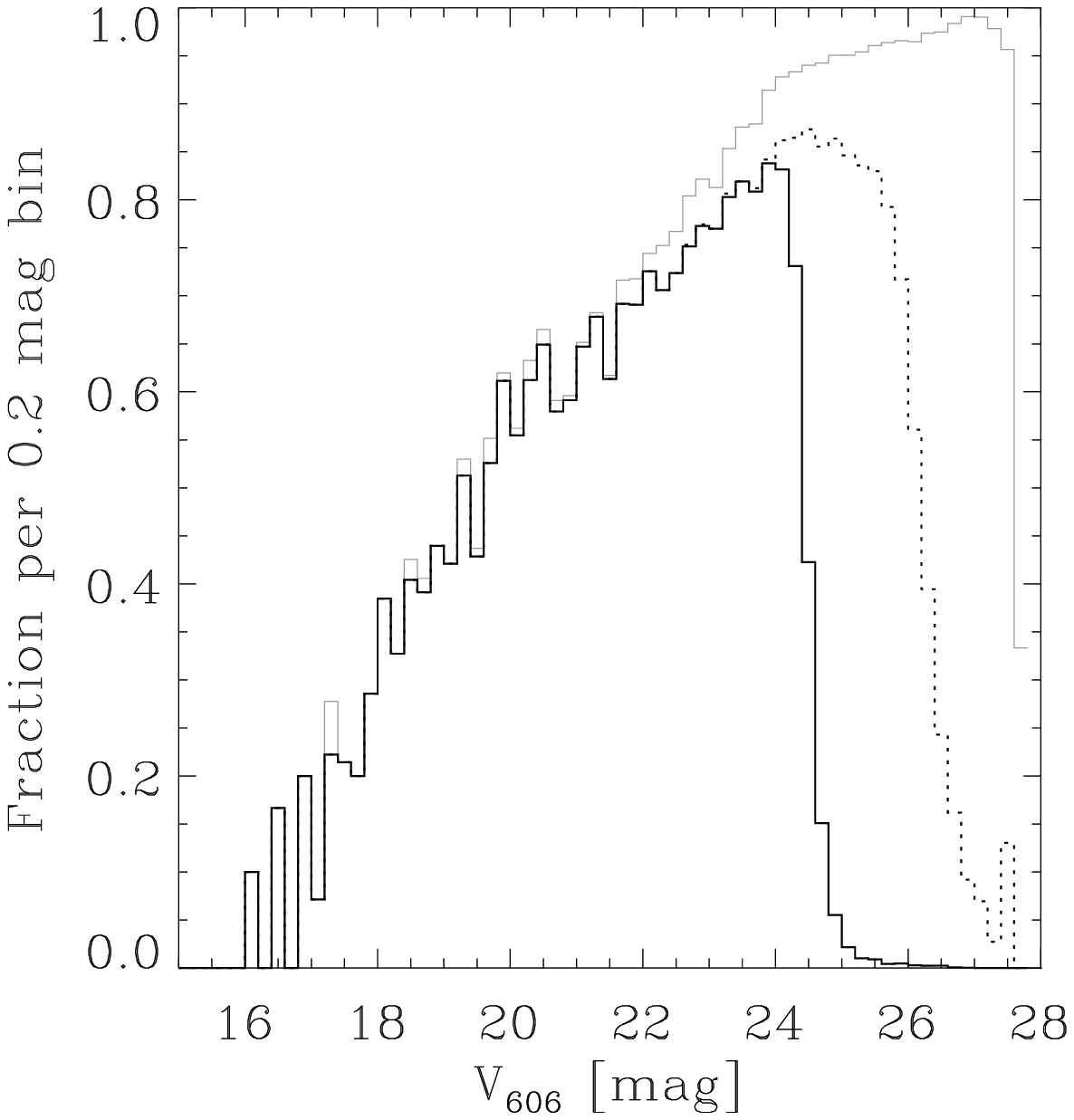,width=0.9\columnwidth}}
\caption{Fraction of extended STAGES objects and COMBO-17
counterparts.  The grey line shows the extended source fraction in
STAGES. At bright magnitudes most sources are compact, while at the
faint end almost all are extended. The black dotted line shows
extended sources in STAGES with a COMBO-17 counterpart. At
$V_{606}\sim26$ the COMBO-17 completeness limit is reached. Almost no
fainter sources are found in COMBO-17. The black solid line shows
extended sources in STAGES with a COMBO-17 counterpart having
$R_{\textrm{ap}}<24$. Out to $V_{606}\sim22$ almost every extended
STAGES object has a COMBO-17 counterpart: the cross-correlation
completeness defined with respect to the STAGES catalogue is almost
100\% (i.e.~the ratio of black and grey lines); at $V_{606}\sim24$ it
is $\sim$90\%.  See \S\ref{sec-xcorr} for further discussion.}
\label{fig-xcorrcomp}
\end{figure}

In the Fig.~\ref{fig-xcorrcomp} we display the galaxy fraction as a
function of $V_{606}$ magnitude (grey histogram). Out to $V_{606}\sim
22$ almost every galaxy detection on the HST images has a COMBO-17
counterpart; at the COMBO-17 sample limit $V_{606}\sim 24$ the
matching completeness for STAGES objects is still $\sim$90\%. The
cross-matching between COMBO-17 and the HST data is described in more
detail in \S~\ref{sec-xcorr}, where completeness is defined in
reverse, i.e.  maximizing HST counterparts for COMBO-17 objects.

\begin{table*}
\caption{Dual SExtractor parameter values for STAGES F606W object detection in `cold' and `hot' configurations. }
\label{sex_config}
\begin{tabular}{lrrl}
\hline\hline
Parameter & Cold & Hot & Description\\
\hline
DETECT\_THRESH & 2.8 & 1.5 & detection threshold above background \\
DETECT\_MINAREA & 140 & 45 & minimum connected pixels above threshold \\
DEBLEND\_MINCONT & 0.02 & 0.25 & minimum flux/peak contrast ratio\\
DEBLEND\_NTHRESH & 64 & 32 & number of deblending threshold steps\\
\hline
\end{tabular}
\end{table*}

\subsection{\sersic\ profile fitting}\label{sec-fitting}

To obtain \sersic\ model fits for each STAGES galaxy, the
imaging data were processed with the data pipeline GALAPAGOS (Galaxy
Analysis over Large Areas: Parameter Assessment by GALFITting Objects
from SExtractor; Barden et al., in prep.). GALAPAGOS performs all 
galaxy fitting analysis steps from object detection to catalogue
creation automatically. This includes (i) source detection and
extraction with SExtractor; (ii) preparing all detected objects for
\sersic\ fitting with GALFIT \citep{peng02}: i.e., constructing bad
pixel masks, measuring local background levels, and setting up
starting scripts with initial parameter estimates; (iii) running the
\sersic\ model fits; and (iv) compiling all information into a final
catalogue.  

Based on a single startup script, GALAPAGOS first runs SExtractor in
the dual high dynamic range mode described in \S\ref{sec-detect}. As
no SExtractor setup is ever 100\% optimal, we manually inspected all
80 tiles for unwanted detections or over-deblended objects. GALAPAGOS
allows for the removal of such extraction failures automatically given
an input coordinate list. Additionally, we also composed a list of
detections that are bright enough to influence the fitting of
neighbouring astronomical sources (e.g. diffraction spikes from bright
stars).  Unlike the aforementioned bad detections these are not
removed instantly, but kept in the source catalogue throughout the
fitting process and removed only from the final object
catalogue. Again, GALAPAGOS performs this operation automatically
given a second list of coordinates. Further details on the
process of manual fine-tuning of detection catalogues can be found in
Barden et al. (in prep.).

After the second run GALAPAGOS uses the cleaned output source list
(described in \S\ref{sec-detect}) to cut postage stamps for every
object. Postage stamps are required for efficient \sersic\ profile
fitting with GALFIT. The sizes of the postage stamps are based on a
multiple $m$ of the product of the SExtractor parameters KRON\_RADIUS
and A\_IMAGE. We define a ``Kron-ellipse'' with semi-major axis
$r_{\textrm{K}}$ as
\begin{equation}
 r_{\textrm{K}}=m \times \textrm{KRON\_RADIUS} \times \textrm{A\_IMAGE}.
\end{equation} 

The sky level is calculated for each source individually by evaluating
a flux growth curve. GALAPAGOS uses the full science frame for this
purpose in contrast to simply working on the postage stamp. Although
in principle the background estimate provided by SExtractor could have
been used, tests show that using the more elaborate GALAPAGOS scheme
results in more robust parameter fits \citep{haeussler07}. For a
detailed description of the algorithm we refer to Barden et al. (in
prep.). One might argue that GALFIT allows fitting the sky
simultaneously with the science object. However, this requires the
size of the postage stamp to be matched exactly to the size of the
science object. If the postage stamp is too small, the proper sky
value cannot be found; if it is too big, computation takes
unneccesarily long.  Too many secondary sources would have to be
included in the fit and the inferred sky value might be influenced by
distant sources. Additionally, galaxies may not be perfectly
represented by a \sersic\ fit, and the sky may take on unrealistic
values as a result.  Although this method may be the easiest option
for manual fitting, in the general case of fitting large numbers of
sources automatically the most robust option is to calculate the sky
value beforehand and keep its value fixed when running GALFIT
\citep[as demonstrated in][]{haeussler07}.

Another crucial component for setting up GALFIT is determining which
companion objects should be included in the fit.  In particular, in
crowded regions with many closely neighbouring sources the fit quality
of the primary galaxy improves dramatically when including
simultaneously fitting \sersic\ models to these neighbours rather than
simply masking them out.  GALAPAGOS makes an educated guess as to
which neighbours should be fitted or masked (see Barden et al., in
prep. for further details).  The decision is made by calculating
whether the Kron-ellipses of primary and neighbouring source
overlap. This calculation is performed not only for sources on the
postage stamp, but on all objects on the science frames surrounding
the current one, in order to take objects at frame edges into account
properly. Detections not identified as overlapping secondary sources
are treated as well.  Such non-overlapping companions are masked based
on their Kron-ellipse and thus excluded from fitting.

An additional requirement for fitting with GALFIT is an input PSF.  We
constructed a general high S/N PSF for STAGES by combining all stars
(i.e.\ classified by COMBO-17 photometry and having ACS 
SExtractor stellarity index $>0.85$) in the brightness interval
$19.5<V_{606}\le23.5$ and lying away from the chip edges.  This
selects non-saturated stars that can still contribute signal in their
centres. All stars were visually inspected against binarity,
companions, or defects, which resulted in either a manually created
mask, or the star being excluded if masking would not have been
sufficient to isolate the star. With this selection 1\,024 stars
remained and were combined after subpixel cocentering and local
background removal.

In order to sample the field-variations of the PSF well and
not be dominated by the few brightest stars, we weighted all stars
identically in the centre (where all stars carry information), but
applied a suppression of the noise in the outer parts by a Gaussian
downweighting.  The contribution from fainter stars in this process
was suppressed at smaller radii relative to brighter ones. In this way
we created a high S/N true mean PSF image of 255$\times$255 pixel
centred exactly on the PSF and used this for all galaxy-related (but
not AGN related) analyses.  

In its current version, GALAPAGOS sets up GALFIT to fit a
\sersic\ model \citep{sersic68} for each object. A \sersic\ profile is
a generalised de Vaucouleurs model with variable exponent $n$, the
\sersic\ index:
\begin{equation}
 \Sigma\left(r\right)=\Sigma_e\times\exp\left(-\kappa\left[\left(r/r_e\right)^{
1/n}-1\right]\right),
\end{equation} 
with the effective radius $r_e$, the effective surface density
$\Sigma_e$, the surface density as a function of radius
$\Sigma\left(r\right)$ and a normalisation constant
$\kappa=\kappa\left(n\right)$. An exponential profile has $n=1$ while
a de Vaucouleurs profile has $n=4$. The parameters that go into the
model are the position $[x,y]$, total magnitude $m$, the effective
radius $r_e$, the \sersic\ index $n$, the axis ratio $q$ ($q=b/a$; the
ratio of semi-minor over semi-major half-axis ratio) and the position
angle $\theta$.  Starting guesses for all parameters aside from $n$
and $r_e$ are taken directly from the SExtractor output. GALAPAGOS
converts the FLUX\_RADIUS from SExtractor to estimate the effective
radius as $r_e=10^{-0.79}\textrm{FLUX\_RADIUS}^{1.87}$. This formula
was found empirically to work best for simulated \sersic\ profiles in
the GEMS project \citep{haeussler07}. The \sersic\ index is started at
a value $n=2.5$.

For computational efficiency we apply constraints to the parameter
range during the fitting process. Of course, this procedure is not
advisable when fitting objects manually, yet it is mandatory for an
automated process like GALAPAGOS. Our constraints are listed in Table
\ref{tab-constraints}. Non-zero lower boundaries for $r_e$ and $n$
were imposed for computational reasons.  The maximum for $r_e$ allows
fitting the largest galaxy in the field (750 pix correspond to
$\sim60$ kpc at the cluster distance).  The upper limit for the
\sersic\ index is far from the de Vaucouleurs case and includes even
the steepest profiles.  The magnitude constraint flags catastrophic
disagreements between the two photometry codes, where one of the two
does not return a sensible result.  Such problem objects may include
LSB galaxies, where SExtractor fails to see large fractions of the
total flux; or intrinsically faint objects with a peculiar neighbour
or background structure, where GALFIT tries to remove the excess flux.
Objects whose values stall at the constraint limits are most likely
not well represented by a single \sersic\ profile (e.g. stars or
extreme two-component galaxies with a LSB disk).

\begin{table}
\begin{center}
\caption{GALFIT fitting constraints.}
\label{tab-constraints}
\begin{tabular}{ccc}
\hline\hline
Parameter & Lower limit & Upper limit \\
\hline
$r_e$ & 0.3 & 750 \\
$n$ & 0.2 & 8  \\
$|m_{\rm SEx}-m_{\rm GALFIT}|$ & - & 5 \\
\hline
\end{tabular}
\end{center}
\end{table}

Finally, GALAPAGOS combines the SExtractor and GALFIT results into one
FITS-table. At this stage flagged objects (like stellar diffraction
spikes, etc.) are removed from the table. A very detailed description
of GALAPAGOS including setup and computational efficiency will be
presented together with the publication of the code in
Barden et al. (in prep.).  We note that the GALFIT reported errors are
purely statistical (ie. based on the assumption that Poisson noise
dominates the uncertainties of the fit parameters), and as such
certainly under-represent the true uncertainties.  A more meaningful
measure of uncertainties comes from fitting simulated galaxies, as
shown in \citet{haeussler07} and explored here in detail in
\S\ref{sec-sims}.

\begin{figure}
\psfig{file=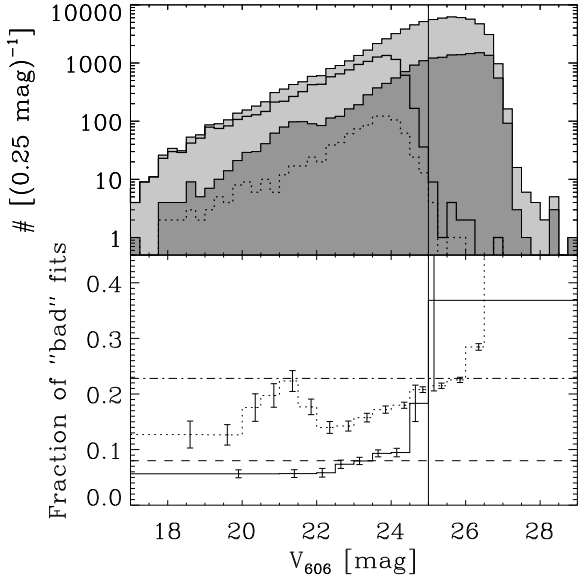,width=\columnwidth}
\caption{GALFIT quality.  Top panel: The two grey histograms show the
total number of fitted galaxies (light grey) and galaxies with `bad'
fits where the fitting procedure failed (dark grey).  The heavy solid
and dotted histograms show the same but for the science sample with
$R_{\textrm{ap}}<24$ (i.e. objects with a COMBO-17 counterpart only)
within the overlap region of STAGES and COMBO-17.
Bottom panel: Fraction of `bad' fits (plus 1$\sigma$ error bars) for
all fitted galaxies (dotted histogram) and those with a COMBO-17 match
and $R_{\textrm{ap}}<24$ (solid histogram). Overall, $\sim 23$\% of
all fits ran into a constraint (dashed dotted line). For the science
objects (STAGES/COMBO-17 cross-matched galaxies with
$R_{\textrm{ap}}<24$) the fraction is considerably lower ($\sim8$\%;
dashed line). The vertical line roughly indicates the surface
brightness completeness limit. The `bump' at $V_{606}\sim21$ possibly
results from merging two SExtractor setups (the `hot' and `cold'
configurations described in \S\ref{sec-detect}).  }\label{fig-badfits}
\end{figure}

With our setup we were able to achieve an overall total of $\sim92$\%
high quality fits for our science targets, i.e. galaxies with a
cross-match in the COMBO-17 catalogue and $R_{\textrm{ap}}<24$. We
define `bad' fits as those where GALFIT stalled at one of the
constraints in Table~\ref{tab-constraints}. In Fig.~\ref{fig-badfits}
we show the fraction of those bad fits as a function of SExtractor
magnitude. At the bright end ($V_{606}<22$), the fraction of failures
is less than 6\% and rises steadily from there. Only when reaching the
(surface brightness) completeness limit (roughly at
$V_{606}\sim24-25$) does the fraction of failed fits reach (and
exceed) 20\%.

\subsection{Completeness and Fit Quality}\label{sec-sims}

To both derive completeness maps and examine fitting quality using
GALAPAGOS, we followed a similar approach as in GEMS and as described
in \cite{haeussler07}, but with a different, more realistic set of
simulated data. Whereas in \cite{haeussler07} a small set of only
1\,600 simulated galaxies was used to find the ideal setup of the
fitting pipeline, we have now decided on a fitting setup using
GALAPAGOS from the start and have carried out much more intensive
tests. We created entire sets of STAGES-like imaging data by
simulating galaxies in all 80 HST/ACS tiles.  Galaxies were
simulated as single-component \sersic\ profiles; multi-component
galaxies or complicated structures such as spiral arms or bars were
not included.

The sample of galaxies to be simulated was derived by using the fits
of real data as described in \S\ref{sec-fitting}. From this superset,
we selected a `galaxy sample' to be simulated by excluding both stars
and those galaxies for which the fit failed. Magnitudes and galaxy
sizes for the simulated galaxies were chosen according to the
probability distribution of this sample. The other simulation
parameters, (e.g. \sersic\ index $n$ and axis ratio $q$) were then
derived by choosing fitting values of real galaxies at approximately
the same magnitude and size.  In this way, the simulated data have
parameters as close as possible to the real galaxy sample.

To cover a larger number of parameter combinations, we slightly
smoothed these values (mag by $\pm$1 mag, $\log(r_e)$ by $\pm$0.25
pix, $n$ by $\pm$0.5 and $q$ by $\pm$0.2).  Care was taken to make
sure that $q$ and $n$ covered sensible values ($0.05 < q < 1$, $0.2 <
n < 8$). We also simulated galaxies two magnitudes fainter than those
found in the real data to be able to derive completeness maps from the
same pipeline. Twenty sets of STAGES-like data (80 tiles each) were
simulated using this setup. In a further 50 sets, we introduced a
uniform distribution of the \sersic\ index over the full range $0.2 <
n < 8$ over all magnitudes and sizes for 5\% of galaxies.  This
imposed pedestal was required in order to fill in gaps in the
parameter space with bad number statistics or no galaxies at all, and
was especially important for galaxies with high $n$-value seen
face-on.  Both position and position angle, $\theta$, were randomly
chosen for each galaxy: thus no clustering was simulated, in contrast
to the real data.  Simulating around 107\,000 objects per dataset, we
were able to derive an object density comparable to the real data with
a mean of 60\,612 galaxies found per dataset.  This compares to
75\,805 galaxies in the original GALAPAGOS output from the real data,
with $\sim$35\,000 objects in the `galaxy' catalogue from which we
draw the input parameters for the simulations.

After choosing the parameters this way, we used the same simulation
script that was described in detail in \cite{haeussler07} to simulate
the galaxies. The images were placed in an empty image which was made
up by empty patches of sky from the STAGES data to resemble the noise
properties of the real data. Convolution was performed using a STAGES
PSF.  In a change to the \cite{haeussler07} setup, we also simulated
galaxies on neighbouring tiles (or closely outside the data area) to
realistically model effects from neighbouring galaxies, as well as to
examine effects of combining the individual SExtractor catalogues
within GALAPAGOS.

\begin{figure}
\psfig{file=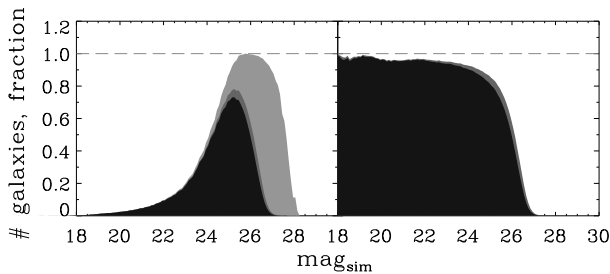,width=\columnwidth}
\caption{\label{fig:fit1} Completeness as a function of
magnitude. {\em Left}: the number of simulated galaxies (light grey),
recovered by SExtractor (dark grey), and subsequently fit successfully
by GALAPAGOS (black) as a function of input magnitude. {\em Right}:
Completeness functions for SExtractor (grey) and GALAPAGOS (black)
output. One can see that GALAPAGOS returns a useful result in most
cases.  Only for relatively faint galaxies does the fit run into
fitting constraints for a fraction of the objects. At $V_{606}\sim26$,
the STAGES profile fitting is therefore 80\% complete.}
\end{figure}

By simulating fainter galaxies than are found in the real data, we
were not only able to test the fitting quality but also the survey
completeness.  Fig. \ref{fig:fit1} shows the completeness as derived
from this data as a function of magnitude.  The left plot shows the
number of galaxies simulated (light grey), the number of galaxies
recovered (dark grey) and the number of galaxies with successful fit
(black; meaning that the fit did not run into any fitting
constraints).  All three histograms are normalized by the value
of the bin containing the maximum number of simulated galaxies. In
total, of the 7\,497\,614 galaxies simulated, 43.4\% were not found in
the data  using the GALAPAGOS and SExtractor setups used to
analyse the real STAGES data. Failed objects in general were too faint
to be detected.  A further 52.5\% were successfully recovered,
identified and fitted, and 4.0\% were recovered but excluded from all
plots as the fit ran into fitting constraints.  For 305 galaxies
(0.004\%), the fit crashed and did not return a result at all.

We additionally find 51\,043 galaxies (0.7\% of simulated galaxies)
that could not be identified by our search algorithm, which looked for
the closest match within 1.0\arcsec. Examination of these galaxies
shows that they are either (a) very low surface brightness galaxies
for which the SExtractor positioning was not very secure, or (b) two
neighbouring LSB galaxies that SExtractor detected as one object, also
resulting in an insecure position.

\begin{figure*}
\centerline{\psfig{file=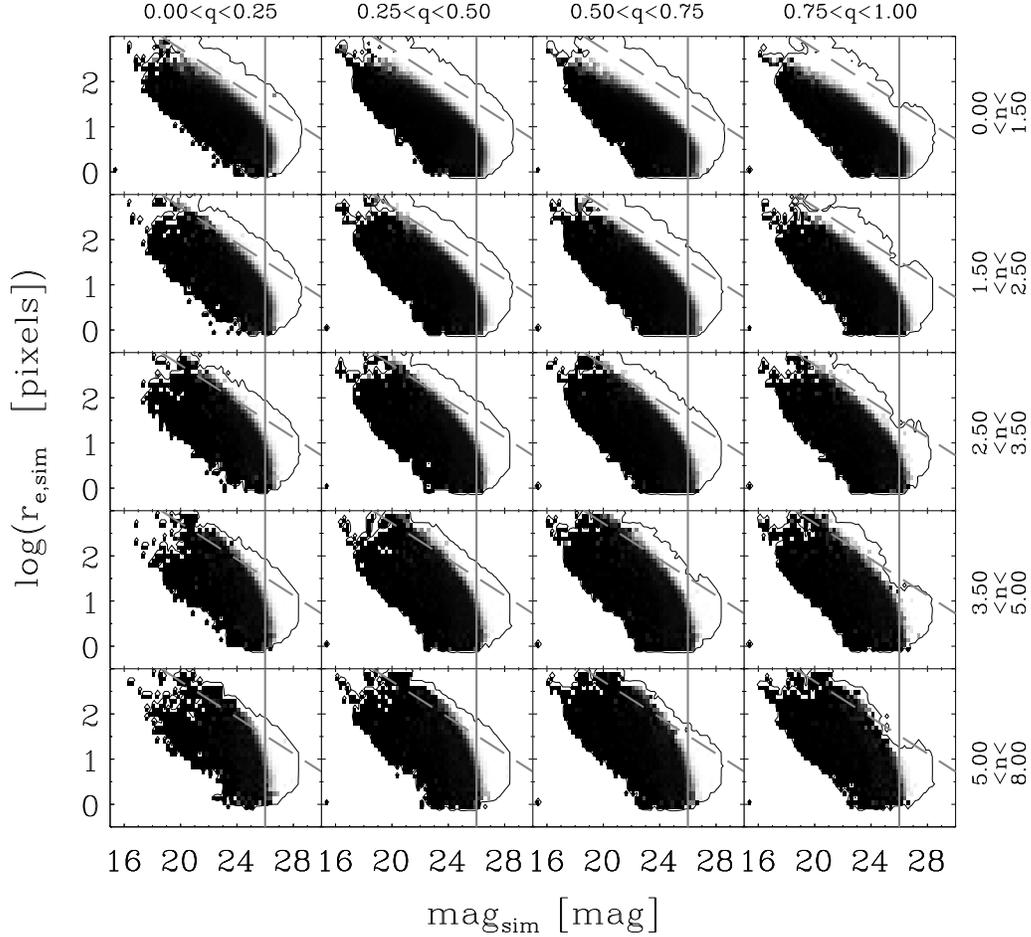,width=0.8\textwidth}}
\caption{\label{fig-fit2} Completeness maps as a function of \sersic\
index $n$ and axis ratio $q$ (as labelled above and to the right of the
plots).  To guide the eye, we overplot a vertical line at mag 26 and a
surface brightness line (diagonal, dashed) at 28 \magarc. As one can
clearly see, the completeness (shown in greyscale, black is complete,
white is incomplete or no data) is a strong function of all magnitude,
size (and therefore surface brightness), $q$ and $n$. The outline
contour shows the region in this plot where galaxies have been
simulated to demonstrate where these plots are reliable.}
\end{figure*}

Using the whole available simulated dataset, we can derive a much more
detailed completeness for STAGES.  Magnitude alone is not a good
estimator for completeness, as the internal light distribution has
great influence on this value.  More concentrated galaxy profiles,
such as elliptical, high-$n$ profiles, are more likely to be detected
by SExtractor than disk-like low-$n$ profiles.  In addition, the
inclination angle plays an important role. As shown in
Fig.~\ref{fig-fit2}, we can divide the galaxies in different bins of
$n$ and $q$ and for each bin can estimate a 2-D completeness map
showing the completeness as a function of both magnitude and galaxy
size. By looking at each bin one can clearly see that the completeness
is indeed a function of magnitude as well as size. The completeness
catalogue from these extensive simulations will be made publicly
available as part of the STAGES data release.  With the large sample
and complete coverage of the parameter space populated by real
galaxies, one could make up customized completeness maps tailored to
the particular sample in question.

\begin{figure*}
\centerline{\psfig{file=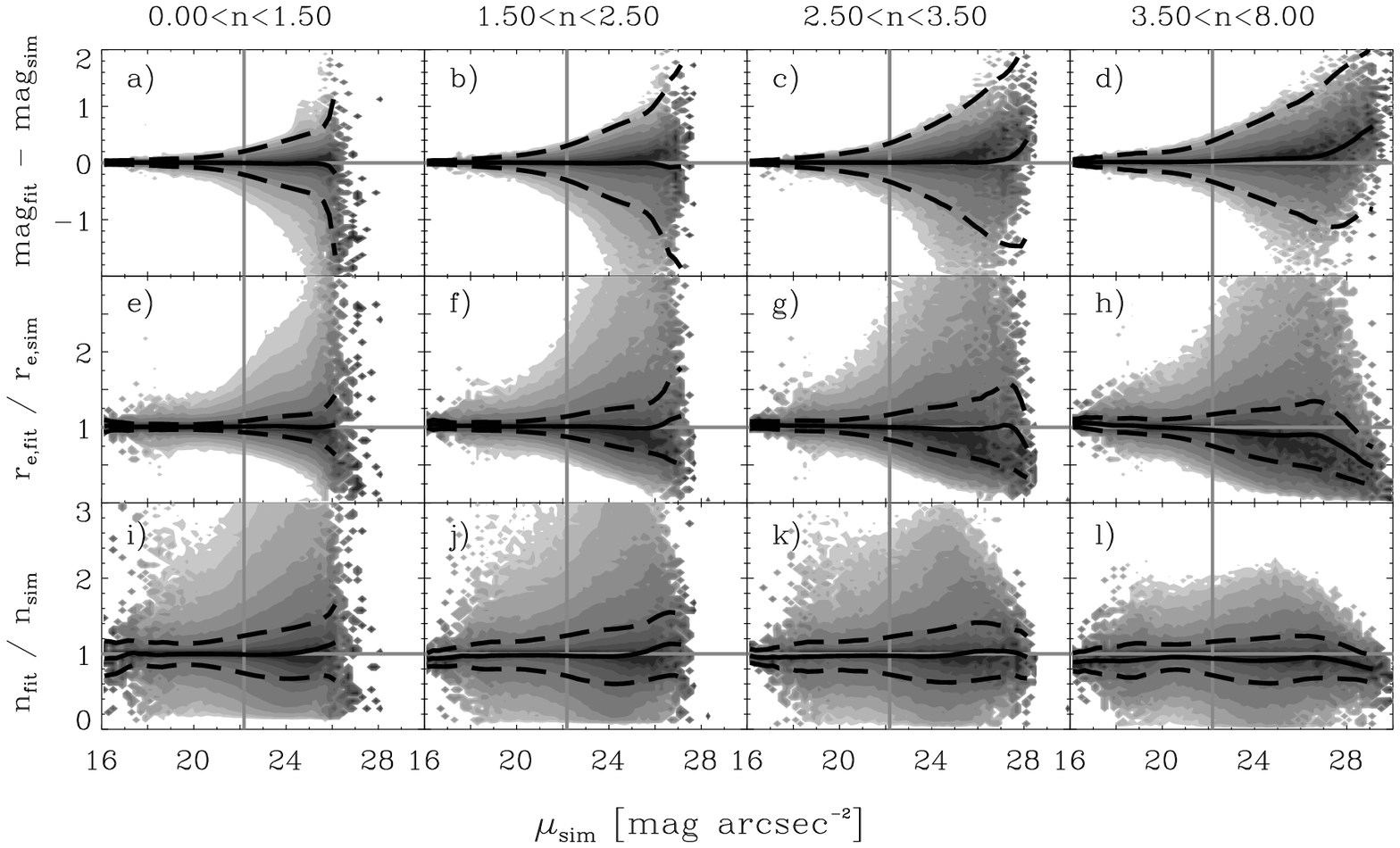,width=0.9\textwidth}}
\caption{\label{fig-fit3} Fit Quality. The deviations of the most
important galaxy parameters as a function of surface brightness. {\em
Top row}: Magnitude deviation (fit - simulated), {\em middle row}:
Size ratio (fit/simulated), {\em bottom row}: \sersic\ index n
(fit/simulated). Contours show the data normalized by the number
of galaxies in each surface brightness bin; the black solid line
shows the mean of the distribution; and the black dashed lines show
the sigma of the distribution ($3\sigma$ in case of magnitudes,
$1\sigma$ in size and \sersic\ index). All plots are shown for
different \sersic\ indices as labelled above the plots. The vertical
grey line represents the mean brightness of the sky background in
STAGES. The magnitude and sizes are less well recovered in high-$n$
galaxies, but the {\em relative} recovery of $n$ is similar in all
cases.}
\end{figure*}

The same is true for the fitting quality. As can be seen from
Fig. \ref{fig-fit3}, the fitting behaviour is a function of both
surface brightness and \sersic\ index. We only show the quality as a
function of \sersic\ index, but again one can determine fitting
quality as a function of any combination of the fitted parameters.
One can see that high-$n$ galaxies are harder to fit than low-$n$
galaxies, e.g. the magnitude deviation $\Delta$ is 0.00 ($\sigma =
0.07$) at around the sky level for galaxies with \mbox{$0<n<1.5$},
while $\Delta=0.03$ ($\sigma = 0.12$) at the highest $n$-bin. The
effect is even larger at fainter galaxies: $\Delta=0.00$ ($\sigma =
0.18$) at 25 \magarc and $\Delta=0.08$ ($\sigma = 0.28$) for low- and
high-$n$ galaxies, respectively. A similar trend can be seen for
galaxy sizes: $\Delta = 0.8\%$ ($\sigma = 7.3\%$) and $\Delta =
-3.7\%$ ($\sigma = 19.0\%$) at the sky level, and $\Delta = -0.4\%$
($\sigma = 18.3\%$) and $\Delta = -10.7\%$ ($\sigma = 36.1\%$) at 25
\magarc. If one examines relative deviations of the \sersic\ index,
there is essentially no trend seen between different bins of $n$.  In
an absolute sense, then, the \sersic\ index is still less well
recovered in the high-$n$ bin.

In general, the systematic deviations are very small except at the
faintest galaxies detectable, and both deviation $\Delta$ and $\sigma$
of the distributions are well understood within STAGES.  As was
pointed out in \cite{haeussler07}, the uncertainties returned by
GALFIT (and therefore GALAPAGOS) underestimate the true uncertainty by
a large amount.  Using a statistical approach therefore returns more
reliable errorbars for the individual parameters.  The simulations and
catalogue presented here allow a flexible means of estimating errors
on profile fitting for any possible subsample of galaxies.

\section{COMBO-17 data}\label{sec-c17}

\subsection{COMBO-17 observations and catalogue}\label{sec-c17cat}

In this section we briefly describe the COMBO-17 data on the A901/2
field, including observations, catalogue entries and object
samples. The corresponding data on the CDFS field were published in
\citet[][hereafter W04]{wolf04}, where further technical details can
be found.

The filter set (Table~\ref{tab-combodata}) contains five broad-band
filters (UBVRI) and 12 medium-band filters covering wavelengths from
350 to 930~nm.  All observations were obtained with the Wide Field
Imager (WFI) at the MPG/ESO 2.2m-telescope on La Silla, Chile. A field
of view of $34\arcmin \times 33\arcmin$ (see Fig.~\ref{fig-field})
is covered by a CCD mosaic consisting of eight 2k $\times$ 4k CCDs
with a scale of $0\farcs238$ per pixel. The observations on the A901/2
field were spread out over three observing runs between January 1999
and February 2001. They encompass a total exposure time of
$\sim$185~ks of which $\sim$20~ks were taken in the $R$-band during
the best seeing conditions. A dither pattern with at least ten
telescope pointings spread by \mbox{$\Delta\alpha$, $\Delta\delta <\pm
72\arcsec$} allowed us to cover the sky area in the gaps of the CCD
mosaic.

\begin{table*}
\caption{COMBO-17 imaging data on the A901/2 field: For all filters we
list total exposure time, average PSF among individual frames, the
10$\sigma$ (Vega) magnitude limits for point sources and the observing
runs (see Tab.~\ref{tab-comboruns}) in which the exposure was
collected. For flux and magnitude conversions we list AB magnitudes
and photon fluxes of Vega in all filters. The $R$-band observations were
taken in the best seeing conditions.
\label{tab-combodata}}
\begin{tabular}{ccrccl|cc}
\hline \hline
  \multicolumn{2}{c}{$\lambda_\mathrm{\mathrm{cen}}$/fwhm} &
  $t_\mathrm{\mathrm{exp}}$ & seeing &
  $m_\mathrm{\mathrm{lim},10\sigma}$ & run code & mag of Vega &
  $F_\mathrm{phot}$ of Vega \\ \multicolumn{2}{c}{(nm)} & (sec) & &
  (Vega mags) & & (AB mags) & $(10^8~\mathrm{photons/m^2/nm/s})$ \\
  \noalign{\smallskip} \hline \noalign{\smallskip} 365/36 &$U$ & 22100
  & 1\farcs10 & 23.7 & G & $+0.77$ & 0.737\\ 458/97 &$B$ & 20500 &
  1\farcs20 & 25.4 & A, G & $-0.13$ & 1.371\\ 538/89 &$V$ & 6000 &
  1\farcs20 & 24.3 & E & $-0.02$ & 1.055\\ 648/160 &$R$ & 20300 &
  0\farcs75 & 25.0 & E & $+0.19$ & 0.725\\ 857/147 &$I$ & 7500 &
  1\farcs00 & 22.7 & E & $+0.49$ & 0.412\\ \noalign{\smallskip} 418/27
  & & 7300 & 1\farcs20 & 24.0 & E & $-0.19$ & 1.571\\ 462/13 & & 10000
  & 1\farcs20 & 23.7 & E & $-0.18$ & 1.412\\ 486/31 & & 5500 &
  1\farcs15 & 24.0 & E & $-0.06$ & 1.207\\ 519/16 & & 6000 & 1\farcs05
  & 23.6 & E & $-0.06$ & 1.125\\ 572/25 & & 5000 & 0\farcs85 & 23.5 &
  E & $+0.04$ & 0.932\\ 605/21 & & 6000 & 0\farcs95 & 23.4 & E &
  $+0.10$ & 0.832\\ 645/30 & & 4950 & 1\farcs30 & 22.7 & E & $+0.22$ &
  0.703\\ 696/21 & & 6600 & 1\farcs00 & 22.7 & E & $+0.27$ & 0.621\\
  753/18 & & 7000 & 1\farcs05 & 22.2 & E & $+0.36$ & 0.525\\ 816/21 &
  & 19200 & 0\farcs85 & 22.8 & A & $+0.45$ & 0.442\\ 857/15 & & 16600
  & 1\farcs15 & 21.7 & E & $+0.56$ & 0.386\\ 914/26 & & 15700 &
  0\farcs95 & 21.9 & E & $+0.50$ & 0.380\\ \noalign{\smallskip} \hline
\end{tabular}
\end{table*}

\begin{table}
\caption{COMBO-17 observing runs with A901/2 imaging. 
\label{tab-comboruns} }
\begin{center}
\begin{tabular}{ll}
\hline \hline
COMBO-17 run code    &  Dates  \\ 
\hline
A           &  11.02.-22.02.1999  \\
E           &  28.01.-11.02.2000  \\
G           &  19.01.-20.01.2001  \\
\hline
\end{tabular}
\end{center}
\end{table}

Flux calibration was done with our own tertiary standard stars based
on \emph{spectrophotometric} observations, a suitable method to
achieve a homogeneous photometric calibration for all 17 WFI filter
bands. Two G stars with $B\simeq 15$ (with COMBO-17 identification
numbers 45811 and 46757) were observed at La Silla with DFOSC at the
Danish 1.54\,m telescope. A wide ($5\arcsec$) slit was used for the
COMBO-17 standards as well as for an external calibrator star.

The object search for the COMBO-17 sample was done with SExtractor
software \citep{bertin96} in default setup, except for choosing a
minimum of 12 significant pixels required for the detection of an
object.  We first search rather deep and then clean the list of
extracted objects of those having a S/N ratio below 4, which
corresponds to $>0\fm2422$ error in the total magnitude MAG\_BEST. As
a result we obtained a catalogue of 63\,776 objects with positions,
morphology, total $R$-band magnitude and its error.  The astrometric
accuracy is better than $0\farcs15$. Using our own aperture photometry
we reach a 5$\sigma$ point source limit of $R\approx 25.7$.

We obtained spectral energy distributions of all objects from
photometry in all 17 passbands by projecting the known object
coordinates into the frames of reference of each single exposure and
measuring the object fluxes at the given locations. In order to
optimize the signal-to-noise ratio, we measure the spectral shape in
the high surface brightness regions of the objects and ignore
potential low surface brightness features at large distance from the
centre. However, this implies that for large galaxies at low redshifts
$z<0.2$ we measure the SED of the central region and ignore colour
gradients.

Also, we suppressed the propagation of variations in the seeing into
the photometry by making sure that we always probe the same physical
footprint outside the atmosphere of any object in all bands
irrespective of the PSF. Here, the footprint $f(x,y)$ is the
convolution of the PSF $p(x,y)$ with the aperture weighting function
$a(x,y)$. If all three are Gaussians, an identical physical footprint
can be probed even when the PSF changes, simply by adjusting the
weighting function $a(x,y)$. We chose to measure fluxes on a footprint
of $1\farcs5$ FWHM outside the atmosphere ($\sim 4.2$ kpc at $z\sim
0.165$).  In detail, we use the package MPIAPHOT \citep{mpiaphot93} to
measure the PSF on each individual frame, choose the weighting
function needed to conserve the footprint and obtain the flux on the
footprint. Fluxes from individual frames are averaged for each object
and the flux error is derived from the scatter. Thus, it takes not
only photon noise into account, but also suboptimal flatfielding and
uncorrected CCD artifacts.

All fluxes are finally calibrated by the tertiary standards in our
field.  The aperture fluxes correspond to total fluxes for point
sources, but underestimate them for extended sources. The difference
between the total (SExtractor-based) and the aperture (MPIAPHOT-based)
magnitude is listed as an aperture correction and used to calculate
e.g. luminosities.  For further details on the observations and the
data processing, see W04.

The A901/2 field is affected by substantial foreground dust reddening at
the level of $E(B-V)\approx 0.06$, in contrast to the CDFS. Hence,
any SED fitting and derivation of luminosities requires dereddened
SEDs.  Therefore, in the catalogue we list three sets of photometry:

\begin{enumerate}
\item $R$-band total and aperture magnitudes as observed for the definition 
of samples and completeness;
\item aperture fluxes $F_{\rm phot}$ in 17 bands, dereddened using 
$A_V=0.18$ and $(A_U,A_B,A_R,A_I)=A_V \times (1.63,1.24,0.82,0.6)$ with
similar numbers for medium-band filters (rereddening with these 
numbers would restore original measurements); and 
\item aperture magnitudes (Vega) in all 17 bands, dereddened, on the
Asinh system \citep{lupton99} that can be used for logarithmic flux
plots with no trouble arising from formally negative flux
measurements.
\end{enumerate}

Fluxes are given as photon fluxes $F_\mathrm{phot}$ in units of 
photons/m$^2$/s/nm, which are related to other flux definitions by
\begin{equation}
   \nu F_\nu = hcF_{\rm phot} = \lambda F_\lambda   ~ .
\end{equation}
Photon fluxes are practical units at the depth of current surveys.  A
magnitude of $V=20$ corresponds to 1~photon/m$^2$/s/nm in all systems
(AB, Vega, ST), provided $V$ is centred on 548~nm. Flux values of an
object are missing in those bands where every exposure was saturated.

The final catalogue contains quality flags for all objects in an
integer column (`phot\_flag'), holding the original SExtractor flags
in bit 0 to 7, corresponding to values from 0 to 128, as well as some
COMBO-17 quality control flags in bits 9 to 11 (values from 512 to
2048). We generally recommend that users ignore objects with flag
values phot\_flag $\ge 8$ for any statistical analysis of the object
population. If an object of particular interest shows bad flags, it
may still have accurate COMBO-17 photometry and could be used for some
purposes. Often only the total magnitude was affected by bright
neighbours, while the aperture SED is valid.

We then employ the usual COMBO-17 classification and redshift
estimation by template fitting to libraries of stars, galaxies, QSOs
and white dwarfs.  There, the error rate increases very significantly
at $R_{\rm ap}>24$. We refer again to W04 for details of the libraries
and known deficiencies of the process, but repeat here (and correct a
misprint in W04) the definition of the classifications (see
Table~\ref{tab-comboclasses}).

\begin{table}
\caption{Definition of entries for the `mc\_class' column and
comparison of object numbers between the COMBO-17 data sets of the
A901/2 and CDFS field. The samples refer to a magnitude range of
$R_{\rm ap}=[16,24]$ and only objects with phot\_flag$<8$. The A901/2
field is richer in stars because of its galactic coordinates. It is
also richer in galaxies due to the cluster, while the CDFS is
underdense at $z=[0.2,0.4]$.  We note that these definitions are based
on the COMBO-17 data SED and morphology; star-galaxy separation
employing morphological information from the HST imaging
(Equation~\ref{eqn-stargal}) is considered separately.
\label{tab-comboclasses}}
\begin{tabular}{llrr}
\hline\hline
Class entry      &  Meaning  &  $N\_{\rm A901/2}$  &  $N\_{\rm CDFS}$ \\
\hline
Star             &  stars 			& 2096	& 992	\\
                 &  (only point sources) \\
WDwarf           &  white dwarf			&  14	&   9	\\
                 &  (only point sources) \\
Galaxy           &  galaxies			& 14555 & 11054	 \\
                 &  (shape irrelevant) \\
Galaxy  (Star?)  &  binary or low-z galaxy &  44	&  46  \\
                 &  (star SED but extended; \\
		 &  ambiguous colour space) \\
Galaxy  (Uncl!)  &  SED fit undecided 		& 316	& 243	\\
                 &  (most often galaxy) \\
QSO              &  QSOs 			&  73	&  66	\\
                 &  (only point sources) \\
QSO     (Gal?)   &  Seyfert-1 AGN or            & 36	&  31	\\
                 &  interloping galaxy	\\
                 &  (AGN SED but extended; \\
		 &  ambiguous colour space) \\
Strange Object   &  unusual strange spectrum 	&   1	&   3	\\
                 &  ($\chi ^2_{\rm red}>30$) \\
\noalign{\smallskip} \hline
\end{tabular}
\end{table}

\begin{figure*}
\includegraphics*[height=\textwidth,angle=270]{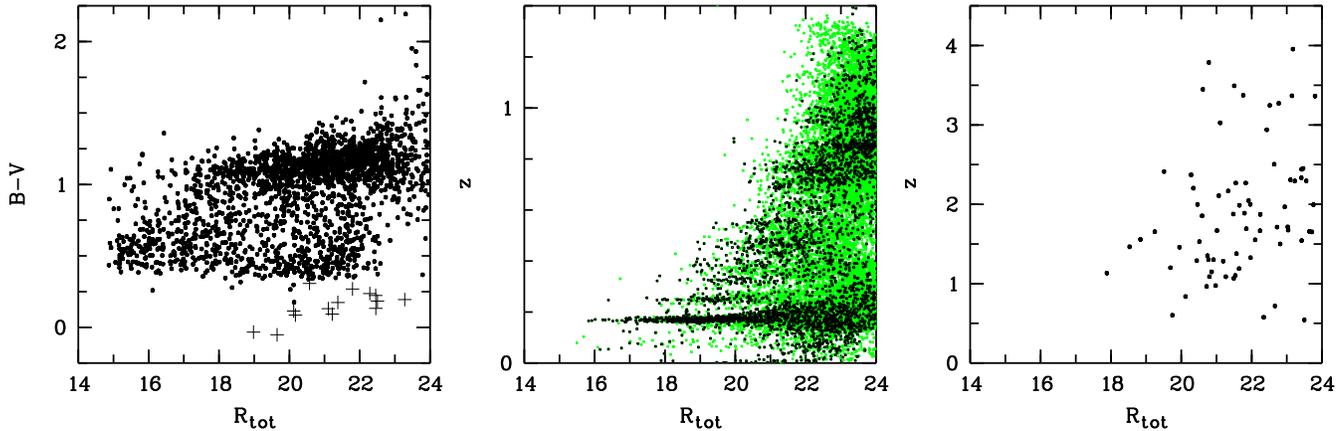}
\caption{ {\it Left panel:} Stars (dots) and white dwarfs (crosses):
$B-V$ colour vs.  $R_{\rm tot}$. The two reddest stars at $R\approx 23$ and
$B-V>2$ are M5-6 stars.  {\it Centre panel:} Red-sequence (black) and
blue-cloud galaxies (green): MEV redshift vs.  $R_{\rm tot}$.  {\it Right
panel:} QSOs: MEV redshift vs.  $R_{\rm tot}$.
\label{fig-CW_class}}
\end{figure*}

We also show in Table~\ref{tab-comboclasses} a comparison of the sample
sizes in different classes between the A901/2 and the CDFS field of
COMBO-17. The main difference is that the A901/2 field contains more
than twice the number of stars given its position at relatively low
galactic latitude ($+33.6\deg$).  Another difference is that it
contains 30\% more galaxies than the CDFS, which is both a consequence
of the cluster A901/2 and the underdensity in the CDFS seen at
$z\sim[0.2,0.4]$. Fig.~\ref{fig-CW_class} shows a colour-magnitude
diagram of the star and white dwarf sample as well as redshift-magnitude
diagrams for galaxies and QSOs.

Redshifts are given as Maximum-Likelihood values (the peak of the
PDF), or as Minimum-Error-Variance values (the expectation value of
the PDF). MEV redshifts have smaller true errors, but are only given
when the width of the PDF is lower than $\sigma_z/(1+z)< 0.125$. If
PDFs are bimodal with modes of sufficiently small width, then both
values are given with the preferred (larger-integral) mode providing
the primary redshift. Our team uses only MEV redshifts (with column
name `mc\_z') for their analyses.

The galaxy sample with MEV redshifts is $>90$\% complete at all
redshifts for $R_{\rm ap}<23$. Near $z\sim 1$, the MEV redshifts are
this complete even at $R_{\rm ap}=24$. Below this cut, increasing
photon noise drives an expansion of the width of the PDF. The error
limit for MEV redshifts then makes the completeness of galaxy samples
with MEV redshifts drop. The 50\% completeness is reached at $R\sim
24$ to $25$ depending on redshift. These results have been determined
from simulations and are detailed in W04.  Completeness maps are
included in the data release and take the form of a 3-D map of
completeness depending on aperture magnitude, redshift and restframe
$U-V$ colour.

To date, the photo-z quality on the A901/2 field has only been
investigated with a comparison to spectroscopic redshifts at the
bright end.  W04 reported results from a sample of 404 bright galaxies
with $R<20$ and $z=[0,0.3]$, 351 of which were on the A901/2 field,
and 249 of which were members of the A901/2 cluster complex
(\S~\ref{sec-2df}). The other 53 objects were observed by the 2dFGRS
on the CDFS and S11 fields
\citep{Colless2001}. There we found that 77\% of the sample had
photo-z deviations from the true redshift
$|\delta_z/(1+z)|<0.01$. Three objects (less than 1\%) deviate by more
than 0.04 from the true redshift.

Currently, we do not have faint spectroscopic samples on the A901/2
field, however a spectroscopic dataset from VVDS exists on the
COMBO-17 CDFS field. From a sample of 420 high-quality redshifts that
are reasonably complete to $R_{\rm ap}<23$, we find a $1\sigma$
scatter in $\delta_z/(1+z)$ of 0.018, but also a mean bias of
$-0.011$. Furthermore, the faint CDFS data show $\sim 5$\% outliers
with deviations of more than 0.06 \citep{HWB08}. From a collection of
spectroscopic samples we modelled the overall 1$\sigma$ redshift
errors at $R\la 24$ and $z\la 1$ in W04 as
\begin{equation}
   \sigma_z/(1+z) \approx 0.005 \times \sqrt{1+10^{0.6 (R_{\rm ap}-20.5)}}  ~ .
   \label{dz_rel}
\end{equation}
Later we use a variant of this approximation to estimate the
completeness of photo-z based selection rules for cluster members.

The template fitting for galaxies produces three parameters,
i.e. redshift as well as formal stellar age and dust reddening
values. The age is encoded in a template number running from 0
(youngest) to 59 (oldest), where we use the same PEGASE
\citep[see][for discussion of an earlier version of the model]{fioc97}
template grid as described in W04. The look back times to the onset of
the $\tau =1$~Gyr exponential burst range from 50~Myr to 15~Gyr.

Restframe properties are derived for all galaxies and QSOs as
described in W04. Table~\ref{tab-rfvega} lists the restframe passbands
we calculate and gives conversion factors from Vega magnitudes to AB
magnitudes and to photon fluxes. The SED shape is defined by the
aperture photometry and the overall normalization is given by the
total SExtractor photometry from the deep $R$-band. However, if a
galaxy has both a steep colour gradient {\it and} a large aperture
correction, then the restframe colours will be biased by the nuclear
SED.

\begin{table}
\caption{The restframe passbands and their characteristics. 
\label{tab-rfvega} }
\begin{tabular}{ll|cc}
\hline \hline
  name  &  $\lambda_\mathrm{\mathrm{cen}}$/fwhm  & 
  mag of Vega  &  $F_\mathrm{phot}$ of Vega \\
     &  (nm)  &  (AB mags)  &  $(10^8~\mathrm{phot/m^2/nm/s})$  \\ 
\noalign{\smallskip} \hline \noalign{\smallskip} 
(synthetic)     &  145/10  & $+2.33$ & 0.447\\ 
(synthetic)     &  280/40  & $+1.43$ & 0.529\\ 
\noalign{\smallskip} \hline \noalign{\smallskip} 
Johnson $U$     &  365/52  & $+0.65$ & 0.820\\ 
Johnson $B$     &  445/101 & $-0.13$ & 1.407\\ 
Johnson $V$     &  550/83  & $+0.00$ & 1.012\\ 
\noalign{\smallskip} \hline \noalign{\smallskip} 
SDSS $u$        &  358/56  & $+0.84$ & 0.704\\ 
SDSS $g$        &  473/127 & $-0.11$ & 1.305\\
SDSS $r$        &  620/115 & $+0.14$ & 0.787\\ 
\hline
\end{tabular}
\end{table}

The column `ApD\_Rmag' contains the magnitude difference between the
total object photometry and the point-source calibrated,
seeing-adaptive aperture photometry:
\begin{equation}
   ApD\_Rmag = Rmag - Ap\_Rmag   ~ .
\end{equation}
On average, this value is by calibration zero for point sources, and
becomes more negative for more extended sources.

\subsection{Cross-correlation of STAGES and COMBO-17 catalogues}\label{sec-xcorr}

Having created separate catalogues from the STAGES
(\S\ref{sec-detect},\S\ref{sec-fitting}) and COMBO-17
(\S\ref{sec-c17cat}) datasets, we next wish to create a combined,
master catalogue.  In GEMS, this was accomplished by applying a
nearest neighbour matching algorithm with a maximum matching radius of
0\farcs75.  The choice of maximum radius is governed by the resolution
of the two datasets (HST: 0\farcs1; COMBO-17: 0\farcs75).

For STAGES we have however chosen to improve over this approach. For most
galaxies, their measured centres do not change if the input image is
smoothed. For example, if the HST image of a normal spiral or
elliptical galaxy is convolved with a Gaussian function to match the
ground-based seeing, the centre estimated from the high-resolution (in
this case STAGES) and the low-resolution (here COMBO-17) images should
coincide. For distorted galaxies or mergers, this may no longer be the
case.  Instead, the brightest peak in the STAGES image, detected as
the object centre by SExtractor, may be relatively far from the centre
in the COMBO-17 image.

\begin{figure*}
\centerline{\psfig{file=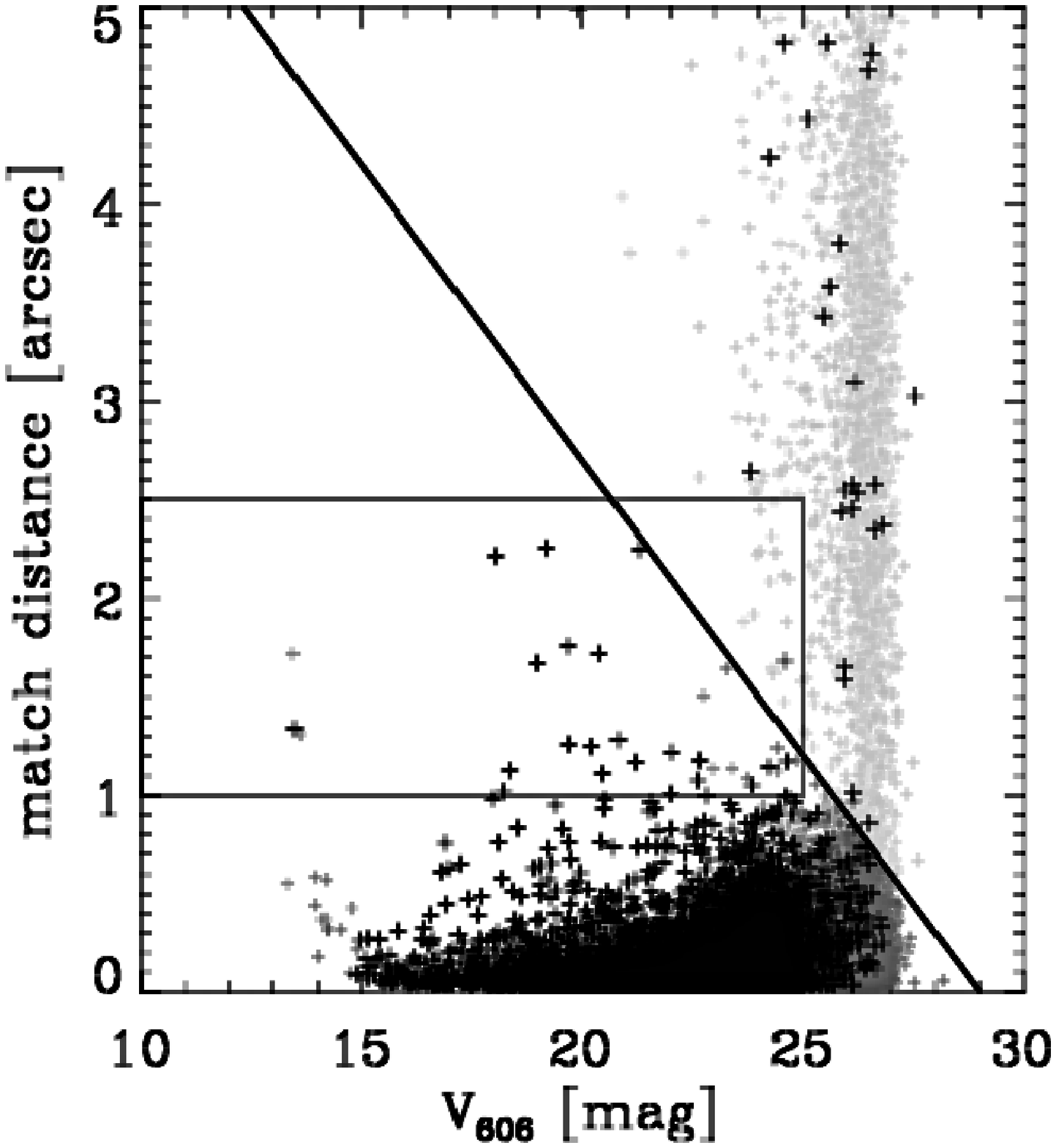,width=0.7\columnwidth}
\hspace{0.1\columnwidth}\psfig{file=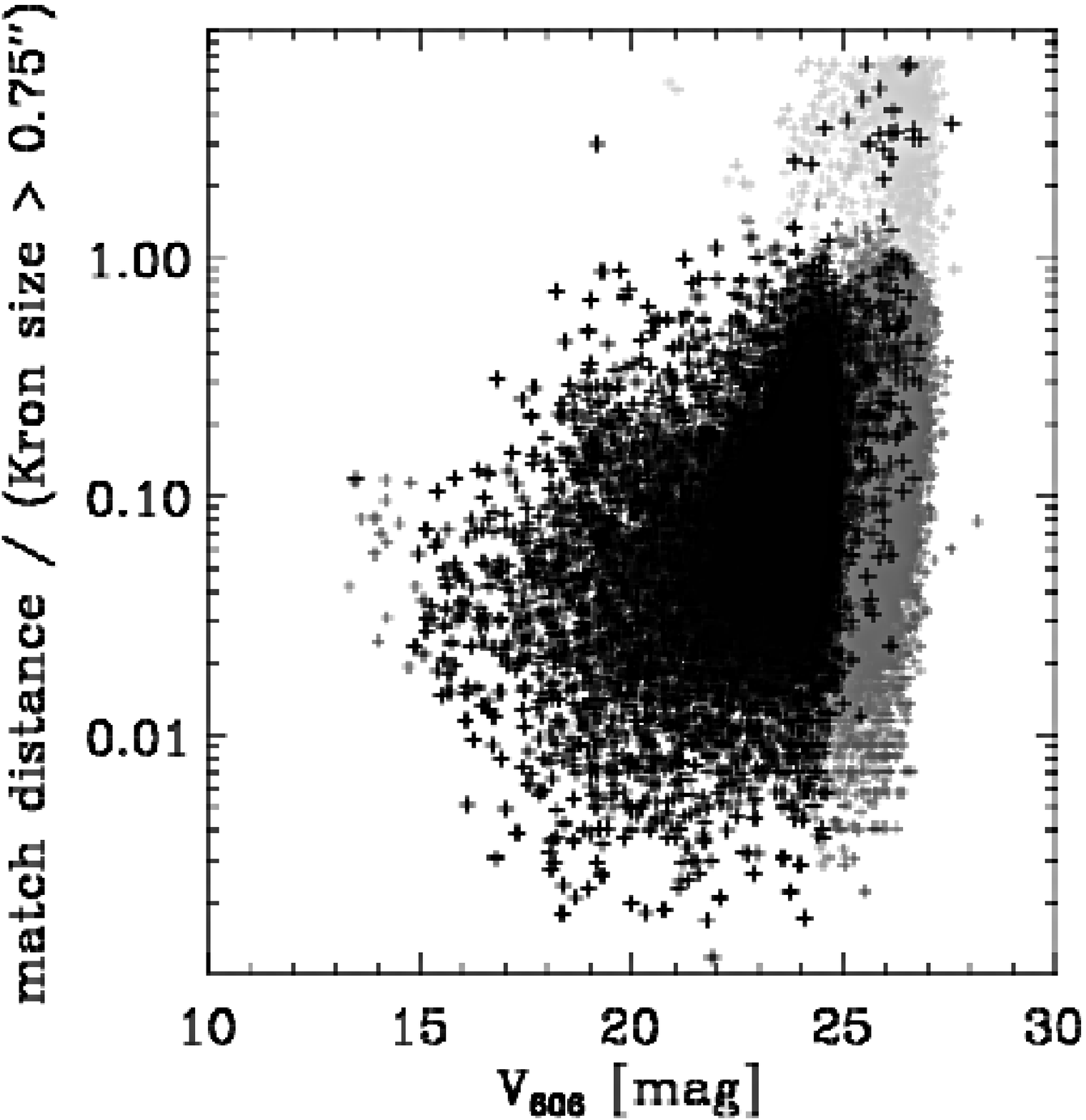,width=0.75\columnwidth}}
\caption{Cross-correlation of HST and COMBO-17 data.  {\it Left:} The
distance to the nearest neighbour within a search radius of 5\arcsec
is plotted as a function of HST magnitude. At the faint end galaxies
are matched to uncorrelated neighbours.  Resolving irregular
structures in the HST images results in detected galaxy centres being
located farther from the COMBO-17 galaxy centre than a seeing
distance. Matching bright objects at large separations while removing
random correlations at faint fluxes requires a cut as indicated by the
diagonal line.  Objects within the box ($V_{606}<25$ and
$1\arcsec<$ match distance $<2.5\arcsec$) were inspected by eye. {\it
Right:} Ratio of matching distance and Kron size as a function of HST
magnitude. Values larger than $\sim1$ imply a matching radius larger
than the object size in the HST image.  Sources with
$R_{\textrm{ap}}<24$ are shown as black symbols; objects with a match
below the cut (diagonal line in left panel) are plotted in dark grey;
the remaining sources with a match within 5\arcsec\ are shown as light
grey symbols.
}\label{fig-xcorr2}
\end{figure*}

\begin{figure}
\centerline{\psfig{file=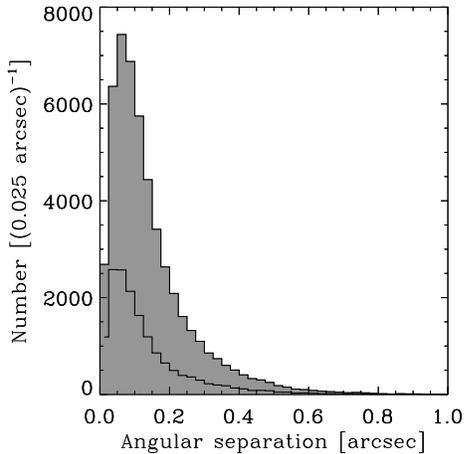,width=0.85\columnwidth}}
\caption{Histogram of matching radii for all objects (outer
  histogram) and $R_{\rm ap}<24$ objects (inner histogram). The
  typical angular separation between a COMBO-17 object with
  $R_{\textrm{ap}}<24$ and its HST counterpart is $\sim 0.12$\arcsec
  $\pm 0.08$\arcsec.}\label{fig-radhist}
\end{figure}

In order to maximise the number of good matches between STAGES and
COMBO-17, in particular at low redshift, i.e.\ A901/2 cluster
distance, we have devised the following scheme. For STAGES the average
source density corresponds to roughly two objects per 5\arcsec-radius
circle. We cross-correlate the STAGES and COMBO-17 catalogues using a
nearest neighbour matching algorithm as described above with a maximum
matching radius of 5\arcsec. The resulting matches we plot in
Fig.~\ref{fig-xcorr2} (left panel). In particular at faint
magnitudes many matches are found that appear unrelated. In
contrast, at brighter magnitudes several sources are correlated at
radii much larger than the COMBO-17 seeing (0.75\arcsec), which still
identify the same object. In Fig.~\ref{fig-xcorr2} we also show a
line that subdivides the plot into two regions:
\begin{equation}
d_m=-0.3\times\left(V_{606}-29\right),
\end{equation}
with the matching radius $d_m$ in arcsec and the STAGES SExtractor
magnitude $V_{606}$. Below the line, objects are considered to be
correlated, while above they are not correlated. This division is
empirically motivated by the requirement to match objects at the faint
end out to the COMBO-17 resolution limit (0.5\arcsec-1.0\arcsec) while
also correlating sources at larger radii at the bright end. The slope
of the curve was determined by visual inspection of the matches inside
the indicated box. Typically, the distance between centroids is
$\sim 0\farcs1$ (Fig.~\ref{fig-radhist}).

Another way of investigating this issue is by calculating whether the
nearest matching neighbour falls within the area covered by the object
in the STAGES image. If the projected COMBO-17 position is beyond
the optical extent of the source in STAGES, it is uncorrelated. From
the STAGES SExtractor data we estimate the `extent' of an object by
its Kron size $K=r_\textrm{K}\times a$, from the Kron radius
$r_\textrm{K}$ and semi-major axis radius $a$. We limit the Kron
size to $K>0.75$\arcsec. A ratio of $d_m/K \ga 1$ indicates that
the matched COMBO-17 source lies outside the region covered by the
object in the STAGES image. In Fig.~\ref{fig-xcorr2} (right panel)
we overplot in grey all sources that were assigned a partner from the
nearest neighbour matching. This provides further evidence for the
improved quality of our new cross-correlation method.

In summary, the combined catalogue contains 88\,879 sources. Of these,
$\sim6\,577$ objects with a COMBO-17 ID are not within the region
covered by the STAGES HST mosaic ($\sim 1\,664$ of these have
$R_{\textrm{ap}}<24$).  Moreover, $\sim 1\,271$ STAGES detections are
outside the COMBO-17 observation footprint.\footnote{The observation
footprint for both STAGES and COMBO-17 is rather difficult to
determine. Therefore, we provide only approximate numbers good to
$\sim50$ objects. A more elaborate scheme than the one used to produce
these numbers is well beyond the scope of this paper.} Inside the
region covered by both surveys, there are $\sim 81\,031$ sources. For
50701 objects the method described above provides a match between
COMBO-17 and STAGES (15760 of these have $R_{\textrm{ap}}<24$). $\sim
23\,833$ sources detected in STAGES do not have counterparts in
COMBO-17; $\sim 6\,497$ sources from the COMBO-17 catalogue are not
matched to STAGES detections. Out of these, only $\sim 79$ objects
have \mbox{$R_{\textrm{ap}}<24$}. We therefore emphasize that for
our science sample of COMBO-17 objects, defined as having
$R_{\textrm{ap}}<24$, 99.9\% have a STAGES counterpart.  The majority
of failures result from confusion by neighbouring objects or simply
non-detections.

\subsection{Selection of an A901/2 cluster sample}

We wish to define a `cluster' galaxy sample of galaxies belonging to
the A901/2 complex for various follow-up studies of our team that are
in progress. These studies may have different requirements for the
{\em completeness} of cluster members and the {\em contamination} by
field galaxies. We therefore quantified how these two key values vary
with both magnitude and width of the redshift interval in order to
inform our choice of definition.

The photo-z distribution of cluster galaxies was assumed to follow a
Gaussian with a width given by the photo-z scatter in
Equation~\ref{dz_rel}.  The distribution of field galaxies was assumed to
be consistent with the average galaxy counts $n(z,R)$ outside the
cluster and varies smoothly with redshift and magnitude assuming no
structure in the field. Samples were then defined by redshift
intervals $z_{\rm phot}=[0.17-\Delta z, 0.17+\Delta z]$, where the
half-width $\Delta z$ was allowed to vary with the
magnitude.\footnote{We use $z_{\rm phot}=0.17$ for the mean cluster
redshift here rather than the spectroscopically confirmed $z_{\rm
spec}\sim0.165$ due to the known bias discussed in
\S\ref{sec-c17cat}.} We calculated completeness and contamination at
all magnitude points simply using the counts of our smooth models.

We found that as long as the half-width in redshift is not much larger
than a couple of Gaussian FWHMs, the contamination changes only
little.  The ratio of selected cluster to field galaxies is almost
invariant as shrinking widths cut into numbers for both origins. Only
enlarging the width significantly over that of the Gaussian increases
contamination by field galaxies. On the contrary, such large widths do
not affect the completeness of the cluster sample much, while
shrinking the width too far eats into the true cluster distribution
and reduces completeness of the cluster sample.

\begin{figure}
\centerline{\includegraphics*[height=0.85\columnwidth,angle=270]{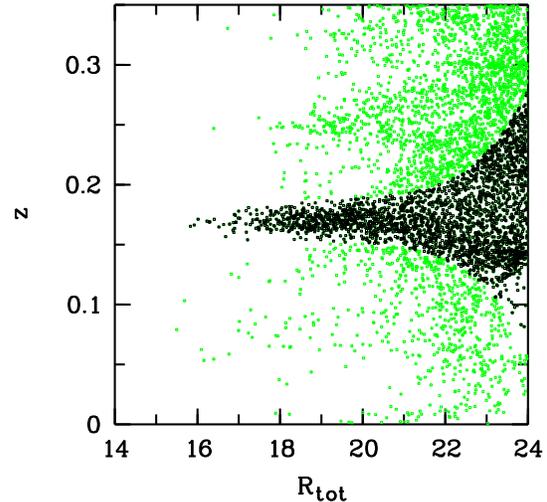}}
\caption{MEV redshift estimate vs. total $R$-band magnitude. The
'galaxy' sample is shown in green, while the sample of `cluster'
galaxies defined by Equation~\ref{eqn-cluster} is shown in black.  The
magnitude-dependent redshift interval guarantees almost constant high
completeness, while the field contamination increases towards faint
levels (Fig.~\ref{fig-CW_comcon}).  We note that at faint
magnitudes there is an apparent asymmetry towards lower redshift at
faint magnitudes within the cluster sample.  The photometric redshifts
may be skewed by systematic effects but the average $-0.02$ offset at
$R_{\rm tot}\sim22.5$ is within the $1\sigma$ error envelope.
\label{fig-CW_clus}}
\end{figure}

\begin{figure}
\centering
\includegraphics*[height=\columnwidth,angle=270]{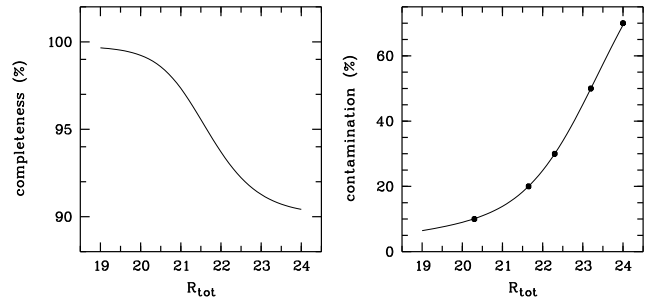}
\caption{{\it Left panel:} Completeness of the cluster sample defined
in Fig.~\ref{fig-CW_clus} and designed to provide high completeness at all 
magnitudes.
{\it Right panel:} The field contamination of the cluster sample 
increases at faint levels due to photo-z dilution of the cluster. 
Narrowing the selected redshift interval would not reduce the 
contamination. Contamination rates are estimated to be
$(10,20,30,50,70)\%$ at $R_{\rm ap}=(20.3,21.65,22.3,23.2,24.0)$.
\label{fig-CW_comcon}}
\end{figure}

For our purposes, we compromised on a photo-z width such that the
completeness is $>90$\% at any magnitude, just before further widening
starts to increase the contamination above its mag-dependent minimum
(see Fig.~\ref{fig-CW_clus} and Fig.~\ref{fig-CW_comcon}, left
panel). For this we chose a half-width of

\begin{equation}\label{eqn-cluster}
   \Delta z (R) = \sqrt{0.015^2+0.0096525^2 
                (1+10^{0.6 (R_{\rm tot}-20.5)})  }  ~ .
\end{equation}

This equation defines a half-width that is limited to 0.015 at the
bright end and expands as a constant multiple of the estimated photo-z
error at the faint end. The floor of the half-width is motivated by
including the entire cluster member sample previously studied by
WMG05. The completeness of this selection converges to nearly 100\%
for bright galaxies, as a result of intentionally including the WGM05
sample entirely.

The right panel of Fig.~\ref{fig-CW_comcon} shows that the
differential contamination increases rapidly towards faint magnitudes,
simply as a result of the photo-z error-driven dilution of the cluster
sample. Here, contamination means the fraction of galaxies that are
field members, as measured in a bin centred on the given magnitude
with width 0.1 mag. Contamination at a given apparent magnitude
translates into contamination at a resulting luminosity at the cluster
distance (except that scatter in the aperture correction smears out
the contamination relation slightly).

Already at $R_{\it ap}=23.2$ the sample contains as many cluster as
field members. This corresponds to $M_V \approx -16.5$ for the average
galaxy, but scatters around that due to aperture corrections.  As we
probe fainter this selection adds more field galaxies than cluster
members. Follow-up studies can now determine an individual magnitude
or luminosity limit given their maximum tolerance for field
contamination. For example, WGM05 selected cluster galaxies at $M_V -
< -17.775$ ($M_V < -17$ for their adopted cosmology with $H_0=100$ km
s$^{-1}$ Mpc$^{-1}$) for an earlier study of the A901/2 system in
order to keep the contamination at the faint end below 20\%.

The cluster sample thus obtained covers quite a range of photo-z
values at the faint end, and restframe properties are derived assuming
these redshifts to be correct. However, if we assume a priori that an
object is at the redshift of the cluster, then we may want to know
these properties assuming a fixed cluster redshift of
$z=0.167$. Hence, the SED fits and restframe luminosities are
recalculated for this redshift and reported in additional columns of
the STAGES catalogue in Table~\ref{tabcolumns} (with \mbox{`\_cl' }
suffix indicating cluster redshift). Of course, if the a-priori
assumption is to believe the redshifts as derived, then the original
set of columns for which we have derived the values is relevant.

\section{Further multiwavelength data and derived
  quantities}\label{sec-multi}

In this section we describe further multiwavelength data for the
A901/2 region taken with other facilities (Fig.~\ref{fig-field}).  We
also present several resulting derived quantities (stellar masses and
star formation rates) that appear as entries in the STAGES master
catalogue.

\subsection{Spitzer}\label{sec-mir}

Spitzer observed a $1\degr \times 0{\fdg}5$ field around the A901/2
system in December 2004 and June 2005 as part of Spitzer GO-3294 (PI:
Bell).  The MIPS 24{\micron} data were taken in slow scan-map mode,
with individual exposures of 10\,s.  We reduced the individual image
frames using a custom data-analysis tool (DAT) developed by the GTOs
\citep{gordon05}.  The reduced images were corrected for geometric
distortion and combined to form full mosaics; the reduction which we
currently use does not mask out asteroids and other transients in the
mosaicing.\footnote{This only minimally affects our analyses because we
match the IR detections to optical positions, and most of the bright
asteroids are outside the COMBO-17 field.}  The final mosaic has a
pixel scale of $1\farcs25$~pixel$^{-1}$ and an image PSF FWHM of
$\simeq 6$\arcsec.  Source detection and photometry were performed
using techniques described in \citet{papovich04}; based on the
analysis in that work, we estimate that our source detection is 80\%
complete at 97~$\mu$Jy\footnote{We note that for previous papers we
used the catalogue to lower flux limits, down to 3$\sigma$;
accordingly, we have included such lower-significance (and more
contaminated) matches in the catalogue.}  for a total exposure of
$\sim 1400$\,s\,pix$^{-1}$.  By detecting artificially-inserted
sources in the A901 24 image, we estimated the completeness of the
A901 24 $\mu$m catalog. The completeness is 80\%, 50\% and 30\% at 5,
4 and 3$\sigma$, respectively.

Note that there is a very bright star at 24{\micron} near the centre
of the field at coordinates $(\alpha,\delta)_{\rm J2000}=(09^h56^m32\fs 4,
-10\degr 01 \arcmin 15\arcsec)$ (see \S\ref{sec-mira} for
details of this object). In our analysis of the 24{\micron} data we
discard all detections less than 4$'$ from this position in order to
minimise contamination from spurious detections and problems with the
background level in the wings of this bright star.  It is to be noted
that there are a number of spurious detections in the wings of the
very brightest sources; while we endeavoured to minimise the incidence
of these sources, they are difficult to completely eradicate without
losing substantial numbers of real sources at the flux limit of the
data.

To interpret the observed 24{\micron} emission, we must match the
24{\micron} sources to galaxies for which we have redshift estimates
from COMBO-17.  We adopt a 1\arcsec matching radius.  In the areas of
the A901/2 field where there is overlap between the COMBO-17 redshift
data and the full-depth MIPS mosaic, there are a total of 3506(5545)
24{\micron} sources with fluxes in excess of 97(58)$\mu$Jy.  Roughly
62\% of the 24{\micron} sources with fluxes $> 58 \mu$Jy are detected
by COMBO-17 in at least the deep $R$-band, with $R \la 26$.  Some 50\%
of the 24{\micron} sources have bright $R_{\rm tot} < 24$ and have
photometric redshift $z<1$; these 50\% of sources contain nearly 60\%
of the total 24{\micron} flux in objects brighter than 58$\mu$Jy.
Sources fainter than $R \ga 24$ contain the rest of the $f_{24} >
58\mu$Jy 24{\micron} sources; investigation of COMBO-17 lower
confidence photometric redshifts, their optical colours, and results
from other studies lends weight to the argument that essentially all
of these sources are at $z>0.8$, with the bulk lying at $z > 1$
(e.g. \citealt{lefloch2004}, \citealt{papovich04}; see
\citealt{lefloch2005} for a further discussion of the completeness of
redshift information in the CDFS COMBO-17 data).

Observations with IRAC \citep[Infrared Array Camera;][]{Fazio2004} at
3.6, 4.5, 5.8 and 8.0{\micron} were also taken as part of this Spitzer
campaign: those data are not discussed further here, and will be
described in full in a future publication.

\subsection{Star formation rates}

We provide estimates of star formation rate, determined using a
combination of 24{\micron} data (to probe the obscured star formation)
and COMBO-17 derived rest-frame 2800{\AA} luminosities (to probe
unobscured star formation).  Ideally, we would have a measure of the
total thermal IR flux from 8--1000{\micron}; instead, we have an
estimate of IR luminosity at one wavelength, 24{\micron},
corresponding to rest-frame 22--12{\micron} at the redshifts of
interest $z=0.1-1$.  Local IR-luminous galaxies show a tight
correlation between rest-frame 12--15{\micron} luminosity and total IR
luminosity \citep[e.g.,][]{spi95,cha01,rou01,papovich02}, with a
scatter of $\sim 0.15$ dex.\footnote{Star-forming regions in local
galaxies appear to follow a slightly non-linear relation between
rest-frame 24{\micron} emission and SFR, with SFR $\propto L_{24\mu
m}^{0.9}$ \citep{calzetti07}, although note that this calibration is
between 24{\micron} emission and SFR (not total IR luminosity).}
Following \citet{papovich02}, we choose to construct total IR
luminosity from the observed-frame 24{\micron} data.  We use the Sbc
template from the \citet{dev99} SED library to translate
observed-frame 24{\micron} flux into the 8--1000{\micron} total IR
luminosity.\footnote{Total 8--1000{\micron} IR luminosities are $\sim
0.3$ dex higher than the 42.5--122.5{\micron} luminosities defined by
\citet{Helou1988}, with an obvious dust temperature
dependence.}  The IR luminosity uncertainties are primarily
systematic.  Firstly, there is a natural diversity of IR spectral
shapes at a given galaxy IR luminosity, stellar mass, etc.; one can
crudely estimate the scale of this uncertainty by using the full range
of templates from \citet{dev99}, or by using templates from, e.g.,
\citet{dale01} instead.  This uncertainty is $\la$0.3\,dex (this
agrees roughly with the scatter seen between 24{\micron} luminosity
and SFR seen in \citealp{calzetti07}).  Secondly, it is possible that
a significant fraction of $0.1<z<1.0$ galaxies have IR spectral energy
distributions not represented in the local Universe: while it is
impossible to quantify this error until the advent of Herschel Space
Telescope, current results suggest that the bulk of intermediate--high
redshift galaxies have IR spectra similar to galaxies in the local
universe \citep{appleton04,elbaz05,yan05,zheng07}.

We estimate SFRs using the combined directly-observed UV light from young 
stars and the dust-reprocessed IR emission of the sample galaxies
\citep[e.g.,][]{fluxrat}.  Following \cite{bell05}, 
we estimate SFR $\psi$ using a calibration 
derived from PEGASE assuming a 100\,Myr-old stellar population 
with constant SFR and a \citet{chabrier03} IMF: 
\begin{equation}
\psi / ({\rm M_{\sun}\,yr^{-1}}) = 9.8 \times 10^{-11} \times
       (L_{\rm IR} + 2.2L_{\rm UV}),  \label{eqn:sfr}
\end{equation}
where $L_{\rm IR}$ is the total IR luminosity (as estimated above) and
$L_{\rm UV} = 1.5 \nu l_{\nu,2800}$ is a rough estimate of the total
integrated 1216{\AA}--3000{\AA} UV luminosity, derived using the
2800{\AA} rest-frame luminosity from COMBO-17 $l_{\nu,2800}$.  The
factor of 1.5 in the 2800{\AA}-to-total UV conversion accounts for the
UV spectral shape of a 100 Myr-old population with constant SFR, and
the UV flux is multiplied by a factor of 2.2 before being added to the
IR luminosity to account for light emitted longwards of 3000{\AA} and
shortwards of 1216{\AA} by the unobscured young stars.  This SFR
calibration is derived using identical assumptions to \citet{k98}, and
the calibration is consistent with his to within 30\% once different
IMFs are accounted for.  Uncertainties in these SFR estimates are a
factor of two or more in a galaxy-by-galaxy sense, and systematic
uncertainty in the overall SFR scale is likely to be less than a
factor of two \citep[see, e.g.,][for further discussion of
uncertainties]{bellsfr,bell05}.  The adopted calibration assumes
that the infrared luminosity traces the emission from young stars
only; contributions from potential AGN can be identified and excluded
by cross-matching with the X-ray and optical data as in
\citet{gilmour2007} and \citet{gallazzi08}.

Again, for galaxies in the `cluster' sample, we present also SFR
estimates assuming that the galaxies are at the cluster redshift with
the suffix `\_cl' in added to the column name.

\subsection{Stellar Masses}

\citet{borch06} estimated the stellar masses of galaxies in COMBO-17
using the 17-passband photometry in conjunction with a template
library derived using the PEGASE stellar population model.  The
non-evolving template stellar populations had an age/metallicity
combination equivalent to roughly solar metallicity and $\sim 6$\,Gyr
since the start of star formation.\footnote{Local comparison samples,
e.g., the SDSS, typically adopt template combinations with `older'
ages, potentially leading to offsets between the overall mass scale of
our masses and local masses at a given rest-frame colour.  We make no
attempt to resolve this issue here, and refer the interested reader to
\citet{belldejong} and \citet{Bell07} for further discussion of this
issue.}  \citet{borch06} adopted a \citet{kroupa93} stellar IMF; the
use of a \citet{kroupa01} or \citet{chabrier03} IMF would have yielded
the same stellar masses to within $\sim 10$\%.  Such masses are
quantitatively consistent with those derived using a simple
colour-stellar M/L relation \citep{bell03}, and comparison of stellar
and dynamical masses for a few $z \sim 1$ early-type galaxies yielded
consistent results to within their combined errors (see
\citealt{borch06} for more details).

There are some galaxies for which the 17-band classification failed to
find a satisfactory solution (2\% of the galaxies with redshift
estimates); we choose to adopt in these cases a rest-frame
colour-derived stellar mass, using rest-frame $B$ and $V$ absolute
magnitudes/luminosities, and a $V$-band absolute magnitude of the Sun
of 4.82:
\begin{equation}
\log_{10} M_*/M_{\sun} = -0.728 + 1.305(B-V) + \log_{10} L_V/L_{\sun}.
\end{equation}

As with restframe photometric properties, we also present estimates of
stellar mass assuming that the galaxy is at the cluster redshift
(denoted in the catalogues by the suffix `\_cl' in the column names).
Random stellar mass errors are estimated to be $\sim 0.1$\,dex on a
galaxy-by-galaxy basis in most cases, and systematic errors in the
stellar masses (setting the overall mass scale and its redshift
evolution) were argued to be at the 0.1\,dex level for galaxies
without ongoing or recent major starbursts; for galaxies with strong
bursts, masses could be overestimated by $\la 0.5$\,dex.

Finally, we note potential aperture effects on stellar masses and
SEDs for some objects. The colours are estimated within an aperture
but are normalized by the total light in the deep $R$-band image
alone.  For small objects or particularly large objects without colour
gradients this has no consequence. But if large size, low
concentration and strong colour gradients are combined, the total SED
will deviate from the aperture SED underlying the $M/L$ estimate.  In
a companion paper studying properties of spiral galaxies in the
supercluster, Wolf et al. (MNRAS, accepted) have investigated this effect by
examining the total colours across a wide parameter space in the
sample.  In most cases the aperture values are similar to the total
ones, but they identify an issue for morphologically-classified spiral
galaxies in the supercluster and eliminate the highest-mass regime
with $\log M_*/M_{\sun} > 11$ from their study.

\subsection{GALEX}

The Abell 901/902 field was observed by GALEX in the far-UV ($f,
\lambda_{\rm eff}\sim 1528$\AA ) and near-UV ($n, \lambda_{\rm
eff}\sim 2271$\AA) bands.\footnote{Unlike all other datasets detailed
here, the GALEX observations were not led by members of the STAGES
team.  We list the publicly archived data products here for
completeness.} Individual observations (or single orbit `visits')
between the dates 12 February 2005 and 25 February 2007 were coadded
by the GALEX pipeline \citep[GR4 version][]{Morrissey2007} to produce
images with net exposure times of 57.18 ks in $n$ (47 visits) and
50.19 ks in $f$ (40 visits).  The GALEX field of view in both bands is
a 0.6\degr radius circle, and the average centre of the visits (the
GALEX field centre) is $(\alpha,\delta)_{\rm J2000} = (9^{\rm
h}56^{\rm m}20\fs7, -10\degr6\arcmin21\farcs6)$. The GALEX PSF near
the field centre has $\sim 4.2$\arcsec\ FWHM at $f$ and $\sim
5.3$\arcsec\ FWHM at $n$, both of which increase with distance from
the field centre (variations in the PSF that are not a function of
distance from the field centre are smoothed out by the distribution of
roll angles of the visits). The astrometric accuracy is $\sim
0.7$\arcsec, and $>97$\% of catalogued source positions are within
2\arcsec~of their true positions. The photometric calibration is stable
to $0.02$ mag in $n$ and $0.045$ mag in $f$ \citep{Morrissey2007}.

Source detection and photometry is via the GALEX pipeline code, which
employs a version of SExtractor \citep{bertin96} modified for use with
low-background images. Magnitudes are measured both in fixed circular
apertures and in automatic Kron elliptical apertures, and in isophotal
apertures. The 5$\sigma$ point-source sensitivities in the Abell
901/902 field are $f \sim 24.7$ mag (AB) and $n \sim 25.0$ mag (AB),
though there are spatial variations across field, especially a
slightly decreasing sensitivity towards the edge of the field.  At
these levels source confusion in the $n$ band becomes an issue, and
the $n$ band fluxes of faint objects ($n\ga 23$ mag) are likely to
be overestimated.  GALEX data products include intensity, background,
and relative response (i.e., effective exposure time) maps in both
bands as well as source catalogues in both bands and a band-merged
source catalogue.

\subsection{2dF spectroscopy}\label{sec-2df}

Spectra of cluster galaxies were obtained using the 2dF
instrument on the AAT in March 2002 and March 2003.  A total of 86
galaxies were observed using the 1200B grating (spanning the observed
wavelength range 4000--5100 \AA) in a single fibre configuration during
the 2002 run.  Three fibre configurations using the lower resolution
600V grating (spanning 3800--5800 \AA) were observed during the 2003
run: fibres were placed on 368 objects, with 47 repeated from 2002.
The primary selection function assigned higher priority to those
galaxies selected by photometric redshift to be within the
supercluster redshift slice and having $R<20$, with additional fibres
being allocated to secondary targets (including fainter galaxies and a
small number of white dwarfs and QSOs) when available.  Data reduction
was performed with the standard {\tt 2dfdr} (v2.3) pipeline package.

In total, spectra were obtained for 407 unique objects.  Redshifts
were determined by two independent means: firstly by manual line
profile fitting of the Ca H and K features in absorption and secondly
by cross-correlation with template spectra using the XCSAO task within
IRAF (Kurtz \& Mink 1998).  Comparison of the two measurements showed
no cause for concern, with $\Delta_z=0.00149\pm0.00006$. After
eliminating non-galaxy and poor quality spectra, we have redshifts for
353 galaxies in total.  

The 2dF spectroscopic data have previously been used to quantify the
reliability of the COMBO-17 redshifts in W04 (see also \S\ref{sec-c17}),
to verify cluster membership for the matched X-ray point sources
\citep{gilmour2007}, and to create composite spectra for three
photometric classes of cluster galaxies in WGM05.  A dynamical
analysis of the the clusters using the 2dF redshifts will be presented
in Gray et al. (in prep.).

\subsection{XMM-Newton} \label{sec-xray}

X-ray data for the A901/2 region is desirous both to detect
point-source emission from cluster members (star-formation or AGN) and
the extended intracluster medium (ICM).  A 90~ks XMM image of the
A901/2 field was taken on May 6/7 2003 using the three EPIC cameras
(MOS1, MOS2 and PN) and a thin filter, under program 14817 (PI: Gray).
The level 1 data were taken from the supplied pipeline products, and
reduced with SAS v5.4 and the calibration files available in May 2003.
Final exposure times were $\sim67$ ks for MOS and $\sim$61 ks for PN
following the removal of time intervals suffering from soft proton
flares.  Four energy bands were used: 0.5-2 keV (soft band), 2-4.5 keV
(medium band), 4.5-7.5 keV (hard band) and 0.5-7.5 keV (full band).

The creation of the point-source catalogue using wavelet detection
methods is described in detail elsewhere \citep{gilmour2007}.  A total
of 139 significant sources were found.  The presence of an X-ray
luminous Type-I AGN near the centre of A901a (see
Appendix~\ref{sec-agnnote}) complicated the detection of the underlying
extended cluster emission.  A maximum-likelihood technique was used to
match this catalogue to COMBO-17 resulting in 66 secure counterparts
with photometric redshifts. \cite{gilmour2007} used these data to
examine the local environments of the cluster AGN and their host
properties.

To isolate the remaining extended emission coming from the clusters, a
separate conservative point-source catalogue was constructed.  Care
was taken to remove both the cosmic background and spatial variations in
the non-cosmic background.  The background subtracted images were
weighted by appropriate energy conversion factors to create flux
images for each detector.  These flux images were masked and summed
together to create merged background-subtracted images in each band.

Point source regions were removed and replaced with the local
background value selected randomly from a source free area within 10 pixels
(or 20 pixels if there were not enough background pixels within the
smaller radius). Smoothed images were created in each band using a
Gaussian kernel of radius 4 pixels.  Maps of the extended emission and
an examination of the global X-ray properties of the clusters will be
presented in Gray et al. (in prep.).

\subsection{GMRT}

The A901/2 field was observed on 2007 March 25th and 26th March with
the Giant Metrewave Radio Telescope (GMRT, see
\citealt{Ananthakrishnan2005} for further details).  The field was
centred at $(\alpha,\delta)_{\rm J2000}=(09^{\rm h} 56^{\rm m} 17^{\rm
s}, -10\degr 01\arcmin 28\arcsec)$ and observed at 610 and 1280~MHz on
respective nights. The GMRT is an interferometer, consisting of thirty
antennas, each 45~m in diameter. The bright sources 3C147 and 3C286
were observed at the start and end of each observing session, in order
to set the flux density scale. During the observations a nearby
compact source 0943$-$083 was observed for about 4 minutes at roughly
30 minutes intervals, to monitor and correct any antenna-based
amplitude and phase variations.

The total integration time on the field was $\sim$6.5 hours at
each frequency. The observations covered two 16~MHz sidebands,
positioned above and below the central frequency. Each sideband was
observed with 128 narrow channels, in order to allow narrow band
interference to be identified and efficiently removed. The observed
visibility data were edited and calibrated using standard tasks with
the AIPS package, and then groups of ten adjacent channels were
averaged together, with some end channels discarded. This reduced the
volume of the visibility data, whilst retaining enough channels so
that chromatic aberration is not a problem \citep[e.g., see][for further
details of GMRT analysis]{Garn2007}. Given the relatively large field
of view of the GMRT compared with its resolution, imaging in AIPS
requires several `facets' to be imaged simultaneously, and then be
combined. Preliminary imaging results, after several
iterations of self-calibration, have produced images with resolutions
of about 5\arcsec and 2\farcs5 at 610 and 1280-MHz respectively, with
r.m.s.\ noises of approximately 25 and 20 $\mu$Jy~beam$^{-1}$ in the
centre of the fields, before correction for the primary beam of the
GMRT. The primary beam -- i.e.\ the decreasing sensitivity away from
the field centres due to sensitivity of individual 45-m antennas -- is
approximately Gaussian, with a half-power beam width (HPBW) of
approximately 44\arcmin and 26\arcmin at 610 and 1280-MHz
respectively. These images are among the deepest images made at these
frequencies with the GMRT.  Further analysis and the source catalogue
will be presented in Green et al.(in prep.).

\subsection{Simulations and mock galaxy catalogues}\label{sec-mocks}

In order to facilitate the interpretation of the observational results
and to study the physical processes of galaxy evolution, N-body,
hydrodynamic, and semi-analytic simulations that closely mimic the
A901/2 system are being produced (van Kampen et al., in prep.).  We
constrain initial conditions using the method of \citet{hoffman1991}
to take into account the gross properties of A901a, A901b, A902, the
SW group, and the neighbouring clusters A868 and A907 (outside the
observed field).  The simulations produce a range of mock large-scale
structures to test three basic formation scenarios: a 'stationary'
case, where A901(a,b) and A902 will not merge within a Hubble time,
and a pre- as well as a post-merger scenario.  When the likelihood of
each scenario is understood, one can further test the models for the
detailed physical processes known to be operating on galaxies in and
around such clusters.

\section{Summary and Data Access}\label{sec-summary}

We have presented the multiwavelength data available for the A901/2
supercluster field as part of the STAGES survey: high-resolution HST
imaging over a wide area, extensive photometric redshifts from
COMBO-17, and further multiwavelength observations from X-ray to
radio.  These data have already been used to create a high resolution
mass map of the system using weak gravitational lensing
\citep{heymans08}.  
Further work by the STAGES team to study galaxy evolution and
environment is ongoing and includes the following:

\begin{itemize}

\item \citet{gallazzi08} explore the amount of obscured star-formation
as a function of environment in the A901/2 supercluster and associated
field sample by combining the UV/optical SED from COMBO-17 with the
Spitzer 24$\mu$m photometry in galaxies with $M_*>10^{10} M_{\sun}$.
Results indicate that while there is an overall suppression in the
fraction of star-forming galaxies with density, the small amount of
star formation surviving the cluster environment is to a large extent
obscured.

\item Wolf et al. (MNRAS, accepted) investigate the properties of optically
passive spiral and dusty red galaxies in the supercluster and find
that the two samples are largely equivalent.  These galaxies form
stars at a substantial rate that is only a factor of four times lower
than blue spirals at fixed mass, but their star formation is more
obscured and has weak optical signatures. They constitute over half of
the star forming galaxies at masses above $\log M_*/M_{\sun}=10$ and
are thus a vital ingredient for understanding the overall picture of
star-formation quenching in cluster environments.

\item Marinova et al. (ApJ, submitted) identify and characterize bars in
bright ($M_{V} \le -18$) cluster galaxies through ellipse-fitting. The
selection of moderately inclined disk galaxies via three commonly used
methods, visual classification, colour, and \sersic\ cuts, shows that
the latter two methods fail to pick up many red, bulge-dominated disk
galaxies in the clusters.  However, all three methods of disk
selection yields a similar global optical bar fractions ($f_{\rm
bar-opt}\sim0.3)$, averaged over all galaxy types.  When host galaxy
properties are considered, the optical bar fraction is found to be a
strong function of both the luminosity and morphological property
(bulge-to-disk ratio) of the host galaxy, similar to trends recently
reported in field galaxies.  Furthermore, results indicate that the
global optical bar fraction for bright galaxies is not a strong
function of local environment.

\item Heiderman et al. (in prep.) identify interacting galaxies in the
supercluster using quantitative analysis and visual classifications.
Their findings include that $4.9\pm 1.3\%$ of bright ($M_{V} \le
-18$), intermediate mass ($M_{*} \ge 1 \times 10^{9} M_{\sun}$)
galaxies are interacting. The interacting galaxies are found to lie
outside the cluster cores and to be concentrated in the region between
the cores and virial radii of the clusters.  Explanations for the
observed distribution include the large galaxy velocity dispersion in
the cluster cores and the possibility that the outer parts of the
clusters are accreting groups, which are predicted to show a high
probability for mergers and strong interactions. The average star
formation rate is enhanced only by a modest factor in interacting
galaxies compared to non-interacting galaxies, similar to conclusions
reported in the field by \citet{Jogee08}.  Interacting galaxies only
contribute $\sim$~20\% of the total SFR density in the A901/902
clusters.

\item Boehm et al. (in prep.) are utilizing the stability of the PSF on
the STAGES images for a morphological comparison between the hosts of
~20 type-1 AGN and ~200 inactive galaxies at an average redshift
$\left < z \right >\sim0.7$. This analysis includes extensive simulations of the
impact of a bright optical nucleus on quantitative galaxy morphologies
in terms of the CAS indices and Gini/$M_{20}$ space. We find that the
majority of the hosts cover parameters typical for disk+bulge systems
and mildly disturbed galaxies, while evidence for strong gravitational
interactions is scarce.

\item Bacon et al. (in prep.) are examining the higher order lensing
properties of the STAGES data. They construct a shapelets catalogue
\citep{refregier03} for the STAGES galaxies; this is then used to
estimate the gravitational flexion \citep{bacon06} at each galaxy
position. Galaxy--galaxy flexion is measured, leading to estimates of
concentration and mass for STAGES galaxies; constraints on cosmic
flexion are also found, showing very good containment of systematic
effects. The ability of flexion to improve convergence maps is also
discussed.

\item Robaina et al. (in prep.) make use of a combined GEMS and STAGES
sample of $0.4<z<0.8$ galaxies to find that interacting and merging
close pairs of massive galaxies ($>10^{10} M_{\odot}$) show a modest
enhancement of their star formation rate; in particular, less than
15\% of star formation at $0.4<z<0.8$ is triggered by major
interactions and mergers.

\item Barden et al. (in prep.) are exploring both the GEMS and STAGES
data sets to investigate the evolution of structural parameters of
disc galaxies as a function of luminosity and stellar mass over a wide
range of environments and morphologies. In the process, GALAPAGOS will
be extended to perform bulge/disc decomposition.

\item McIntosh et al. (in prep.) are using both quantitative and
qualitative morphologies to explore the morphological mix of red
sequence galaxies as a function of stellar mass over the last seven
billion years from the combined STAGES + GEMS sample.

\end{itemize}

It is our intention that the data products described here should be
publicly available for use by the wider community for those interested
in the supercluster itself or for data-mining the entire survey
volume.  To that end, the reduced HST images (both tiles and
individual galaxy postage stamps) are available for download at the
Multimission Archive at Space
Telescope\footnote{\texttt{http://archive.stsci.edu}} (MAST).
Furthermore, the complete STAGES catalogue described in this paper is
available from the STAGES
website,\footnote{\texttt{http://www.nottingham.ac.uk/astro/stages}}
including all HST-derived parameters; GALFIT profile fitting results;
COMBO-17 photometry, SEDs and photometric redshifts; and stellar
masses and star-formation rates.  The multiwavelength data available
there includes the Spitzer/MIPS 24\micron\ images and catalogue; the
X-ray point source catalogue \citep{gilmour2007} and the gravitational
lensing mass maps \citep{heymans08}.  GALEX data and catalogues are
available via MAST.  The X-ray maps, 2dF spectra and radio catalogue
and mocks will be also be placed on the website with the publication
of their associated papers, or may be made available upon request.
Table~\ref{tab-data} contains a summary of the available data
products.

\begin{table*}
\caption{Description of all available A901/902 data products.}
\begin{tabular}{lll}
\hline
Data product & Date of release & Reference \\
\hline
\hline
HST F606W imaging, reduced:  tiles, thumbnails, colour jpegs & immediate &
this paper\\
STAGES master catalogue: SExtractor, GALFIT, COMBO-17, stellar masses,
SFRs& immediate & this paper\\
COMBO-17 SEDS and completeness tables & immediate & this paper\\
GALFIT profile fitting completeness from simulations & immediate & this paper\\
Spitzer 24\micron\ imaging and catalogue & immediate & this paper\\
HST-derived weak lensing mass map & immediate & Heymans et al 2008\\
XMM point source catalogue & immediate & Gilmour et al 2007\\
GALEX imaging and catalogues (from the GALEX archive) &  immediate & this paper\\
X-ray imaging & on request &  Gray et al. (in prep.)\\
2dF spectroscopy & on request  & Gray et al. (in prep.)\\
GMRT catalogue & TBC & Green et al. (in prep.)\\
constrained simulations and mock galaxy catalogues & TBC & van Kampen et al. (in prep.)\\
\hline
\end{tabular}\label{tab-data}
\end{table*}

\section{Acknowledgements}
The STAGES team would like to thank Hans-Walter Rix for his crucial
support in bringing this project to fruition.  We also thank Alfonso
Arag\'{o}n-Salamanca, Anna Gallazzi, Amanda Heiderman, Irina Marinova,
and Aday Robaina for their work in exploiting the STAGES dataset.
Support for STAGES was provided by NASA through GO-10395 from STScI
operated by AURA under NAS5-26555.  MEG and CW were supported by STFC
Advanced Fellowships.  CH acknowledges the support of a European
Commission Programme 6th framework Marie Cure Outgoing International
Fellowship under contract MOIF-CT- 2006-21891. CYP was supported by
the NRC-HIA Plaskett Fellowship, and the STScI Institute/Giacconi
Fellowship. EFB and KJ are grateful for support from the DFG's Emmy
Noether Programme of the Deutsche Forschungsgemeinschaft, AB by the
DLR (50 OR 0404), MB and EvK by the Austrian Science Foundation FWF
under grant P18416, SFS by the Spanish MEC grants AYA2005-09413-C02-02
and the PAI of the Junta de Andaluc´a as research group FQM322, SJ by
NASA under LTSA Grant NAG5-13063 and NSF under AST-0607748 and DHM by
NASA under LTSA Grant NAG5-13102.


\bibliographystyle{mn2e}
\bibliography{ref}

\appendix

\section{Notes on individual objects}\label{app-notes}

\begin{figure}
\centering
\includegraphics[clip,angle=270,width=0.95\hsize]{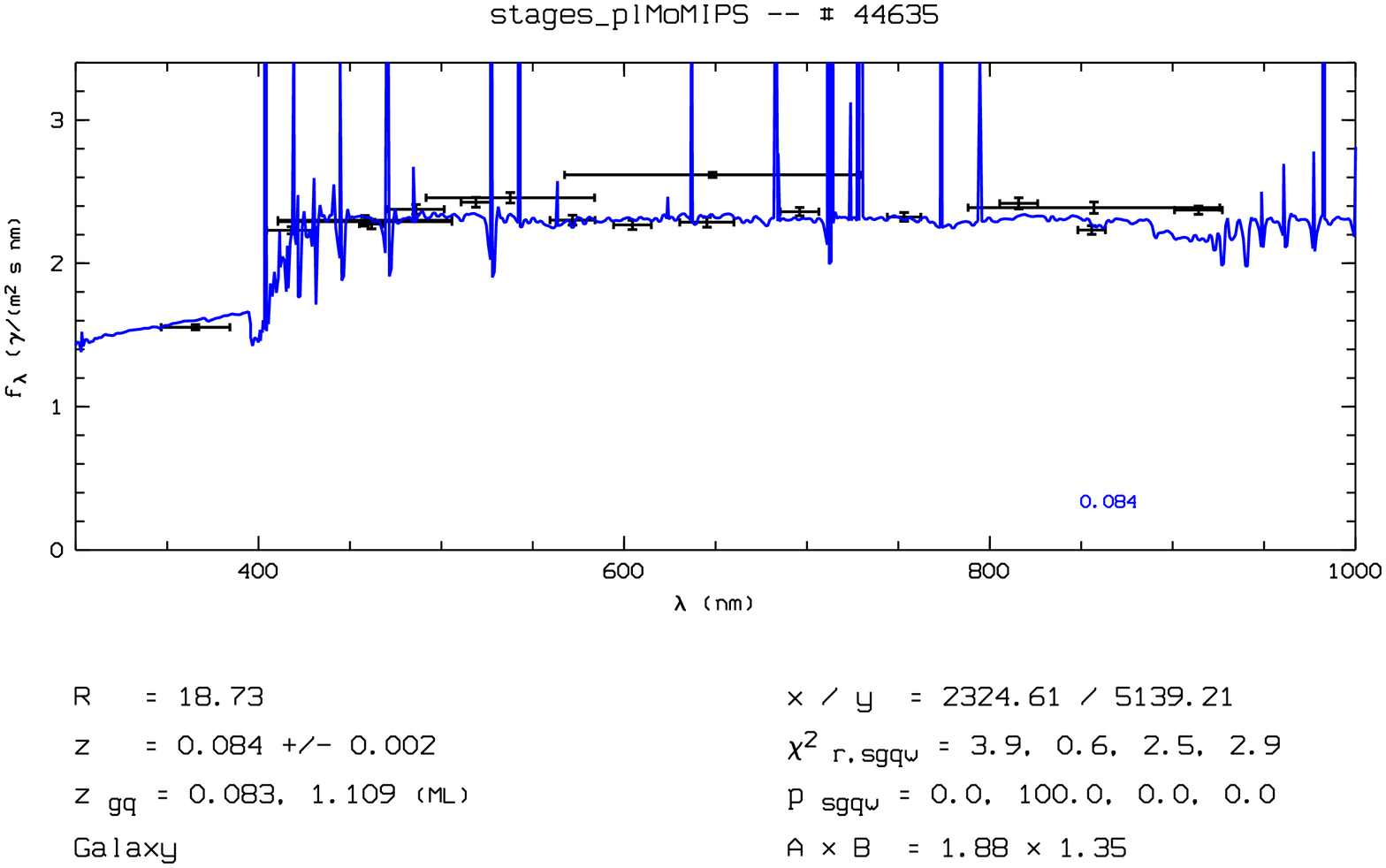}
\includegraphics[clip,angle=270,width=0.95\hsize]{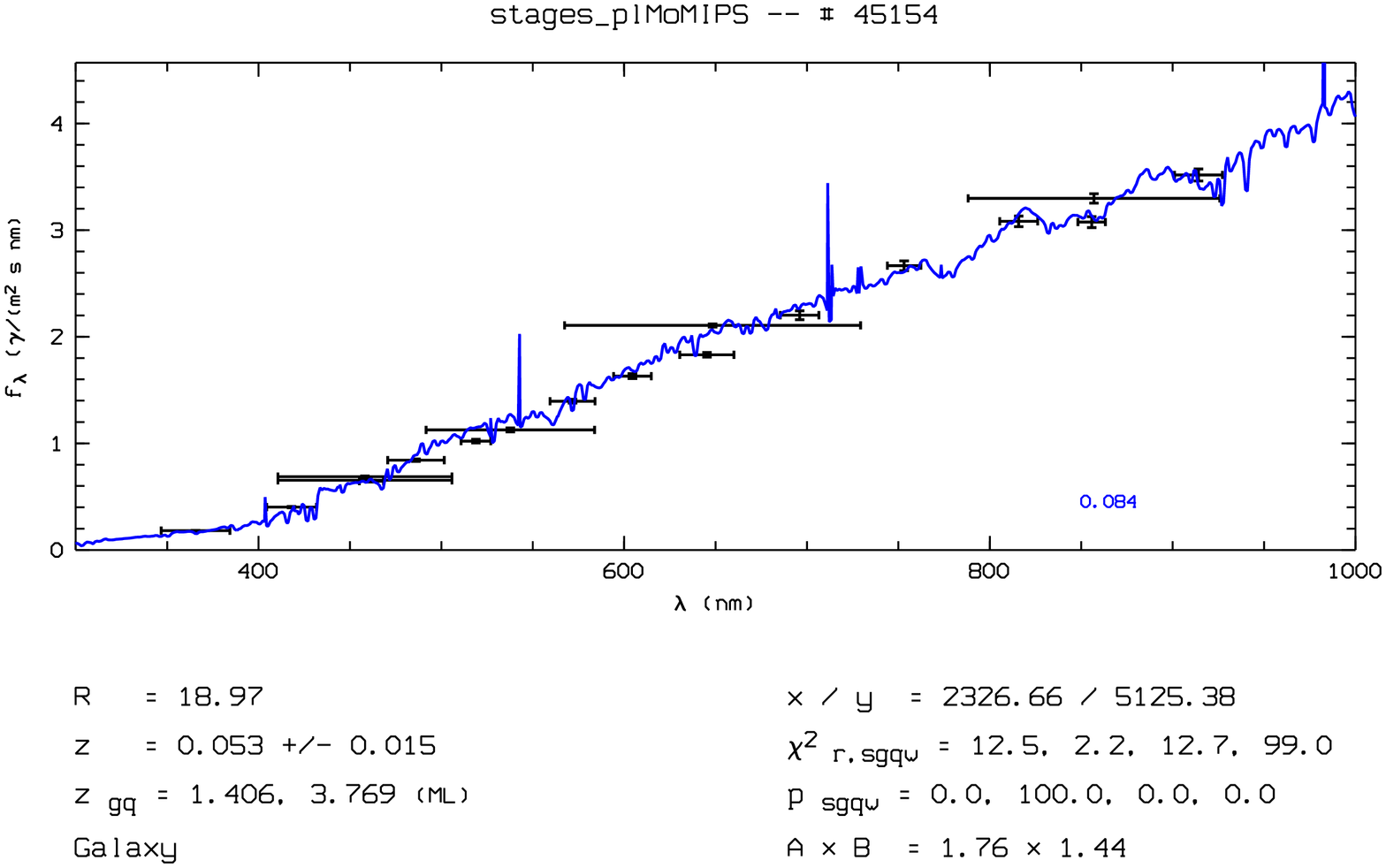}
\caption{The COMBO-17 SEDs of the merging system 44635 ({\it top}) and 
45154 ({\it bottom}). The latter case is a dust-reddened fit by eye to 
$z=0.08$, the likely redshift of the system.
\label{SED_merger}}
\end{figure}
\begin{figure}
\centering
\hbox{
\includegraphics[clip,width=0.49\hsize]{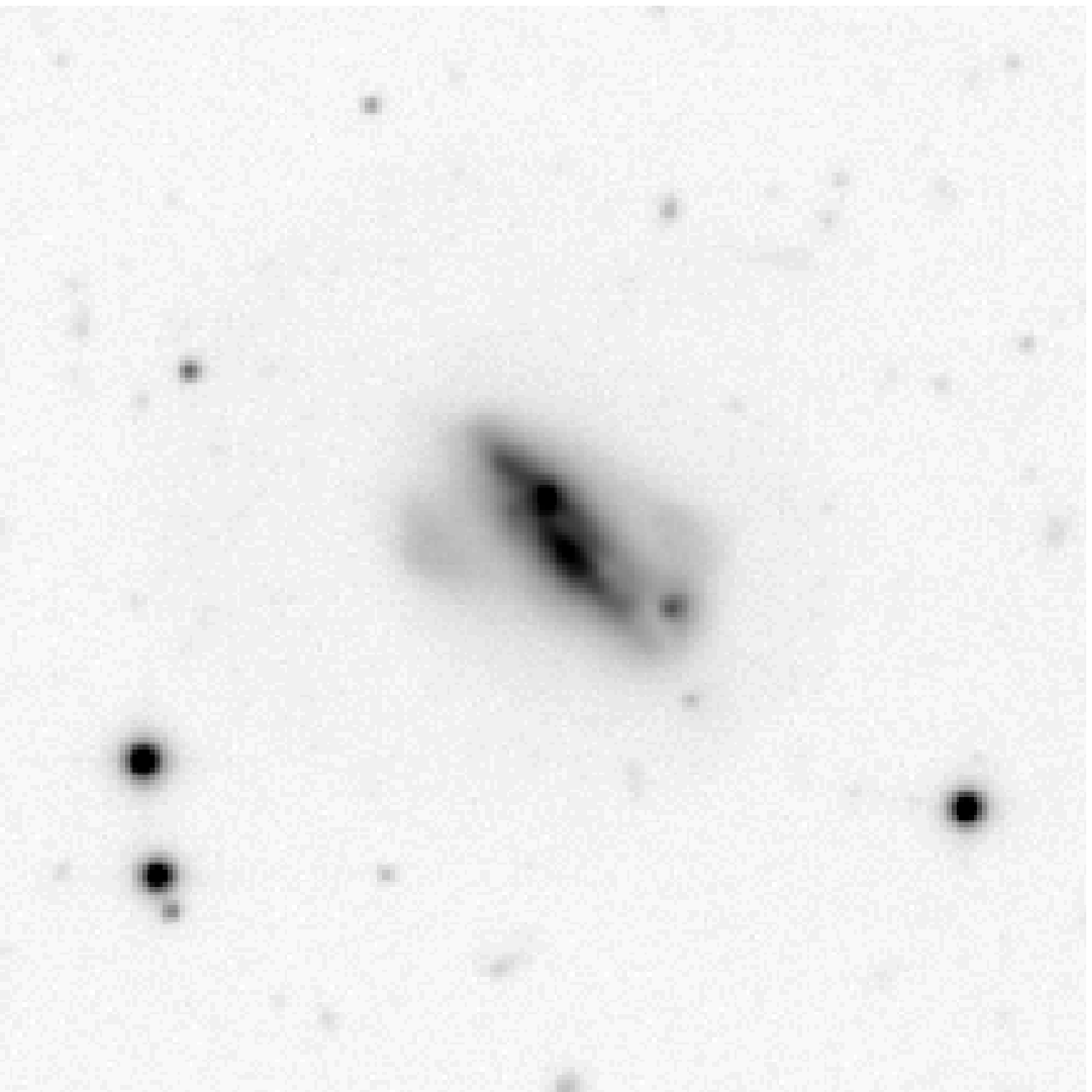}
\includegraphics[clip,width=0.49\hsize]{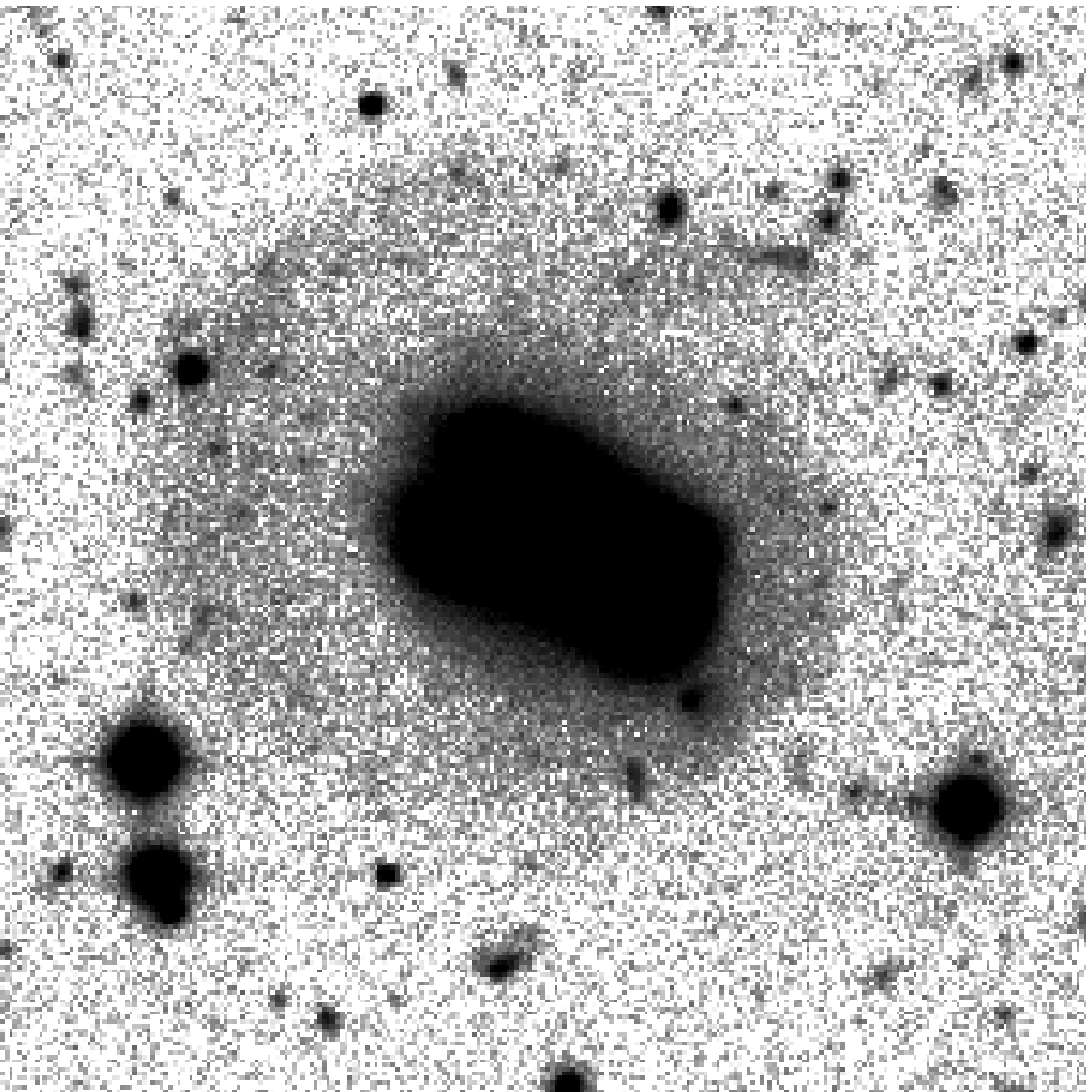}}
\caption{{\it Left panel:} The 20~ks COMBO-17 $R$-band image of the merging 
system that is with $\sim 50$~mJy the brightest extragalactic 24$\mu$ source in 
the field (objects 44635 and 45154, size of image $1\arcmin \times 1\arcmin$, 
N is up, E is left) and missing from the matched catalogue.
{\it Right panel:} The same image in hard cuts reveals a tidal arm with
1/5000th of the surface brightness of the central disks. This arm is too
faint to be visible in the STAGES/HST images.
\label{Im_merger}}
\end{figure}

\begin{figure*}
\centering
\includegraphics[clip,height=4.5cm]{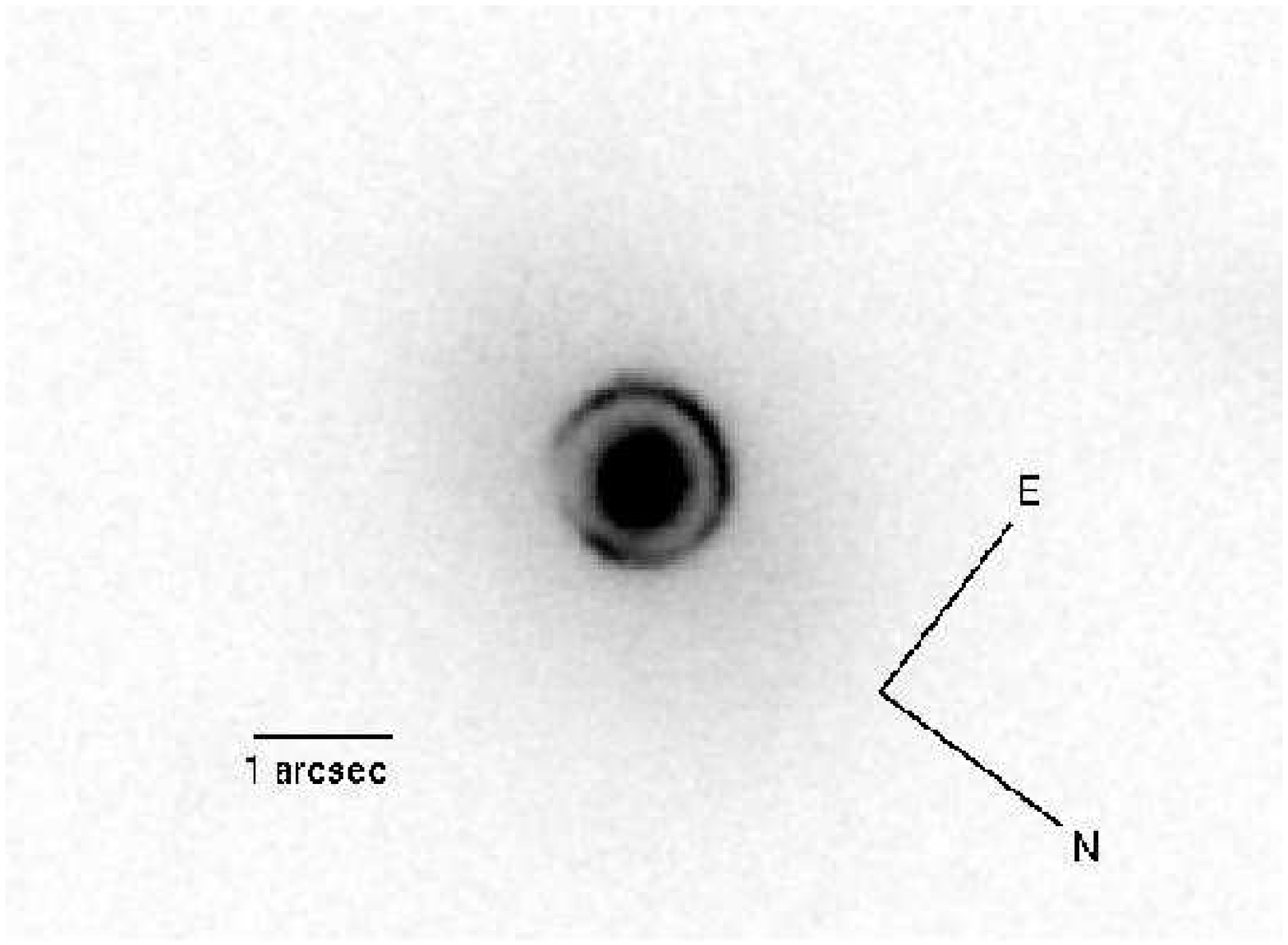}
\hspace{0.cm}\includegraphics[clip,height=4.5cm]{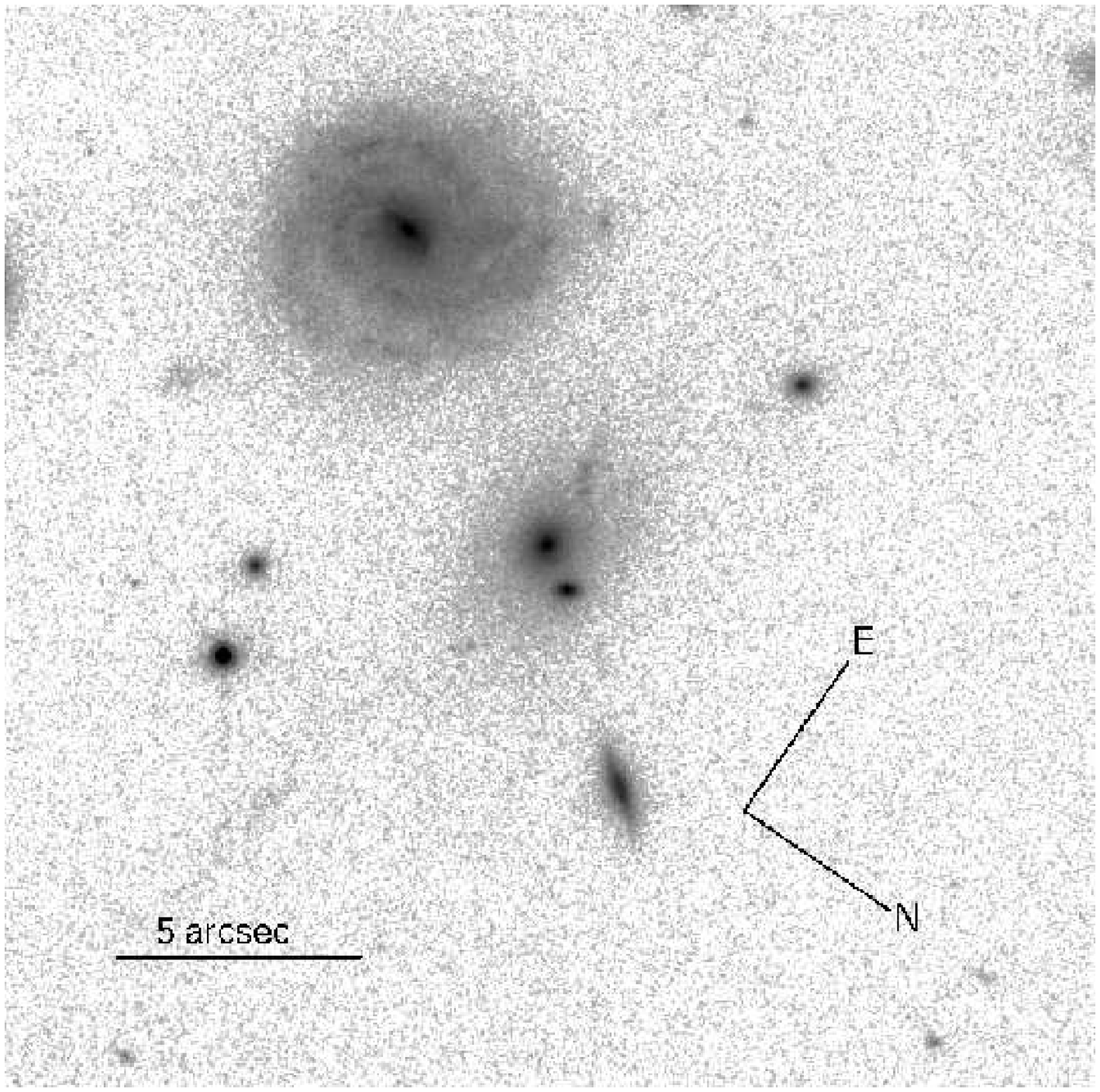}
\hspace{0.cm}\includegraphics[clip,height=4.5cm]{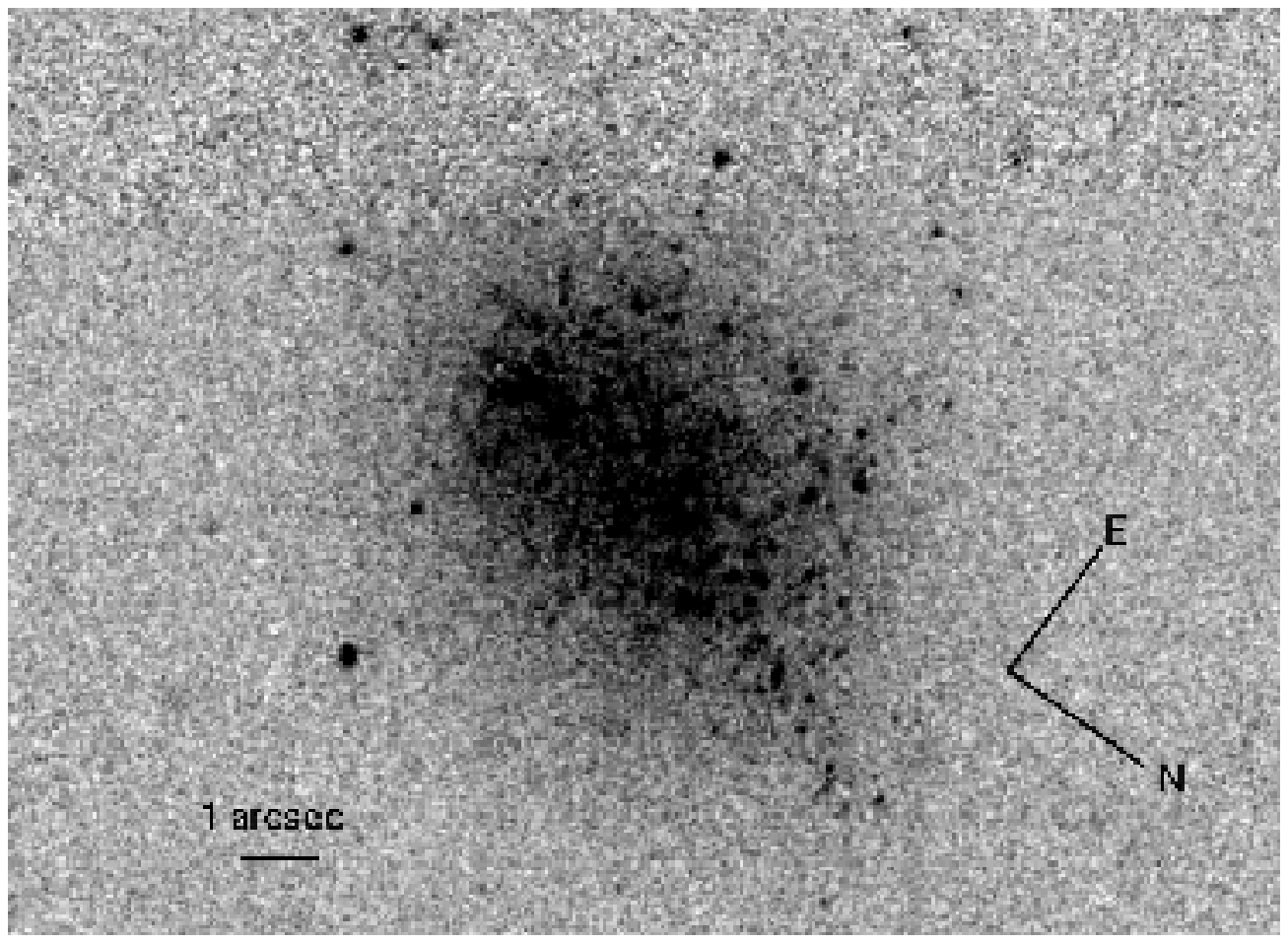}
\caption{{\em Left:} The Einstein ring on an S0 cluster member.  {\em
Centre:} The cD galaxy in CB I at $z\approx 0.47$ is the central
object, while the bright spiral to the upper left is a member of
A902. {\em Right:} The nearby dwarf irregular STAGES I.}
\label{Im_3}
\end{figure*}

\begin{figure}
\centering
\includegraphics[clip,angle=270,width=0.95\hsize]{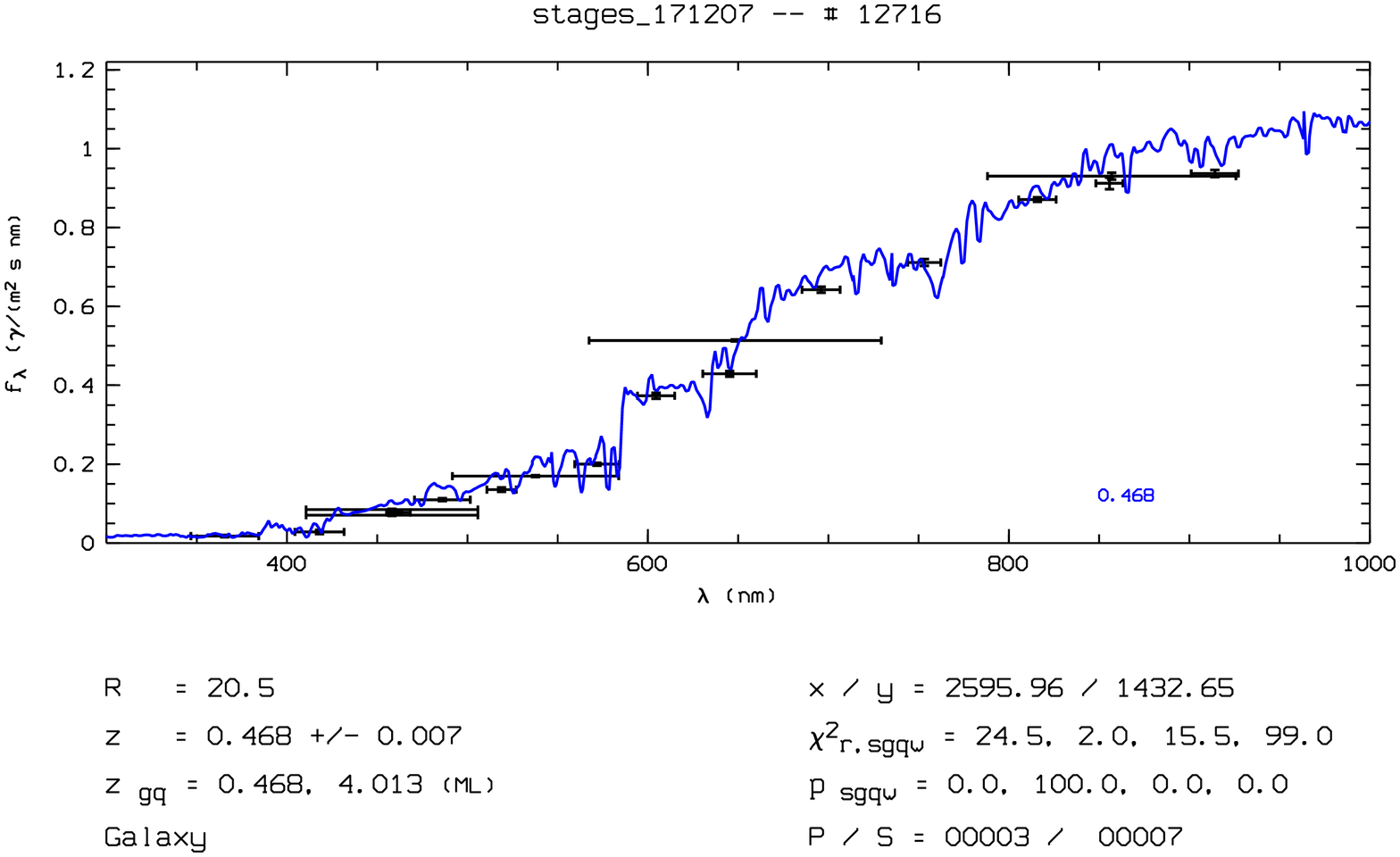}
\caption{The COMBO-17 SED of object 12716, the central dominating
(cD) galaxy of CB I, the cluster at $z\approx 0.47$ in the background of
A902.
\label{SED_cD}}
\end{figure}

\begin{figure}
\centering
\includegraphics[clip,angle=270,width=0.95\hsize]{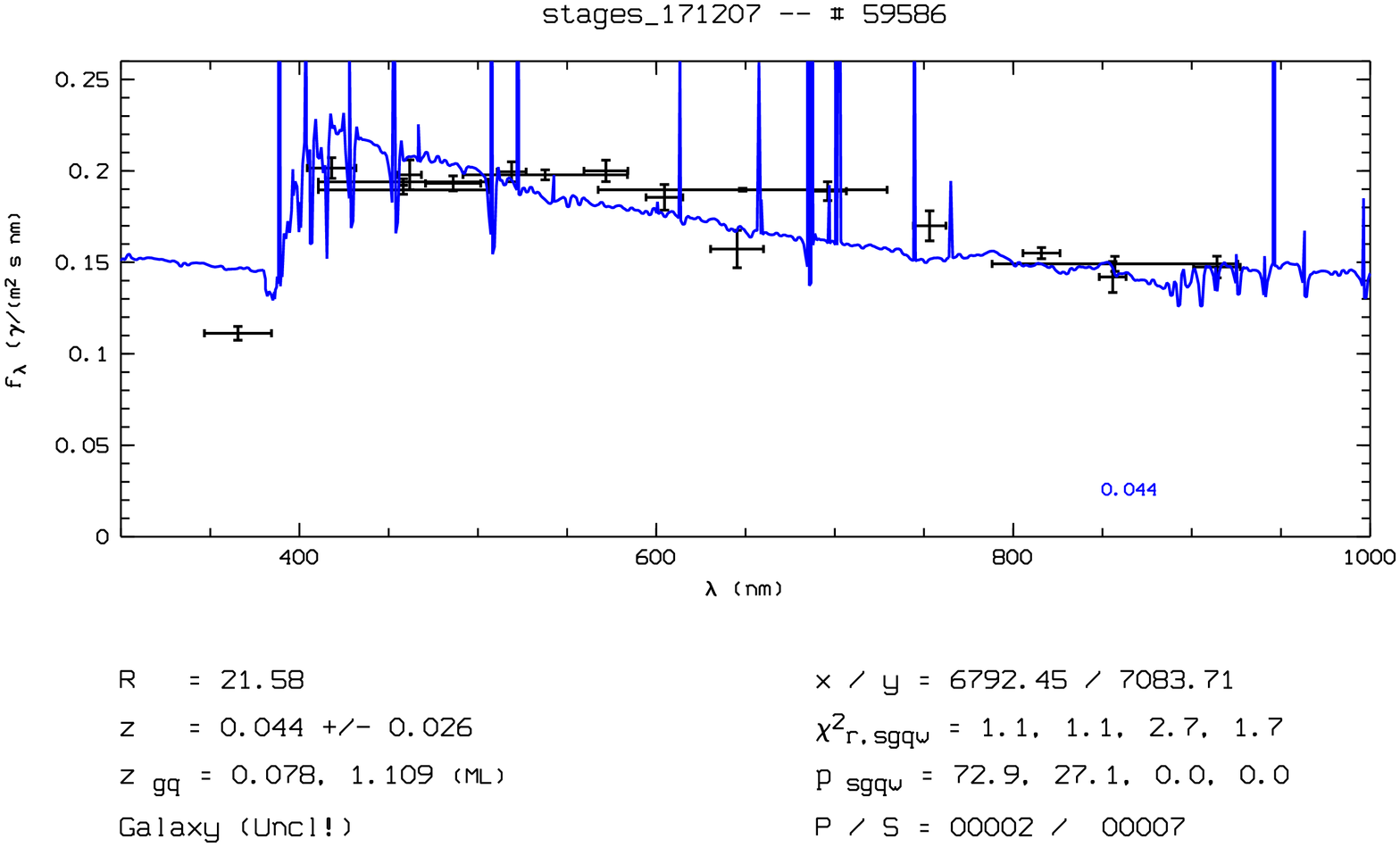}
\caption{The COMBO-17 SED of STAGES I (object 59586), a 
dwarf irregular at $z_{\rm phot}\approx 0.04$ (but likely $z<0.01$).
\label{SED_dSph}}
\end{figure}

\begin{figure}
\centering
\includegraphics[clip,angle=270,width=0.95\hsize]{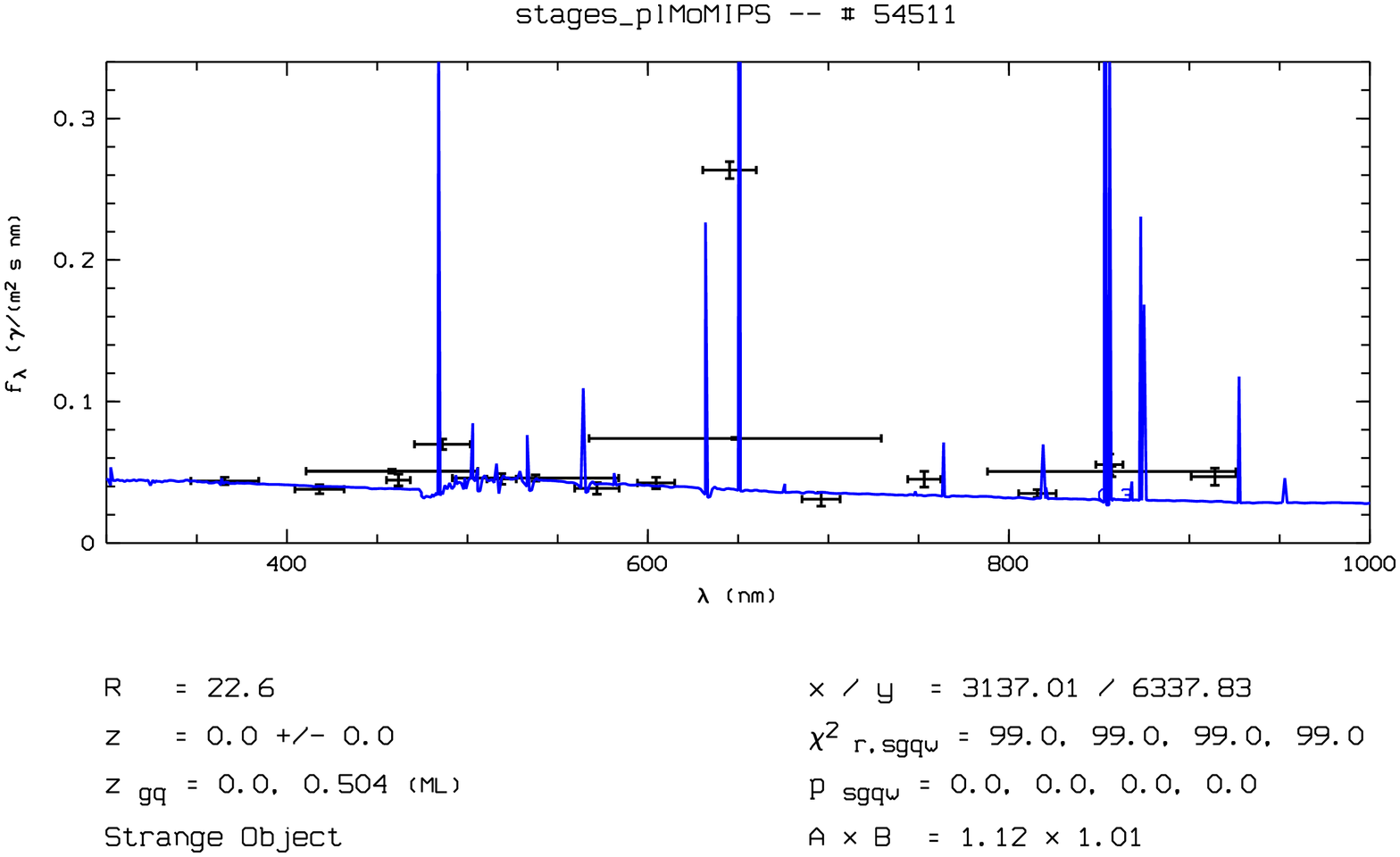}
\caption{The COMBO-17 SED of the only photometrically-classified
`strange' object in the dataset: galaxy 54511 is at $z\approx 0.3$ and
has extremely strong emission lines (OIII+H$\beta$ with $EW\approx
150$~nm).
\label{SED_StrObj}}
\end{figure}

\begin{figure}
\centering
\includegraphics[clip,angle=270,width=0.95\hsize]{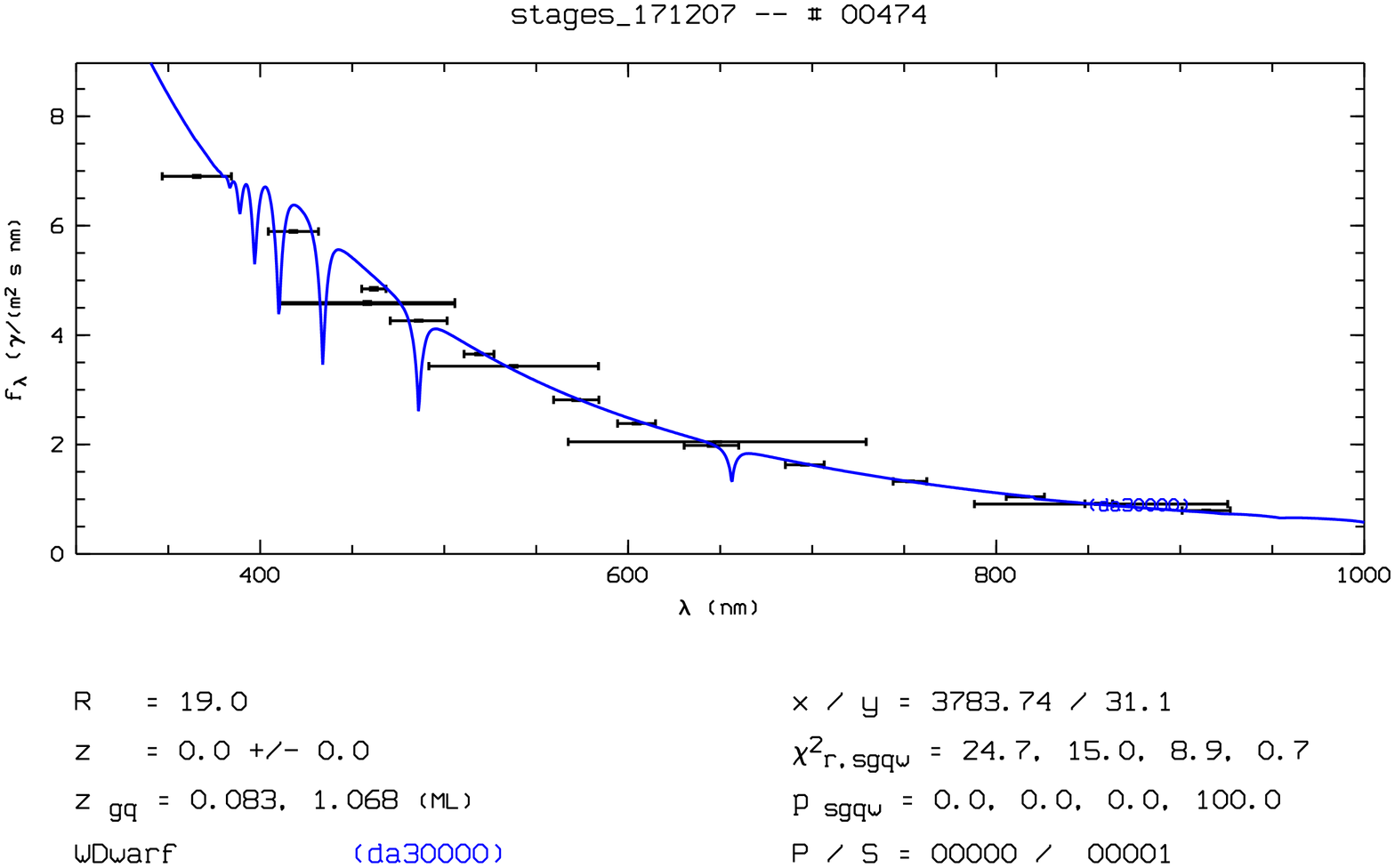}
\caption{The COMBO-17 SED of object 474, the bluest white dwarf 
in the field. The lack of H$\beta$ absorption (see 485 filter) makes it a 
DB white dwarf. The best-fitting temperature is $\sim 30,000$~K.
\label{SED_WD}}
\end{figure}

\begin{figure}
\centering
\includegraphics[clip,angle=270,width=0.95\hsize]{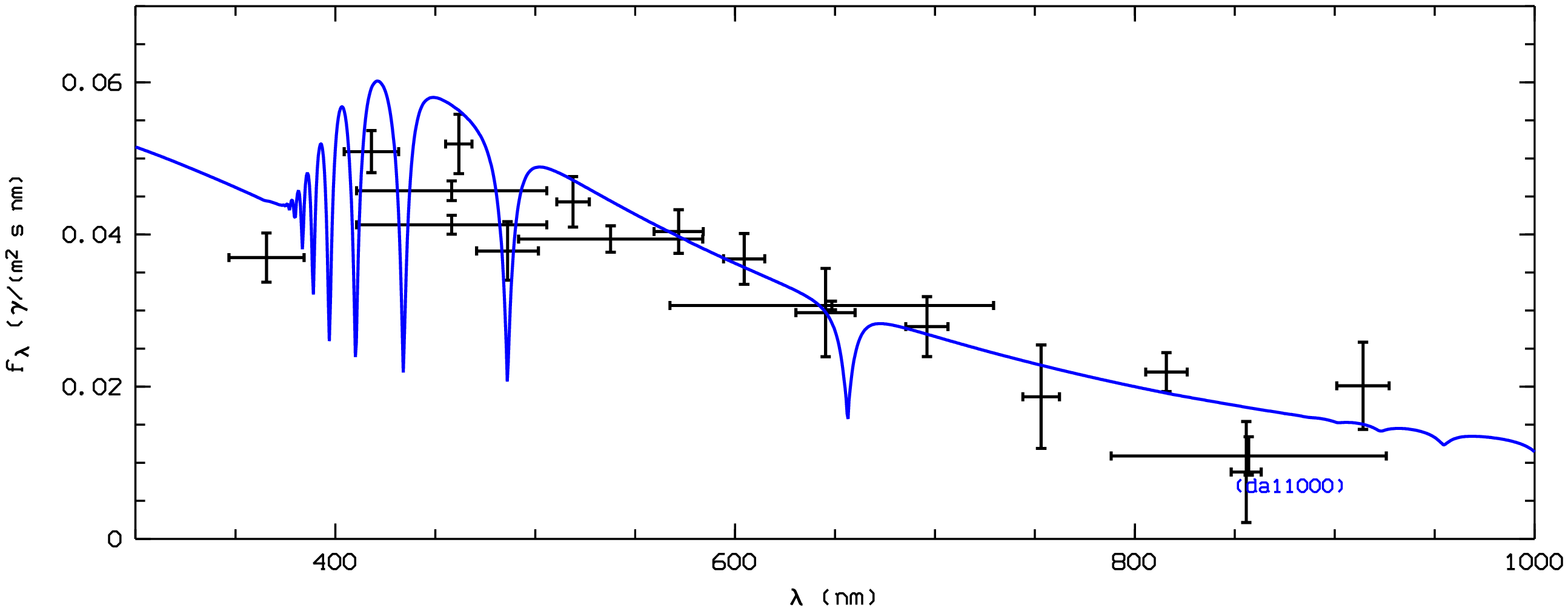}
\caption{The COMBO-17 SED of object 33783, the faintest white dwarf 
in the field. The strong H$\beta$ absorption line in the 485 filter allows 
its classification even at this faint level ($R=23.4$, $T_{\rm eff} \approx 
11,000$~K).
\label{SED_WDfaint}}
\end{figure}

Here, we collect some details on ten noteworthy individual objects
that have either extreme properties or are intrinsically rare and
found only by chance in a field of this size. They are drawn from the
COMBO-17 sample and are identified here via their COMBO-17 object
numbers.

\subsection{The brightest near-infrared source: a Mira variable}\label{sec-mira}

The object with the COMBO-17 number 35250 is classified as a very red
star of spectral type M8~III in the {\it 13th General Catalogue of MK
Spectral Classifications} \citep{Bus98}. It is also known as the IRAS
point source $09540-0946$, located at $(\alpha,\delta)_{\rm
J2000}=(09^h56^m32\fs 4,-10\degr 01 \arcmin 15\arcsec)$, and it
is a ROSAT All-Sky Survey Bright source \citep{voges1999}. It has a very red
SED with $(B,R,J,K)\approx (16,13.3,7.25, 5.75)$ and is the brightest
object in the field at $\lambda>1~\micron$.  However, it has a large
variability amplitude and was identified as a long-period pulsating
Mira star in a search for high-redshift QSOs \citep{KHI97}. The area
around this object had to be excluded from the Spitzer IRAC imaging
due to its high brightness.

\subsection{The brightest far-infrared galaxy: a merger}

The brightest 24$\micron$ galaxy is a system of two merging disk galaxies
with a total magnitude of $R\approx 16$. The Northern system (44635)
has a very blue SED $(U-V)_{\rm rest}=0.14$ and implies a very strong
H$\alpha$ line given its elevated $R$-band flux (see
Fig.~\ref{SED_merger}). The Southern system (45154) has an extremely
red SED $(U-V)_{\rm rest}=1.97$ and implies strong
dust-reddening. Their reshifts are estimated as $z_{\rm phot}=0.084$
and $0.053$, but the blue SED is better constrained by emission lines.
Assuming $z=0.08$ for both objects, the projected separation between
their two nuclei of $3\farcs5$ translates into 5~kpc.

The 20~ks $R$-band image of COMBO-17 shows tidal features with very low
surface brightness (Fig.~\ref{Im_merger}). The arm that reaches once
around the entire galaxy has 5000$\times$ lower surface brightness
than the main disks of the two merging galaxies. The system is also a
strong radio source (NVSS J095643-095544) and was seen by IRAS. In our
Spitzer MIPS images it shows $\sim 50$~mJy of flux, but such bright
FIR measurements are missing from our matched catalogue due to
matching difficulties.  Preliminary analysis of the GMRT data
reveals a strong radio detection at both 1280 MHz and 610 MHz with
total flux S(1280)=$5.63\pm0.05$ mJy/bm and S(610)=$13.68\pm0.05$ mJy.
The radio souce is partially resolved with a deconvolved size of
$4.5\arcsec\times1.9\arcsec$ at 1280 MHz and
$4.1\arcsec\times1.9\arcsec$ at 610 MHz, and a position angle of
40\degr at both frequencies.

\subsection{The brightest X-ray source: a type-I AGN in A901a}\label{sec-agnnote}

Object 41435 is a massive red-sequence elliptical with excess blue
light in its SED \citep[see Fig.~7 of][]{gilmour2007} that has biased
the redshift estimation. While it has $z_{\rm phot}\approx 0.33$, it
is almost certainly a cluster member and a $z=0.16$ template fitted by
hand works well and leaves over some room for AGN light. It is the
brightest X-ray source in the STAGES field observed by XMM and a point
source with a luminosity (assuming $z=0.16$) of $L_X=1.55\times
10^{44}$~erg/s.  It is also the brightest radio source at 1280
MHz and is unresolved with total flux $46.33\pm0.01$ mJy/bm.  At 610
MHz it is partially resolved with an integrated flux density of
$171.3\pm0.1$ mJy.

\subsection{An S0 galaxy with a full Einstein ring}

Object 14049 (Fig.~\ref{Im_3}, left) is an S0 galaxy displaying a full optical
Einstein ring.  It has $z_{\rm phot}=0.23$, but 2dF spectroscopy confirms
it is a cluster member with $z_{\rm spec}=0.168$, implying that the SED
is contaminated by light from the lensed galaxy.
Subsequent targetted spectroscopy revealed a source redshift $z_s=1.5$
(Arag\'{o}n-Salamanca et al, in prep.).

\subsection{A galaxy cluster in projection behind A902: CBI}

Examination of the redshift distribution along the line-of-sight to
the A902 cluster revealed the presence of a massive background cluster
at $z\sim0.47$, subsequently designated CBI (Fig.~\ref{Im_3}, centre).
A 3D lensing approach \citep{taylor04} was used to constrain the
masses of the two clusters beyond the 2D mass reconstruction of
\cite{gray02}. Object 12716 ($R=19.1$) is the central cD galaxy of CBI
and is detected as an unresolved object in the preliminary
analysis of the GMRT data with $S_{\rm
int}(1280\rm{MHz})=2.07\pm0.02$mJy/bm and $S_{\rm
int}(610\rm{MHz})=6.70\pm0.04$ mJy.  Its brighter and bluer close
neighbour ($R=17.8$) is an actual member of A902 (see Fig.~\ref{Im_3},
centre).

\subsection{The dwarf irregular galaxy STAGES I}

The object with the COMBO-17 number 59586 is a nearby dwarf irregular
galaxy (see Fig.~\ref{Im_3}, right and Fig.~\ref{SED_dSph}) estimated
at $z_{\rm phot} = 0.044 \pm 0.026$ (consistent with $z=0$ at
1.6$\sigma$).  At the estimated redshift it would have $M_V\approx
-16.7$ and $\log M_*/M_\odot \approx 8.7$; however given the
brightness of the resolved point sources it is most likely at
$z<0.01$.  It has a \sersic\ index of $n = 0.55$ and shows clear signs
of irregularity besides a blue colour.

\subsection{The galaxy with the strongest emission lines}

The COMBO-17 catalogue contains only one object classified as `strange'
as a result of having a $\chi^2_{\rm red}>30$ for its best template fit,
while having good flags: object 54511 is a galaxy with extremely strong
emission lines and $R\approx 22.5$. The emission-line flux in the $R$-band
and the 646-band both suggest $EW\approx 150$~nm, which would need to be
the combined H$\beta$ and OIII lines. A line in the filter 485/30 shows
$EW\approx 14$~nm and is possibly OII. The redshift of the object appears to
be constrained to $0.27<z_{\rm lines}<0.32$ by a third line signal in the
filter 855 (H$\alpha$; see Fig.~\ref{SED_StrObj}).

\subsection{The bluest white dwarf: $U-B<-1$}

Object 474 is the bluest white dwarf with a satisfying fit to our DA template 
library, although the SED (see Fig.~\ref{SED_WD}) shows clearly no H$\beta$
absorption line, rendering this object a DB. The best-fitting temperature is 
$\sim 30,000$~K.

\subsection{The faintest white dwarf we could identify}

Object number 33783 is the faintest white dwarf our classification can
identify with $U=23.3$ and $R=23.4$. At this magnitude level, the WD
selection is already highly incomplete, but the strong H$\beta$
absorption still constrains the template fit (see Fig.~\ref{SED_WDfaint}).

\section{STAGES master catalogue}

The tables in this section contain information relating to the
publically-available STAGES master catalogue.  Table~\ref{tabcolumns}
lists and defines the column names containing STAGES and COMBO-17 data
and derived stellar masses and star-formation rates.
Table~\ref{sampleflags} details the three sample flags in the catalogue
and describes how they are to be used to select relevant populations
from the overlap between the HST, COMBO-17, and Spitzer datasets.

\begin{table*}
\caption{Column entries in the published FITS catalogue, their headers
and meanings. Some restframe luminosities are extrapolated in some
redshift ranges. We give the redshift intervals, where no
extrapolation errors are expected.
\label{tabcolumns}}
\begin{tabular}[t]{ll}
\hline
\multicolumn{2}{c}{STAGES information} \\
\hline
st\_number &  object number \\
st\_x\_image &  x-position from SExtr in [pix] on tile \\
st\_y\_image &  y-position from SExtr in [pix] on tile \\
st\_cxx\_image &  ellipse parameter from SExtr in [pix] \\
st\_cyy\_image &  ellipse parameter from SExtr in [pix] \\
st\_cxy\_image &  ellipse parameter from SExtr in [pix] \\
st\_theta\_image &  pos. angle from SExtr in [deg] in image \\
	& coordinates (measured from right to up) \\
st\_theta\_world &  pos. angle in [deg] in world coordinates \\
st\_ellipticity &  ellipticity from SExtr \\
st\_kron\_radius &  Kron radius in units of [st\_a\_image] \\
st\_a\_image &  semi-major half-axis from SExtr in [pix] \\
st\_b\_image &  semi-minor half-axis from SExtr in [pix] \\
st\_alpha\_J2000 &  right ascension from SExtr in [deg] \\
st\_delta\_J2000 &  declination from SExtr in [deg] \\
st\_ background &  background value from SExtr in [counts] \\
st\_flux\_best &  ``best'' flux from SExtr in [counts] \\
st\_fluxerr\_best & error of st\_flux\_best \\
st\_mag\_best &  ``best'' magnitude from SExtr in [AB mag] \\
st\_magerr\_best & error of st\_mag\_best \\
st\_flux\_radius &  half-light radius from SExtr in [pix] \\
st\_isoarea\_image &  isophotal area from SExtr in [pix$^2$]\\
st\_fwhm\_image &  FWHM from SExtr in [pix] \\
st\_flags &  SExtr quality flags \\
st\_class\_star &  SExtr stellarity estimator \\
st\_org\_image &  postage stamp image file name \\
st\_file\_galfit &  GALFIT output filename containing fit data \\
st\_X\_galfit &  x-position on postage stamp in [pix] \\
st\_Xerr\_galfit & error of st\_X\_galfit \\
st\_Y\_galfit &  y-position from GALFIT in [pix] \\
st\_Yerr\_galfit & error of st\_Y\_galfit \\
st\_MAG\_galfit &  total magnitude from GALFIT in [AB mag] \\
st\_MAGerr\_galfit & error of st\_MAG\_galfit \\
st\_RE\_galfit &  half-light radius from GALFIT in [pix] \\
st\_REerr\_galfit & error of st\_RE\_galfit \\
st\_N\_galfit &  \sersic\ index from GALFIT \\
st\_Nerr\_galfit & error of st\_N\_galfit \\
st\_Q\_galfit &  major-to-minor axis ratio from GALFIT \\
st\_Qerr\_galfit & error of st\_Q\_galfit \\
st\_PA\_galfit &  pos. angle in [deg] measured from up to left \\
st\_PAerr\_galfit & error of st\_PA\_galfit \\
st\_sky\_galfit &  sky value from GALAPAGOS \\
st\_tile       &  tile number in STAGES mosaic \\
\hline 
\multicolumn{2}{c}{COMBO-17 general information} \\
\hline 
COMBO\_nr       &  COMBO-17 A901/2 field object number \\
ra             &  right ascension (J2000) \\
dec            &  declination (J2000) \\
xpix              &  x-position on COMBO-17 $R$-frame in pixels \\
ypix              &  y-position on COMBO-17 $R$-frame in pixels \\
Rmag           &  total $R$-band magnitude \\
e\_Rmag        &  1-$\sigma$ error of total $R$-band mag \\
ap\_Rmag       &  aperture $R$-band magnitude in run E \\
apd\_Rmag      &  difference total to aperture (point source $\sim 0$) \\
\hline
\multicolumn{2}{c}{Various flags for sample selection} \\ 
\hline
phot\_flag     &  COMBO-17 photometry flags (see Sect.~3.5) \\
combo\_flag   &  COMBO-17 sample flag (see Table~\ref{sampleflags}) \\
stages\_flag   &  STAGES sample flag (see Table~\ref{sampleflags}) \\
mips\_flag   &  MIPS sample flag (see Table~\ref{sampleflags}) \\
\hline
\end{tabular}
\begin{tabular}[t]{ll}
\hline
\multicolumn{2}{c}{COMBO-17 classification results} \\
\hline 
chi2red         &  $\chi^2/N_f$ of best-fitting template \\
chi2reds        &  $\chi^2/N_f$ of best-fitting star template \\
chi2redg        &  $\chi^2/N_f$ of best-fitting galaxy template \\
chi2redq        &  $\chi^2/N_f$ of best-fitting QSO template \\
chi2redw        &  $\chi^2/N_f$ of best-fitting WD template \\
chi2redg\_cl    &  $\chi^2/N_f$ of best-fitting galaxy template at $z=0.167$\\
mc\_class       &  multi-colour class (see Table~\ref{tab-comboclasses}) \\
mc\_z           &  mean redshift in distribution $p(z)$ \\
e\_mc\_z        &  standard deviation (1-$\sigma$) in distribution $p(z)$ \\
mc\_z2          &  alternative redshift if $p(z)$ bimodal \\
e\_mc\_z2       &  standard deviation (1-$\sigma$) at alternative redshift \\
mc\_z\_ml       &  peak redshift in distribution $p(z)$ \\
mc\_Ebmv          &  mean $E(B-V)$ in distribution $p(z)$ \\
e\_mc\_Ebmv       &  standard deviation (1-$\sigma$) in distribution $p($E(B-V)$)$\\
mc\_Ebmv\_ml      &  peak value in distribution $p(E(B-V))$ \\
mc\_age           &  mean template age index  \\
e\_mc\_age        &  standard deviation (1-$\sigma$) of template age index \\
mc\_age\_ml       &  peak in template age index distribution \\
mc\_z\_cl         &  redshift assuming cluster membership \\
mc\_Ebmv\_cl      &  mean $E(B-V)$ assuming cluster membership \\
e\_mc\_Ebmv\_cl   &  standard deviation in $p(E(B-V))$ if cluster member \\
mc\_age\_cl      &  mean age index assuming cluster membership \\
e\_mc\_age\_cl   &  standard deviation in age index if cluster member \\
\hline
\multicolumn{2}{c}{total galaxy restframe luminosities} \\
\hline 
S280Mag         &  $M_{\rm abs,gal}$ in 280/40 ($z\approx [0.25,1.3]$)\\
e\_S280Mag      &  1-$\sigma$ error of $M_{\rm abs,gal}$ in 280/40 \\
UjMag           &  $M_{\rm abs,gal}$ in Johnson $U$ (ok at all $z$)\\
e\_UjMag        &  1-$\sigma$ error of $M_{\rm abs,gal}$ in Johnson $U$ \\
BjMag           &  $M_{\rm abs,gal}$ in Johnson $B$ ($z\approx [0.0,1.1]$)\\
e\_BjMag        &  1-$\sigma$ error of $M_{\rm abs,gal}$ in Johnson $B$ \\
VjMag           &  $M_{\rm abs,gal}$ in Johnson $V$ ($z\approx [0.0,0.7]$)\\
e\_VjMag        &  1-$\sigma$ error of $M_{\rm abs,gal}$ in Johnson $V$ \\
usMag           &  $M_{\rm abs,gal}$ in SDSS $u$ (ok at all $z$)\\
e\_usMag        &  1-$\sigma$ error of $M_{\rm abs,gal}$ in SDSS $u$ \\
gsMag           &  $M_{\rm abs,gal}$ in SDSS $g$ ($z\approx [0.0,1.0]$)\\
e\_gsMag        &  1-$\sigma$ error of $M_{\rm abs,gal}$ in SDSS $g$ \\
rsMag           &  $M_{\rm abs,gal}$ in SDSS $r$ ($z\approx [0.0,0.5]$)\\
e\_rsMag        &  1-$\sigma$ error of $M_{\rm abs,gal}$ in SDSS $r$ \\
\hline 
\multicolumn{2}{c}{restframe luminosities at cluster distance}\\
\hline 
S280Mag\_cl         &  $M_{\rm abs,gal}$ in 280/40 (if cluster member)\\
e\_S280Mag\_cl      &  1-$\sigma$ error of $M_{\rm abs,gal}$ in 280/40 \\
UjMag\_cl           &  $M_{\rm abs,gal}$ in Johnson $U$ (if cluster member)\\
e\_UjMag\_cl        &  1-$\sigma$ error of $M_{\rm abs,gal}$ in Johnson $U$ \\
BjMag\_cl           &  $M_{\rm abs,gal}$ in Johnson $B$ (if cluster member)\\
e\_BjMag\_cl        &  1-$\sigma$ error of $M_{\rm abs,gal}$ in Johnson $B$ \\
VjMag\_cl           &  $M_{\rm abs,gal}$ in Johnson $V$ (if cluster member)\\
e\_VjMag\_cl        &  1-$\sigma$ error of $M_{\rm abs,gal}$ in Johnson $V$ \\
usMag\_cl           &  $M_{\rm abs,gal}$ in SDSS $u$ (if cluster member)\\
e\_usMag\_cl        &  1-$\sigma$ error of $M_{\rm abs,gal}$ in SDSS $u$ \\
gsMag\_cl           &  $M_{\rm abs,gal}$ in SDSS $g$ (if cluster member)\\
e\_gsMag\_cl        &  1-$\sigma$ error of $M_{\rm abs,gal}$ in SDSS $g$ \\
rsMag\_cl           &  $M_{\rm abs,gal}$ in SDSS $r$ (if cluster member)\\
e\_rsMag\_cl        &  1-$\sigma$ error of $M_{\rm abs,gal}$ in SDSS $r$ \\
\hline 
\multicolumn{2}{c}{QSO restframe luminosities} \\
\hline 
S145Mag         &  $M_{\rm abs,QSO}$ in 145/10 ($z\approx [1.4,5.2]$)\\
e\_S145Mag      &  1-$\sigma$ error of $M_{\rm abs,QSO}$ in 145/10 \\
\hline
\end{tabular}
\end{table*}

\begin{table*}
\contcaption{}
\begin{tabular}[t]{ll}
\hline  
\multicolumn{2}{c}{observed seeing-adaptive aperture fluxes}  \\
\hline 
W420f        &  photon flux in filter 420  \\
e\_W420f     &  1-$\sigma$ photon flux error in 420  \\
W462f        &  photon flux in filter 462  \\
e\_W462f     &  1-$\sigma$ photon flux error in 462  \\
W485f        &  photon flux in filter 485  \\
e\_W485f     &  1-$\sigma$ photon flux error in 485  \\
W518f        &  photon flux in filter 518  \\
e\_W518f     &  1-$\sigma$ photon flux error in 518  \\
W571f        &  photon flux in filter 571  \\
e\_W571f     &  1-$\sigma$ photon flux error in 571  \\
W604f        &  photon flux in filter 604  \\
e\_W604f     &  1-$\sigma$ photon flux error in 604  \\
W646f        &  photon flux in filter 646  \\
e\_W646f     &  1-$\sigma$ photon flux error in 646  \\
W696f        &  photon flux in filter 696  \\
e\_W696f     &  1-$\sigma$ photon flux error in 696  \\
W753f        &  photon flux in filter 753  \\
e\_W753f     &  1-$\sigma$ photon flux error in 753  \\
W815f        &  photon flux in filter 815  \\
e\_W815f     &  1-$\sigma$ photon flux error in 815  \\
W856f        &  photon flux in filter 856  \\
e\_W856f     &  1-$\sigma$ photon flux error in 856  \\
W914f        &  photon flux in filter 914  \\
e\_W914f     &  1-$\sigma$ photon flux error in 914  \\
Uf           &  photon flux in filter $U$  \\
e\_Uf        &  1-$\sigma$ photon flux error in $U$  \\
Bf\_A        &  photon flux in filter $B$ in run A \\
e\_Bf\_A     &  1-$\sigma$ photon flux error in $B$/A  \\
Bf\_G        &  photon flux in filter $B$ in run G \\
e\_Bf\_G     &  1-$\sigma$ photon flux error in $B$/G  \\
Vf           &  photon flux in filter $V$  \\
e\_Vf        &  1-$\sigma$ photon flux error in $V$  \\
Rf           &  photon flux in filter $R$  \\
e\_Rf        &  1-$\sigma$ photon flux error in $R$  \\
If           &  photon flux in filter $I$  \\
e\_If        &  1-$\sigma$ photon flux error in $I$  \\
\hline 
\multicolumn{2}{c}{observed aperture Asinh Vega magnitudes}  \\
\hline 
W420magA    &  magnitude in filter 420  \\
e\_W420magA &  1-$\sigma$ magnitude error in  420  \\
W462magA    &  magnitude in filter 462  \\
e\_W462magA &  1-$\sigma$ magnitude error in  462  \\
W485magA    &  magnitude in filter 485  \\
e\_W485magA &  1-$\sigma$ magnitude error in  485  \\
W518magA    &  magnitude in filter 518  \\
e\_W518magA &  1-$\sigma$ magnitude error in  518  \\
W571magA    &  magnitude in filter 571  \\
e\_W571magA &  1-$\sigma$ magnitude error in  571  \\
W604magA    &  magnitude in filter 604  \\
e\_W604magA &  1-$\sigma$ magnitude error in  604  \\
W646magA    &  magnitude in filter 646  \\
e\_W646magA &  1-$\sigma$ magnitude error in  646  \\
W696magA    &  magnitude in filter 696  \\
e\_W696magA &  1-$\sigma$ magnitude error in  696  \\
W753magA    &  magnitude in filter 753  \\
e\_W753magA &  1-$\sigma$ magnitude error in  753  \\
W815magA    &  magnitude in filter 815  \\
e\_W815magA &  1-$\sigma$ magnitude error in  815  \\
W856magA    &  magnitude in filter 856  \\
e\_W856magA &  1-$\sigma$ magnitude error in  856  \\
W914magA    &  magnitude in filter 914  \\
e\_W914magA &  1-$\sigma$ magnitude error in  914  \\
\hline
\end{tabular}
\begin{tabular}[t]{ll}
\hline 
\multicolumn{2}{c}{observed aperture Asinh Vega magnitudes (cont.)}  \\
\hline 
UmagA       &  magnitude in filter $U$  \\
e\_UmagA    &  1-$\sigma$ magnitude error in  $U$  \\
BmagA\_A    &  magnitude in filter $B$ in run A \\
e\_BmagA\_A &  1-$\sigma$ magnitude error in  $B$/A  \\
BmagA\_G    &  magnitude in filter $B$ in run G \\
e\_BmagA\_G &  1-$\sigma$ magnitude error in  $B$/G  \\
VmagA       &  magnitude in filter $V$  \\
e\_VmagA    &  1-$\sigma$ magnitude error in  $V$  \\
RmagA       &  magnitude in filter $R$  \\
e\_RmagA    &  1-$\sigma$ magnitude error in  $R$  \\
ImagA       &  magnitude in filter $I$  \\
e\_ImagA    &  1-$\sigma$ magnitude error in  $I$  \\
\hline 
\multicolumn{2}{c}{stellar masses and star formation rates}  \\
\hline 
logmass		&  log10 of stellar mass \\
logmass\_cl	&  log10 of stellar mass if cluster member \\
flux24		&  MIPS 24$\mu$ flux in microJy \\
tir			&  IR luminosity in $L_\odot$ \\
tuv			&  UV luminosity in $L_\odot$ \\
tir\_cl		&  IR luminosity in $L_\odot$ if cluster member \\
tuv\_cl		&  UV luminosity in $L_\odot$ if cluster member \\
sfr\_det		&  SFR from UV $+$ IR if IR detected \\
sfr\_lo		&  SFR lower limit from UV alone \\
			& (if IR non-detected) \\
sfr\_hi		&  SFR upper limit (if IR non-detected) \\
sfr\_det\_cl	&  SFR if IR detected (if cluster member) \\
sfr\_lo\_cl		&  SFR lower limit from UV alone \\
			& (if no-IR, if cluster member) \\
sfr\_hi\_cl		&  SFR upper limit (if no-IR, if cluster member) \\
sed\_type	&  1=old red, 2=dusty red, 3=blue cloud \\
sed\_type\_cl	&  1=old red, 2=dusty red, 3=blue cloud (if cluster member) \\
\hline
\end{tabular}
\end{table*}

\begin{table*}
\caption{Sample flags in the public FITS catalogue and their
meaning. Note that due to a manual reinspection of COMBO-17
photometric quality flags for this work, the 'WGM05' sample contains 9
fewer objects than the actual published sample of \citet{wgm05}.
However, we retain the name for simplicity.  As an example, to select
objects that are defined by COMBO-17 photometry as galaxies and also
have extended morphologies on the HST imaging, one would require that
combo\_flag $\ge 3$ and stages\_flag $\ge 3$.}
\label{sampleflags} 
\begin{tabular}{lclr}
\hline \hline
Flag & Value & Definition &  \multicolumn{1}{c|}N\\
\hline
STAGES\_FLAG & 0 & not in STAGES footprint (only in COMBO-17) & 6577\\
&1 & in STAGES footprint, but not detected by STAGES (only in COMBO-17)
& 6497\\
&2 & detected by STAGES, but not HST extended source & 5061\\
&3 & HST extended source, but GALFIT ran into constraint & 16123\\
&4 & HST extended source, but GALFIT successful & 54621\\
\hline
COMBO\_FLAG & 0 & not in COMBO-17 footprint (only in STAGES) & 1271\\
&1 & in COMBO-17 footprint, but not detected by COMBO-17 (only in
STAGES) & 23833\\
&2 & detected by COMBO-17, but neither galaxy, nor cluster, nor WGM05
& 48860\\
&3 & galaxy but neither cluster, nor WGM05 & 12625\\
&4 & cluster galaxy, but not WGM05 & 1504\\
&5 & cluster galaxy in WGM05 & 786\\
\hline
MIPS\_FLAG & 0 & detected only by STAGES  & 25104\\
&1 & detected by COMBO-17, but outside MIPS footprint & 11858\\
&2 & detected by COMBO-17 and inside MIPS footprint, but not detected
by MIPS & 48885\\
&3 & detected by COMBO-17 and detected by MIPS & 3032\\
\hline
\end{tabular}
\end{table*}

\bsp \label{lastpage} \end{document}